\documentclass{aa}

\usepackage{amsmath, amssymb, amsfonts, amscd}
\usepackage{mathtools} 
\usepackage{array} 
\usepackage[varg]{txfonts} 
\usepackage{multirow}

\usepackage{graphicx} 

\usepackage{natbib}
\bibpunct{(}{)}{;}{a}{}{,} 

\begin{document}

\title{Scaling laws to understand tidal dissipation\\ in fluid planetary regions and stars\\
I - Rotation, stratification and thermal diffusivity}

\author{
P. Auclair Desrotour\inst{1,2}
\and S. Mathis\inst{2,3}
\and C. Le Poncin-Lafitte\inst{4}
}

\institute{
IMCCE, Observatoire de Paris, CNRS UMR 8028, 77 Avenue Denfert-Rochereau, 75014 Paris, France\\
\email{pierre.auclair-desrotour@obspm.fr}
\and Laboratoire AIM Paris-Saclay, CEA/DSM - CNRS - Universit\'e Paris Diderot, IRFU/SAp Centre de Saclay, F-91191 Gif-sur-Yvette, France
\and LESIA, Observatoire de Paris, CNRS UMR 8109, UPMC, Univ. Paris-Diderot, 5 place Jules Janssen, 92195 Meudon, France\\
\email{stephane.mathis@cea.fr}
\and SYRTE, Observatoire de Paris, CNRS UMR 8630, UPMC, LNE, 61 Avenue de l'Observatoire, 75014 Paris, France \\
\email{christophe.leponcin@obspm.fr}
}

\date{Received ... / accepted ...}

\abstract
{Tidal dissipation in planets and stars is one of the key physical mechanisms driving the evolution of star-planet and planet-moon systems. Several signatures of its action are observed in planetary systems thanks to their orbital architecture and the rotational state of their components.}
{Tidal dissipation inside the fluid layers of celestial bodies are intrinsically linked to the dynamics and the physical properties of the latter. This complex dependence must be characterized.}
{We compute the tidal kinetic energy dissipated by viscous friction and thermal diffusion in a rotating local fluid Cartesian section of a star/planet/moon submitted to a periodic tidal forcing. The properties of tidal gravito-inertial waves excited by the perturbation are derived analytically as explicit functions of the tidal frequency and local fluid parameters (i.e. the rotation, the buoyancy frequency characterizing the entropy stratification, viscous and thermal diffusivities) for periodic normal modes.}
{The sensitivity of the resulting possibly highly resonant dissipation frequency-spectra to a control parameter of the system is either important or negligible depending on the position in the regime diagram relevant for planetary and stellar interiors. For corresponding asymptotic behaviors of tidal gravito-inertial waves dissipated by viscous friction and thermal diffusion, scaling laws for the frequencies, number, width, height and contrast with the non-resonant background of resonances are derived to quantify these variations.}
{We characterize the strong impact of the internal physics and dynamics of fluid planetary layers and stars on the dissipation of tidal kinetic energy in their bulk. We point out the key control parameters that really play a role and demonstrate how it is now necessary to develop ab-initio modeling for tidal dissipation in celestial bodies.}


\titlerunning{Scaling laws to understand tidal dissipation in fluid planetary regions and stars}
\authorrunning{Auclair-Desrotour, Mathis, Le Poncin-Lafitte}

\maketitle


\section{Introduction and context}


Tides have a strong impact on the evolution of star-planet and planet-moon systems over secular time scales. Indeed, because of the dissipation of the kinetic energy of flows and displacements they induce in celestial bodies, they drive their rotational and orbital evolutions (see e.g. \cite{Ogilvie2014} and references therein for the cases of stars and giant planets and \cite{Correia2003,CLL2008} for telluric planets). In the case of the Earth-Moon system, their causes and effects are now strongly constrained and disentangled thanks to satellite altimeter high-precision observations of ocean tides \citep{EgbertRay2000,EgbertRay2001,Rayetal2001}. In the Solar System, the actions of tides are detected and estimated from high-precision geodesic or/and astrometric observations (see e.g. \cite{KY1996} for Venus, \cite{Williamsetal2014} for the Moon, \cite{Lainey2007,Jacobson2010,K2011} for Mars, \cite{Lainey2009,Lainey2012} for Jupiter and Saturn and the attempt by \cite{EN2013} for Uranus). Finally, a new extremely important astronomical laboratory to explore and constrain the physics of tides is constituted by the numerous exoplanetary systems discovered over the last twenty years \citep{MQ1995,Perryman2011}. Indeed, they are composed of a large diversity of planets (from hot Jupiters to super-Earths) and host stars while their orbital architecture and the configuration of planetary and stellar spins strongly differ from the one observed in our Solar system \citep[e.g.][]{Albrecht2012,Fabryckyb2012,VR2014}. In this context, the understanding of the tidal formation and evolution of planetary systems is one of the most important problems of modern dynamical astronomy \citep[e.g.][]{Laskar2012,Bolmont2012,FM2013} while the needed understanding and quantitative prediction of tidal dissipation in celestial bodies is still a challenge \citep[e.g.][for complete reviews]{MR2013,Ogilvie2014}.\\



We owe the first theoretical work about a tidally deformed body to Lord Kelvin \citep{Kelvin1863}. Then, a physical formalism has been elaborated by Love, who introduced the so-called Love numbers \citep{Love1911}. The second-order Love number ($ k_2 $) measures the inverse of the ratio between the tidal potential and the resulting linear perturbation of the gravitational potential at the surface of a body hosting a companion. It quantifies its quadripolar hydrostatic elongation along the line of centers. The introduction and estimation of tidal dissipation then became more and more important with the introduction of the  tidal quality factor $ Q $ for the Solar system \citep[e.g.][]{MacDonald1964,Kaula1964,GS1966}. This general parameter is a simple and useful quantity that allows us to simplify the numerical simulations of planetary systems over secular time scales. In particular, the constant $Q$ model, the so-called Kaula's model \citep{Kaula1964}, allows us to take into account internal dissipation with one parameter only. The factor $ Q $ is applied to rocky/icy and fluid bodies and correspond to a simplified Maxwell's model, which assimilates a celestial body hosting a companion to a forced damped oscillator. It evaluates the ratio between the maximum energy stored in the tidal distortion during an orbital period and the energy dissipated by the friction. Therefore, the global response to a tidal perturbation depends on a mean elasticity, which models the restoring force acting on tidal displacements, and a friction, $ 1/Q $ representing the damping coefficient of the system \citep[i.e. the imaginary part of the second-order Love number; for a detailed study, see][]{Greenberg2009}. From this approach, other models have been developed in order to take into account the possible dependence of $ Q $ on the main tidal frequency $ \chi $, such as the constant time lag model \citep{Darwin1879,Alexander1973,Singer1968,Hut1981,Leconte2010}.\\

As the tidal quality factor is inversely proportional to the internal dissipation in a body, it is determined by the physical properties of this latter. Thus, the internal structure of stars, planets and satellites is determinant for the study of tidal effects. In this context, recent works have derived the tidal quality factor as a function of $ \chi $ and rheological parameters. Examples for rocky/icy planetary layers can be found in \cite{EL2007} and \cite{Henning2009} \citep[see also][]{Tobie2005,Efroimsky2012,Remusanelastic2012,Correiaetal2014}. \cite{EL2007} illustrate the important role played by the internal properties of the perturbed host body in the orbital evolution of its system. They focus on the \textit{equilibrium tide} in the case of a circular orbit, which designates the component of the strain rotating with the perturber at the frequency $ \chi = 2\left(n -\Omega\right)$ in the frame attached to the central body, $ n $ being the mean motion of the perturber in a non-rotating frame and $\Omega$ the spin frequency of the primary. The other components corresponding to eigenmodes of oscillation excited by tides constitute the \textit{dynamical tide}. It is particularly important for fluid planetary regions and stars in which it can dominate the equilibrium tide. It leads to a resonant tidal dissipation that varies over several orders of magnitude as a function of their physical properties, evolutionary states and $\chi$ \citep[see][and references therein]{Ogilvie2014}. Applied to an orbital system, a tidal quality factor following this resonant behavior makes the orbital parameters evolve erratically along secular timescales \citep{ADLPM2014}. An interesting example is given by the case of the Earth-Moon system. Indeed, most of tidal dissipation on the Earth is due to tidal waves in oceans \citep{EgbertRay2000} where the barotropic tide is converted non-resonantly into internal waves because of non-trivial bottom topography \citep{Pedlosky1982}. The corresponding frequency-spectrum of the dissipation is then an highly complex function of $\chi$ while its dependence on the rotation of the planet allows us to match tidal evolution results with geological data. This important result would not be possible with constant $Q$ or constant time lag models \citep{Webb1980,Webb1982}.
During the past five decades, numerous theoretical studies have been carried out to characterize and quantify the tidal energy dissipated in fluid celestial bodies \citep[see e.g.][]{Zahn1966a,Zahn1966b,Zahn1966c,Zahn1975,Zahn1977,Zahn1989a,OL2004,Wu2005,OL2007,RMZ2012,Cebron2012,Cebron2013}. Most of them focused on stars and the envelopes of giant gaseous planets \citep[see][]{Ogilvie2014}. They study global models for tidal waves and dissipative mechanisms in fluid regions excited by a tidal perturber that result from complex actions and couplings of rotation, stratification, viscosity and thermal diffusion. The kinetic energy of tidal waves is thus dissipated by viscous friction and thermal diffusion. Magnetism also intervenes in stars and some planets that introduces Ohmic diffusion in addition to the two previous dissipative mechanisms.\\ 

Therefore, tidal perturbations take the form of waves in fluid regions:
acoustic waves driven by compressibility,
inertial waves driven by the Coriolis acceleration,
gravity waves driven by the Archimedian force,
Alfv\'en waves driven by magnetic forces.
Given their high frequencies, acoustic waves are only weakly excited by low-frequency tidal forcing and shall be ignored. 
Tides will rather excite gravito-inertial waves. These mixed waves combine the second and third families enumerated above. Inertial waves are caused by rotation. The Coriolis acceleration acts as a restoring force and their frequencies are smaller than the inertial one $ 2 \Omega $. Gravity waves propagate in stably-stratified fluid regions, with frequencies bounded by $ N $ (the Brunt-V\"ais\"al\"a frequency) that depends on the gradients of the specific entropy. Their restoring force is buoyancy. Alfv\'en waves result from the presence of a magnetic field and propagate in stars and planets, gravito-inertial waves thus becoming magneto-gravito-inertial waves. From now on, magnetism is neglected in a first step and they are thus not studied in this work.\\


Given the complex dynamical resonant tidal response of fluid planetary layers and stars described above and its important astrophysical consequences, it should be understood and characterized systematically. To reach this goal, two approches can be adopted. First, global realistic models constitute an efficient tool to explore the physics of dissipation and to quantify it \citep[e.g.][]{OL2004,OL2007}. Simultaneously, local simplified models allow to understand in detail complex physical mechanisms in action and to explore the large domain of possible parameters in astrophysics \citep[e.g.][]{O2005,JO2014,Barker2013,Barker2014}. In this framework, \cite{OL2004} proposed in an appendix of their work devoted to quantify global tidal dissipation in spherical shells for planetary interiors a reduced model based on a Cartesian fluid box. The objective of this simplified model was to understand the behavior of tidal dissipation caused by the (turbulent) viscous friction acting on inertial waves in planetary and stellar convective regions. Similarly, \cite{GS2005} explored the complex physics of gravito-inertial waves \citep[see also][]{MNT2014}. Following these previous studies, this work develops a new local Cartesian model that generalizes the one by \cite{OL2004} by taking into account the relative inclination between gravity and rotation, a possible stable entropy stratification or a super-adiabaticity established by convection and viscous and thermal diffusions. By studying this set-up, we answer to the following questions:
\begin{itemize}
   \item[$ \bullet $] how does the dissipation due to the viscous friction and heat thermal diffusion depends on the fluid parameters (i.e. rotation, stratification, viscosity and thermal diffusivity)?
   \item[$ \bullet $] how do the properties of the corresponding dissipation frequency spectrum vary as a function of the latter?
   \item[$ \bullet $] what are the asymptotical behaviors relevant for planetary and stellar interiors? 
\end{itemize}

In section 2, the local Cartesian model is presented. Dynamical equations are derived and solved analytically. It allows us to get the energy dissipated per unit mass in the box over a rotation period. Physical quantities are written as a series of resonant normal modes corresponding to the harmonics of the tidal forcing allowing us to study the spectral response of the box. In section 3, it is studied on the one hand to identify the four asymptotic regimes of tidal waves. On the other hand, we characterize the corresponding resonant dissipation frequency spectrum, i.e. the positions, width, heights, number of peaks, the level of the non-resonant background and the \textit{sharpness ratio} (defined as the ratio between the height of the main peak and the level of the non-resonant background). Each of these properties is expressed as a function of the fluid parameters and forcing frequency. These scaling laws give a global overview of the properties of the dissipation. In section 4, we examine the particular case of super-adiabaticity in convective regions for which the square buoyancy (Brunt-V\"ais\"al\"a) frequency is negative. In section 5, we discuss our results. Finally, in section 6, we give our conclusions and highlight new questions for further studies. 

\section{Forced waves in stars and fluid planetary layers}

\subsection{The local model to be studied}

\cite{OL2004} proposed a local Cartesian model describing the behavior of a fluid box submitted to tidal perturbations for convective regions where inertial waves propagate. This approach is of great interest because it yields an analytical expression for the viscous dissipation of the tidal kinetic energy, which is expressed as a function of the fluid properties and particularly of the Ekman number, $ N_{\rm Ek} = \nu / \left( 2 \Omega L^2 \right) $, $ \Omega $ being the angular velocity of the body, $\nu$ the local kinematic viscosity of the fluid, and $ L $ a characteristic length of the convective layer. So, it allows us to understand how dissipation depends on these physical parameters. Here, in addition to inertial waves driven by the Coriolis acceleration, we also take into account the possibility of a stable stratification that introduces buoyancy as an additional restoring force. This allows us to generalize the previous model to study the more general case of gravito-inertial waves \citep[e.g.][]{GS2005}. Moreover, we also take into account thermal diffusion of heat in addition to viscous friction to dissipate tidal kinetic energy.\\ 

\begin{figure}[htb]
\centering
{\includegraphics[width=0.475\textwidth]
{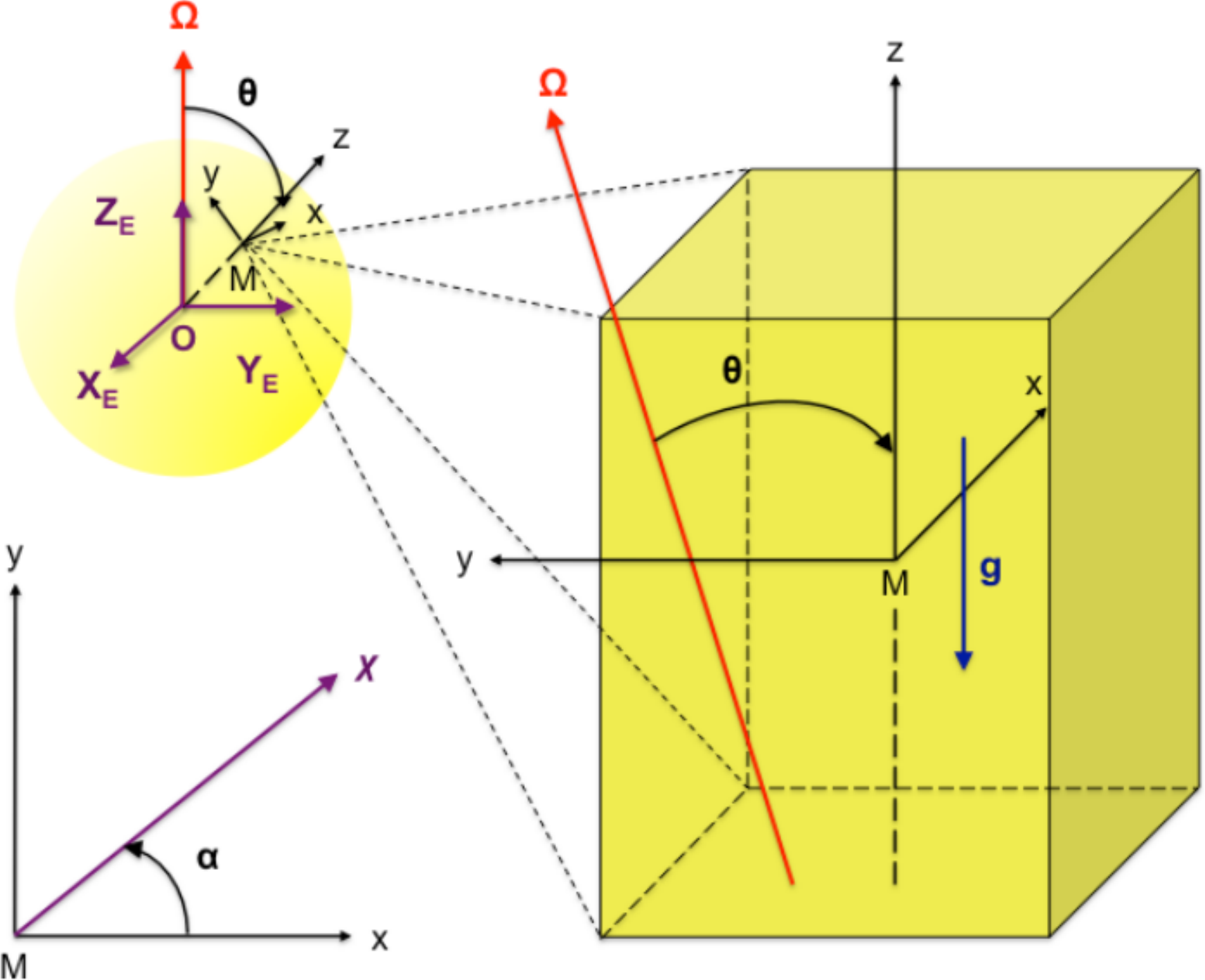}
\textsf{\caption{\label{fig:box} The fluid box, its reference frame and its position in the planet relatively to the spin axis.} }}
\end{figure}

We consider a local fluid Cartesian section belonging to a planet or a star tidally perturbed at a frequency $ \chi $: a box of length $ L $, such that $ L \ll R $, where $ R $ is the radius of the body. The fluid is newtonian, of density $ \rho $, kinematic viscosity $ \nu $ and thermal diffusivity $ \kappa $. The box vertical direction (and gravity ${\vec g}$) is inclined relatively to the spin vector of the body ${\vec \Omega}$ with an angle $ \theta $. The angular velocity of the fluid ($\Omega$) and the local gravity $g$ are assumed to be uniform and constant. The center of the body and of the box are denoted $ O $ and $ M $ respectively. We use two reference frames (Fig. \ref{fig:box}). The global one $ \mathcal{R}_O~:~\left\{ O, \textbf{X}_{\rm E}, \textbf{Y}_{\rm E}, \textbf{Z}_{\rm E}\right\} $ rotates with the body and its natural spherical associated unit-vector basis is denoted $ \left( \textbf{e}_r , \textbf{e}_\theta , \textbf{e}_\varphi  \right) $. In this frame, $ \boldsymbol{\Omega} = \Omega \textbf{Z}_{\rm E}$ and the coordinates of $ M $ are $ \left( r , \theta, \varphi \right) $ in spherical geometry. Then, we introduce the frame fixed to the fluid section, $ \mathcal{R}~:~\left\{ M, \textbf{e}_x , \textbf{e}_y , \textbf{e}_z  \right\} $ whose unit-vectors are $ \textbf{e}_z = \textbf{e}_r  $, $ \textbf{e}_y = -\textbf{e}_\theta $, $\textbf{e}_x = \textbf{e}_\varphi $ and ${\vec g}=-g\textbf{e}_z$. Stratification implies a new frequency, specific to gravito-inertial waves, the Brunt-V\"ais\"al\"a frequency denoted $ N $ and defined by

\begin{equation}
N^2 = - g \left[ \dfrac{d \log \rho}{dz}  - \frac{1}{\gamma}  \dfrac{d \log P}{ dz }    \right],
\end{equation}

where $ \gamma = \left( \partial \ln P / \partial \ln \rho \right)_S $ is the adiabatic exponent ($S$ being the specific macroscopic entropy), and $ P $ and $ \rho $ the radial distributions of pressure and density of the background respectively. These equilibrium quantities are supposed to vary smoothly with the radial coordinate $ z $ compared to the perturbation. For this reason, the Brunt-V\"ais\"al\"a frequency is taken as a constant parameter. Assuming the hydrostatic equilibrium, it becomes 

\begin{equation}
N^2 = - \frac{g}{\rho} \left[ \dfrac{d \rho }{dz} + \frac{g \rho^2}{\gamma P} \right]. 
\end{equation}

The regions studied are stably stratified ($ N^2 > 0 $) or convective ($ N^2 \approx 0 $ or $ N^2 < 0 $). 

\begin{figure}[htb]
\centering
{\includegraphics[width=0.495\textwidth]
{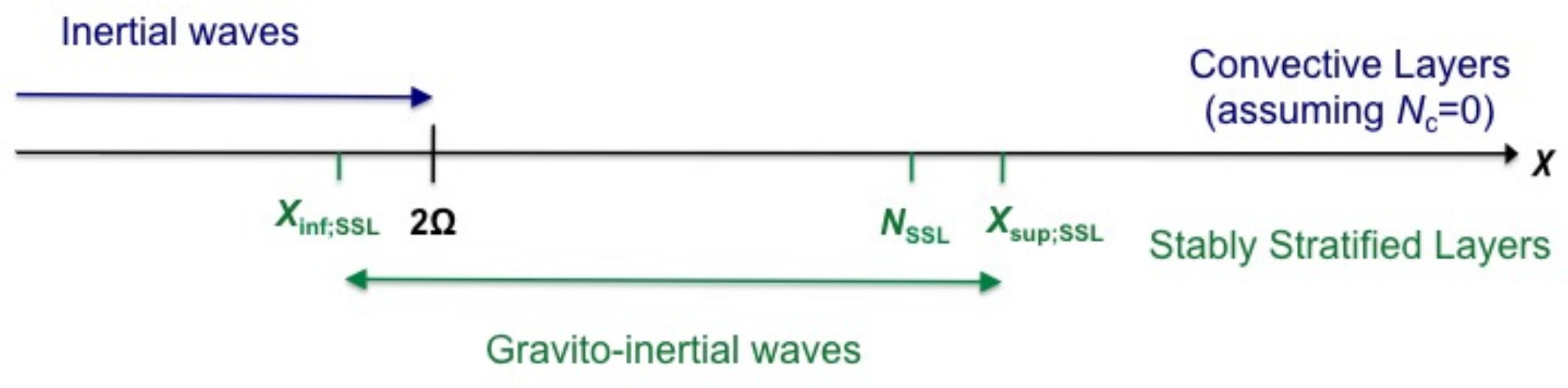}
\textsf{\caption{\label{fig:spectre_ondes} Gravito-inertial waves frequency range in Convective and in Stably Stratified Layers ($N_{\rm c}=0$ and $N_{\rm SLL}>0$ are the corresponding buoyancy frequencies). The expressions of the boundaries $ \chi_{\rm inf;SSL} $ and $ \chi_{\rm sup;SSL} $ are explicitly given in Eq. (\ref{omega_GS}).} }}
\end{figure}

\subsection{Dynamical equations}

Submitted to a tidal forcing massic force ${\vec F}=\left(F_x ,F_y,F_z\right)$, where ${\vec x}\equiv(x,y,z)$ are the local coordinates of space, the fluid moves in the box periodically and dissipates energy through viscous friction and thermal diffusion. To obtain the equations that govern the dynamics of tidal gravito-inertial waves, we write the linearized equations of motion\footnote{In this work, we do not take into account the nonlinear interactions between tidal waves while those with turbulent flows such as convection are treated as an effective viscosity \citep[][]{GK1977,Zahn1989a,OL2012}.} of the stratified fluid on the non-traditional $f$-plane assuming the Boussinesq\footnote{Tidal frequencies are most of the time small compared to the Lamb's frequency. This allows us to filter out acoustic waves and to adopt the anelastic approximation in which $\vec\nabla\cdot\left(\rho\,{\vec u}\right)=0$. In the case where the background density $\rho$ is uniform and constant, it reduces to the Boussinesq approximation in which ${\vec\nabla}\cdot{\vec u}=0$.} and the Cowling approximations \citep{GS2005,Cowling1941}. First, we introduce the wave velocity field $\vec u=\left(u,v,w\right)$, where $u$, $v$ and $w$ are its components in the local azimuthal, latitudinal, and vertical directions. Next, we define the fluid buoyancy
\begin{equation}
{\vec B}=B{\vec e}_{z}=-{g}\frac{\rho^{'}\left(\vec x,t\right)}{\rho}{\vec e}_{z}\,,
\label{buoyancy}
\end{equation}
where $\rho^{'}$ and $\rho$ are the density fluctuation and the reference background density, and $t$ is the time. The three linearised components of the momentum equation are given by
\begin{equation}
\left\lbrace
\begin{array}{lcl}
\displaystyle{\partial_{t}w-2\Omega\sin\theta\,u=-\frac{1}{\rho}\partial_{z}p^{'}+B+\nu\nabla^{2}w+F_{z}}\\
\displaystyle{\partial_{t}v+ 2\Omega\cos\theta\,u=-\frac{1}{\rho}\partial_{y}p^{'}+\nu\nabla^{2}v+F_{y}}\\
\displaystyle{\partial_{t}u- 2\Omega\cos\theta\,v+2\Omega\sin\theta\,w=-\frac{1}{\rho}\partial_{x}p^{'}+\nu\nabla^{2}u+F_{x}}
\end{array}\right.\,,
\label{eq:Dyn}
\end{equation}
where $p^{'}$ is the pressure fluctuation and $\nabla^2\equiv\partial_{x,x}+\partial_{y,y}+\partial_{z,z}$. Next, we write the continuity equation in the Boussinesq approximation
\begin{equation}
{\vec\nabla}\cdot{\vec u}=\partial_{z}w+\partial_{y}v+\partial_{x}u=0.
\label{continuity}
\end{equation}
Finally, we get the equation for energy conservation
\begin{equation}
\partial_{t}B+N^{2}w=\kappa\nabla^{2}B.
\label{eq:Therm}
\end{equation}

By taking the scalar product of the momentum equation (Eq. \ref{eq:Dyn}) with $\vec u$ and multiplying the heat equation (Eq. \ref{eq:Therm}) by $B/N^2$, we obtain the energy equation for the box
\begin{equation}
\partial_{t}\left(E_c+E_p\right)=-\int_{V}\vec\nabla\cdot\left(p^{'}\vec u\right){\rm d}V+D^{\rm visc}+D^{\rm therm}+E^{\rm forcing},
\label{eq:Energy}
\end{equation}
where we have introduced 
\begin{itemize}
\item[$ \bullet $] the kinetic energy
\begin{equation}
E_c=\frac{1}{2}\int_{V}\rho\,{\vec u}^2 {\rm d} V,
\end{equation}
\item[$ \bullet $] the potential energy associated with stratification
\begin{equation}
\begin{cases}
E_p=\displaystyle{\frac{1}{2}\int_{V}\rho\,\frac{B^2}{N^2}{\rm d} V}\quad\hbox{if}\quad N^2\ne0\\
E_p=0\quad\hbox{if}\quad N^2=0
\end{cases},
\end{equation}
\item[$ \bullet $] the energy dissipated by viscous friction
\begin{equation}
D^{\rm visc}=\int_{V}\rho\left(\nu\,\vec u\cdot\nabla^2\vec u\right) {\rm d} V,
\label{eq:Dvisc}
\end{equation}
\item[$ \bullet $] the energy dissipated by thermal diffusion
\begin{equation}
\begin{cases}
D^{\rm therm}=\displaystyle{\int_{V}\rho\left(\frac{\kappa}{N^2}B\,{\nabla}^{2}B\right){\rm d} V}\quad\hbox{if}\quad N^2\ne0\\
D^{\rm therm}=0\quad\hbox{if}\quad N^2=0
\end{cases},
\label{eq:Dtherm}
\end{equation}
\item[$ \bullet $] the energy injected by the forcing
\begin{equation}
E^{\rm forcing}=\int_{V}\rho\left(\vec u\cdot\vec F\right){\rm d}V,
\end{equation}
\end{itemize}
where $V$ is the volume of the box. The first term on the right-hand side of Eq. (\ref{eq:Energy}) is the divergence of the acoustic flux.\\

We introduce dimensionless time and space coordinates and tidal frequency, normalized buoyancy and force per unit mass

\begin{equation}
\begin{array}{c c c c c}
   T = 2 \Omega t, & X = \displaystyle \frac{x}{L},& Y=\displaystyle \frac{y}{L}, & Z = \displaystyle \frac{z}{L}, & \omega = \displaystyle \frac{\chi}{2 \Omega},  \\
   \vspace{0.1mm}\\
   \textbf{b} = \displaystyle \frac{\textbf{B}}{2 \Omega}, & \textbf{f} = \displaystyle \frac{ \textbf{F}}{2 \Omega}.\\
\end{array}
\end{equation}
The linearized Navier-Stockes equation becomes 
\begin{equation}
\partial_{T}{\vec u} + \textbf{e}_z \wedge {\vec u} + \frac{1}{2 \Omega L \rho} {\vec\nabla} p^{'} - N_{\rm Ek} \nabla^2 \textbf{u} - \textbf{b} = \textbf{f},
\label{NavierStokes}
\end{equation}
where we recognize the Ekman number 
\begin{equation}
N_{\rm Ek} = \displaystyle \frac{\nu}{2 \Omega L^2},
\label{eq:Ekman}
\end{equation}
which compare the strength of the viscous force to the Coriolis acceleration.\\

The equation of heat is written
\begin{equation}
\partial_{T} b + A w = N_{\rm diff}  \nabla^2 b, 
\label{heat_transport}
\end{equation}
where the right-hand side corresponds to thermal diffusion. We introduce here two new control parameters
\begin{equation}
\begin{array}{ccc}
A = \left( \displaystyle \frac{N}{2 \Omega}  \right)^2 & \mbox{and} & N_{\rm diff} = \displaystyle \frac{\kappa}{2\Omega L^2};
\end{array}
\label{eq:strat}
\end{equation}
$ A $ is the square ratio of the characteristic frequencies of the system ($2\Omega$ and $N$). This parameter is bound to the nature of tidal waves \citep[e.g.][see also \cite{Berthomieuetal1978} and \cite{Provostetal1981}]{GS2005,GS2005b}:
\begin{itemize}
  \item[$ \bullet $] $ A \leq 0 $ (i.e. $N^2\leq 0$) corresponds to inertial waves,\\
  \item[$ \bullet $] $ 0 < A < 1 $, corresponds to quasi-inertial waves,\\
  \item[$ \bullet $] $ A \geq 1 $, corresponds to gravito-inertial waves.\\
\end{itemize}
The parameter $ N_{\rm diff} $ compares the strength of the Coriolis and thermal diffusion terms by analogy with the Ekman number defined above (Eq. \ref{eq:Ekman}). Finally, we introduce the Prandtl number of the fluid $ P_r $ that characterizes the relative strength of viscous friction and thermal diffusion:
\begin{equation}
P_{r} = \frac{\nu}{\kappa} = \frac{N_{\rm Ek}}{N_{\rm diff}}.
\label{eq:pr}
\end{equation}
 Waves are damped by viscous diffusion if $ P_{r} \ge 1 $; otherwise, thermal diffusion dominates.\\
 
From now on, we assume that the unknown quantities vary with $ x $ and $ z $ only ($ x \in \left[ 0 , L \right] $, $ z \in \left[ 0 , L \right] $). This assumption has no impact on qualitative results and allows us to compare our calculations to the ones made by \cite{OL2004} directly.\\ 

\subsection{Velocity field and dissipation}

We now solve our linearized first-order system (Eqs. \ref{NavierStokes}, \ref{continuity} and \ref{heat_transport})  to get the velocity field and the energy dissipated by viscous friction and thermal diffusion. The tidal perturbation being periodic in time, we expand the quantities in Fourier series:
 
\begin{equation}
\begin{array}{c c }
   u_x = \Re \left[ u(X,Z) e^{-i \omega T}  \right], & u_y = \Re \left[ v(X,Z) e^{-i \omega T}  \right], \\ 
   u_z = \Re \left[ w(X,Z) e^{-i \omega T}  \right], & p = \Re \left[ \psi (X,Z) e^{-i \omega T}  \right], \\
   f_x = \Re \left[ f(X,Z) e^{-i \omega T}  \right], & f_y = \Re \left[ g (X,Z) e^{-i \omega T}  \right], \\
   f_z = \Re \left[ h(X,Z) e^{-i \omega T}  \right], & b = \Re  \left[ b(X,Z) e^{-i \omega T}  \right].\\
\end{array}
\end{equation}

The spatial functions are themselves expanded in periodic spatial Fourier series in $X$ and $Z$ as in \cite{OL2004}:

\begin{equation}
\begin{array}{c c }
   u = \displaystyle \sum u_{mn} e^{i 2 \pi \left( m X + n Z  \right) }, & v = \displaystyle \sum v_{mn} e^{i 2 \pi \left( m X + n Z  \right) },\\
   \vspace{0.1mm}\\
   w = \displaystyle \sum w_{mn} e^{i 2 \pi \left( m X + n Z  \right) }, & \psi = \displaystyle \sum \psi_{mn} e^{i 2 \pi \left( m X + n Z  \right) },\\
   \vspace{0.1mm}\\
   f = \displaystyle \sum f_{mn} e^{i 2 \pi \left( m X + n Z  \right) }, & g = \displaystyle \sum g_{mn} e^{i 2 \pi \left( m X + n Z  \right) },\\
   \vspace{0.1mm}\\
   h = \displaystyle \sum h_{mn} e^{i 2 \pi \left( m X + n Z  \right) }, & b = \displaystyle \sum b_{mn} e^{i 2 \pi \left( m X + n Z  \right) }.\\
\end{array}
\label{fourier}
\end{equation}

For the sake of simplicity, the horizontal and vertical wave numbers $ \left( m,n \right) \in \mathbb{Z}^{*2} $ are not written under summations. Note that by choosing periodic boundary conditions, we consider the volumetric excitation of normal modes \citep[][]{Wu2005,BO2015}. The case of periodic wave attractors can be treated by introducing reflexions on boundaries of the box \citep[e.g.][]{JO2014}. The previous system becomes:

\begin{equation}
\left\{
\begin{array}{rcl}
  - i\omega u_{mn} - \cos \theta v_{mn} + \sin \theta w_{mn} & \!\!\!\!\!\!=\!\!\!\!\!\! & - i m \Lambda \psi_{mn} - E \left( m^2 + n^2 \right)u_{mn} \\
   & & + f_{mn}\\
   \vspace{0.1mm}\\
  - i\omega v_{mn} + \cos \theta u_{mn} & \!\!\!\!\!\!=\!\!\!\!\!\! & - E \left( m^2 + n^2 \right)v_{mn} + g_{mn}\\
  \vspace{0.1mm}\\
  - i \omega w_{mn} - \sin \theta u_{mn} & \!\!\!\!\!\!=\!\!\!\!\!\! & - i n \Lambda \psi_{mn} - E \left( m^2 + n^2 \right)w_{mn} \\
  & &  + b_{mn} + h_{mn} \\
  \vspace{0.1mm}\\
  m u_{mn} + n w_{mn} & \!\!\!\!\!\!=\!\!\!\!\!\! & 0\\
  \vspace{0.1mm}\\
  - i \omega b_{mn} + A w_{mn} & \!\!\!\!\!\!=\!\!\!\!\!\! & K \left( m^2 + n^2 \right) b_{mn}
\end{array}
\right.
\label{systeme_eq},
\end{equation}

which is parametrized by the frequency ratio $ A $, the colatitude $ \theta $ and the numbers:

\begin{equation}
\begin{array}{cccc}
 E = \displaystyle \frac{2 \pi^2 \nu}{ \Omega L^2}, & K = \displaystyle \frac{2 \pi^2 \kappa }{\Omega L^2 } & \mbox{and} & \Lambda = \displaystyle \frac{\pi}{\Omega  \rho L}.
\end{array}
\label{parameters}
\end{equation}

$ E $ and $ K $ are both dimensionless diffusivities. $ E $ is proportional to the Ekman number $ N_{\rm Ek} $ (Eq. \ref{eq:Ekman}) of the fluid and $ K $ to its analogue $ N_{\rm diff} $ (Eq. \ref{eq:strat}) for thermal diffusion. Similarly, we get for the Prandtl number (Eq. \ref{eq:pr})

\begin{equation}
P_r =  \displaystyle \frac{E}{K}.
\end{equation}

Finally, $ \Lambda $ weights the pressure variations. This last parameter does not intervene in the expressions of the velocity field and of the perturbation of buoyancy (Eqs. \ref{Umn} and \ref{bmn}). Therefore, both viscous dissipation and thermal diffusion do not depend on it. The equations yield two complex dimensionless frequencies, because of viscous and heat diffusion, that are functions of $ E $ and $ K $: 

\begin{equation}
\begin{array}{ccc}
\tilde{\omega} = \omega + i E\left( m^2 + n^2 \right) & \mbox{and} & \hat{\omega} = \omega + i K \left( m^2 + n^2 \right) .
\end{array}
\end{equation}

Initially, assuming that $ \textbf{f}=\textbf{0} $, we get the dispersion relation of the viscously and thermally damped gravito-inertial modes:

\begin{equation} 
\tilde{\omega}^2 = \frac{n^2 \cos^2 \theta }{m^2 + n^2}  + \frac{m^2 A}{m^2 + n^2} \frac{\tilde{\omega}}{\hat{\omega}}.
\label{dispersion}
\end{equation}

For slightly damped modes ($ E \ll 1 $ and $ K \ll 1 $), we identify in the second member the wave number $ \textbf{k} = \left( k_H,0,k_V \right) $, with $ k_H = m/L $ and $ k_V = n/L $ \citep[see][]{GS2005}. Indeed, neglecting $ E $ and $  K $, it simplifies to

\begin{equation} 
\chi^2 \approx \frac{m^2 N^2 + n^2 \left( 2 \Omega \cos \theta \right)^2 }{m^2 + n^2},
\label{dispersion2}
\end{equation}

that can be written

\begin{equation}
\chi^2 = \left( \frac{2 \boldsymbol{\Omega}.\textbf{k} }{ | \textbf{k} | } \right)^2 + \left(  N \frac{k_H}{ | \textbf{k} | }  \right)^2,
\end{equation}

where we recognize the dispersion relation of gravito-inertial waves with the respective contributions of inertial and gravity waves.

By letting $ m $ and $ n $ tend to infinity successively, we identify the approximative boundaries of the frequency range, $ \chi_{\rm inf} $ and $ \chi_{\rm sup} $. When $ 2 \Omega < N $ (i.e. $ A > 1 $), $ \chi_{\rm inf} \approx 2 \Omega $ and $ \chi_{\rm sup} \approx N $ (see Fig. \ref{fig:spectre_ondes}). The exact expressions of $ \chi_{\rm inf} $ and $ \chi_{\rm sup} $ are given by \cite{GS2005}:

\begin{equation}
\left\{
\begin{array}{lcl}
   \chi_{\rm inf} & = & \displaystyle \sqrt{ \frac{\lambda - \left[ \lambda^2 - \left( 2 \sigma N \right)^2 \right]^{1/2} }{2}  } \\
   \vspace{0.1mm}\\
   \chi_{\rm sup} & = & \displaystyle \sqrt{ \frac{\lambda + \left[ \lambda^2 - \left( 2 \sigma N \right)^2 \right]^{1/2} }{2}  }\\
\end{array}
\right.\,
\label{omega_GS}
\end{equation}

where $ \lambda = N^2 + \sigma^2 + \sigma_s^2 $. Here, $ \sigma = 2 \Omega \sin \theta $ and $ \sigma_s = 2 \Omega \cos \theta \sin \alpha $, $ \alpha=0$ corresponding to the propagation direction of gravito-inertial waves in the horizontal plane along the $x$ axis (see the bottom-left panel of Fig. \ref{fig:box}). At the end, we obtain the coefficients of the velocity field $ \textbf{u} $,

\begin{equation}
\left\{
\begin{array}{ccl}
  u_{mn} & \!\!=\!\! & \displaystyle  n \frac{i \tilde{\omega} \left( n f_{mn} - m h_{mn} \right) - n \cos \theta g_{mn}  }{ \left( m^2 + n^2 \right) \tilde{\omega}^2 - n^2 \cos^2 \theta - A m^2 \displaystyle \frac{\tilde{\omega}}{ \hat{\omega} } }  , \\
  \vspace*{0,1mm}\\
  v_{mn} & \!\!=\!\! & \displaystyle\frac{n \cos \theta \left( n f_{mn} - m h_{mn}  \right) + i \left[ \left( m^2 + n^2 \right) \tilde{\omega} - \displaystyle \frac{A m^2}{\hat{\omega}} \right] g_{mn} }{\left( m^2 + n^2 \right) \tilde{\omega}^2 - n^2 \cos^2 \theta - A m^2 \displaystyle \frac{\tilde{\omega}}{\hat{\omega}} }, \\
  \vspace*{0,1mm}\\
  w_{mn} & \!\!=\!\! & - m \displaystyle \frac{i \tilde{\omega} \left( n f_{mn} - m h_{mn} \right) - n \cos \theta g_{mn} }{\left( m^2 + n^2 \right) \tilde{\omega}^2 - n^2 \cos^2 \theta - A m^2 \displaystyle \frac{\tilde{\omega}}{\hat{\omega}} }, \\
\end{array}
\right.
\label{Umn}
\end{equation}

of the pressure $ p $,

\begin{equation}
\begin{array}{ccl}
  \psi_{mn} & \!\!=\!\! & \displaystyle \frac{1}{\Lambda} \left[ \frac{ \left( \tilde{\omega} f_{mn} + i \cos \theta g_{mn} \right) \left[ n \sin \theta + i m \left( \displaystyle \frac{A}{ \hat{\omega } } - \tilde{\omega} \right) \right] }{ \left( m^2 + n^2 \right) \tilde{\omega}^2 - n^2 \cos^2 \theta -       A m^2 \displaystyle \frac{\tilde{\omega}}{\hat{\omega}}} \right.\\
   &  &  \left. - \displaystyle \frac{ h_{mn} \left[ m \sin \theta \tilde{\omega} + i n \left( \tilde{\omega}^2 - \cos^2 \theta \right) \right] }{ \left( m^2 + n^2 \right) \tilde{\omega}^2 - n^2 \cos^2 \theta -  A m^2 \displaystyle \frac{\tilde{\omega}}{\hat{\omega}}} \right] , \\
\end{array}
\label{pmn}
\end{equation}

and of the buoyancy $ b $,

\begin{equation}
\begin{array}{ccl}
   b_{mn} & \!\!=\!\! & \displaystyle \frac{i A m}{\omega} \frac{i \tilde{\omega} \left( n f_{mn} - m h_{mn} \right) - n \cos \theta g_{mn}  }{ \left( m^2 + n^2 \right) \tilde{\omega}^2 - n^2 \cos^2 \theta -  A m^2 \displaystyle \frac{\tilde{\omega}}{\hat{\omega}} }.  \\
\end{array}
\label{bmn}
\end{equation}

Note that all the coefficients in Eqs.~(\ref{Umn}), (\ref{pmn}) and (\ref{bmn}) have the same denominator. It corresponds to the left-hand side of the forced momentum equation (Eq. \ref{NavierStokes}). The tidal excitation only affects their numerators.\\ 

As pointed out before, the viscous dissipation of gravito-inertial waves over a tidal period ($D^{\rm visc}$) is deduced from the velocity field (Eq. \ref{eq:Dvisc}). It is provided by the quadrature of the local mean dissipation in the whole box. Using our dimensionless coordinates, we define the viscous dissipation per unit mass:

\begin{equation}
\mathcal{D}^{\rm visc}=\frac{D^{\rm visc}}{\rho L^3}=\frac{1}{L^2} \int_0^1 \int_0^1 \left\langle - \textbf{u} \cdot \nu \nabla_{X,Z}^2 \textbf{u} \right\rangle dX\,dZ\,,
\end{equation} 

where $\nabla_{X,Z}^2\equiv(1/L^2)\nabla^2$ and $\left\langle\!\cdot\!\cdot\!\cdot\!\right\rangle$ is the average in time. Using Fourier series expansions given in Eq. (\ref{fourier}), this expression becomes:

\begin{equation}
\mathcal{D}^{\rm visc} = \frac{2 \pi^2 \nu}{ L^2}  \sum_{ (m,n) \in \mathbb{Z^*}^2 } \left( m^2 + n^2 \right) \left( \left| u_{mn}^2 \right| + \left| v_{mn}^2 \right| + \left| w_{mn}^2 \right| \right),
\label{dissipation_visc}
\end{equation}

as demonstrated by \cite{OL2004}. Similarly, the power dissipated by heat diffusion over a tidal period can be derived from the buoyancy perturbation (Eq. \ref{eq:Dtherm}). Using our dimensionless coordinates, we define the heat diffusion per unit mass:

\begin{equation}
\mathcal{D}^{\rm therm}=\frac{D^{\rm therm}}{\rho L^3}=\frac{1}{L^2}K\,A^{-2}\int_0^1\int_0^1\left\langle - b \nabla_{X,Z}^{2} b \right\rangle dX\,dZ\,.
\end{equation}

Using once again Eq. (\ref{fourier}), it becomes:

\begin{equation}
\mathcal{D}^{\rm therm} = \frac{2 \pi^2 \kappa}{ L^2 A^2}  \sum_{ (m,n) \in \mathbb{Z^*}^2 } \left( m^2 + n^2 \right) \left| b_{mn} \right|^2.
\label{dissipation_therm}
\end{equation}

The total dissipated power is the sum of both contributions:

\begin{equation}
\mathcal{D} = \mathcal{D}^{\rm visc} + \mathcal{D}^{\rm therm}.
\end{equation}

Finally, we introduce the energy dissipated over a rotation period of the planet, denoted $ \zeta $, 

\begin{equation}
\zeta = \frac{2 \pi}{\Omega} \mathcal{D},
\label{energy}
\end{equation}

and its viscous and thermal contributions,

\begin{equation}
\zeta^{\rm visc} = 2 \pi E \sum_{ (m,n) \in \mathbb{Z^*}^2 } \left( m^2 + n^2 \right) \left( \left| u_{mn}^2 \right| + \left| v_{mn}^2 \right| + \left| w_{mn}^2 \right| \right),
\label{energy_visc}
\end{equation}

\begin{equation}
\zeta^{\rm therm} = 2 \pi K A^{-2}  \sum_{ (m,n) \in \mathbb{Z^*}^2 } \left( m^2 + n^2 \right) \left| b_{mn} \right|^2,
\label{energy_therm}
\end{equation}

that will be studied instead of the power ($ \mathcal{D} $). The energy ($ \zeta $ of Eq.~\ref{energy}) only depends on the control parameters $ A $, $ E $ and $ K $ (see Eq. \ref{parameters}).

\subsection{Spectral response}

We have computed analytical formulae of the energies dissipated by viscous friction and thermal diffusion. Considering the dispersion relation Eq. (\ref{dispersion}) and the resonances involved, we know that $ \zeta $ significantly varies with the tidal frequency $ \omega $. As shown by Eq. (\ref{energy_visc}), its behavior is determined by  the position $ \theta $ of the box and the dimensionless numbers $ A $, $ E $ and $ K $. To be able to make our results comparable with the model in the appendix of \cite{OL2004}, we choose the same academic forcing:
\begin{equation}
\begin{array}{cccc}
f_{mn} = - \displaystyle \frac{i}{4 \left| m \right| n^2 }, & g_{mn} = 0 & \mbox{and} & h_{mn}=0.
\end{array}
\end{equation}
We verified that the forms of the $\left\{f_{mn},g_{mn},h_{mn}\right\}$ coefficients do not affect properties of the spectra that are governed by the left-hand side of the forced momentum equation (Eq. \ref{NavierStokes}), i.e. the position and the width of the resonances in the frequency-dissipation spectra. However, the spectral dependence on $m$ and $n$ modulate their number, their height, and the amplitude of the non-resonant background. Nevertheless, the contrast between the latter and the height of the main resonance, which will be called the sharpness ratio and evaluates the relative amplitude between dynamical and equilibrium tides, are less affected. Moreover, we verified that the different main asymptotic behaviors of the system remain unchanged. In future works, it would be important to implement a more realistic treatment of the forcing term, as done for example by \cite{Barker2013} for local numerical simulations of the tidal elliptic instability.\\

\begin{figure*}[ht!]
 \centering
 \includegraphics[width=0.48\textwidth,clip]{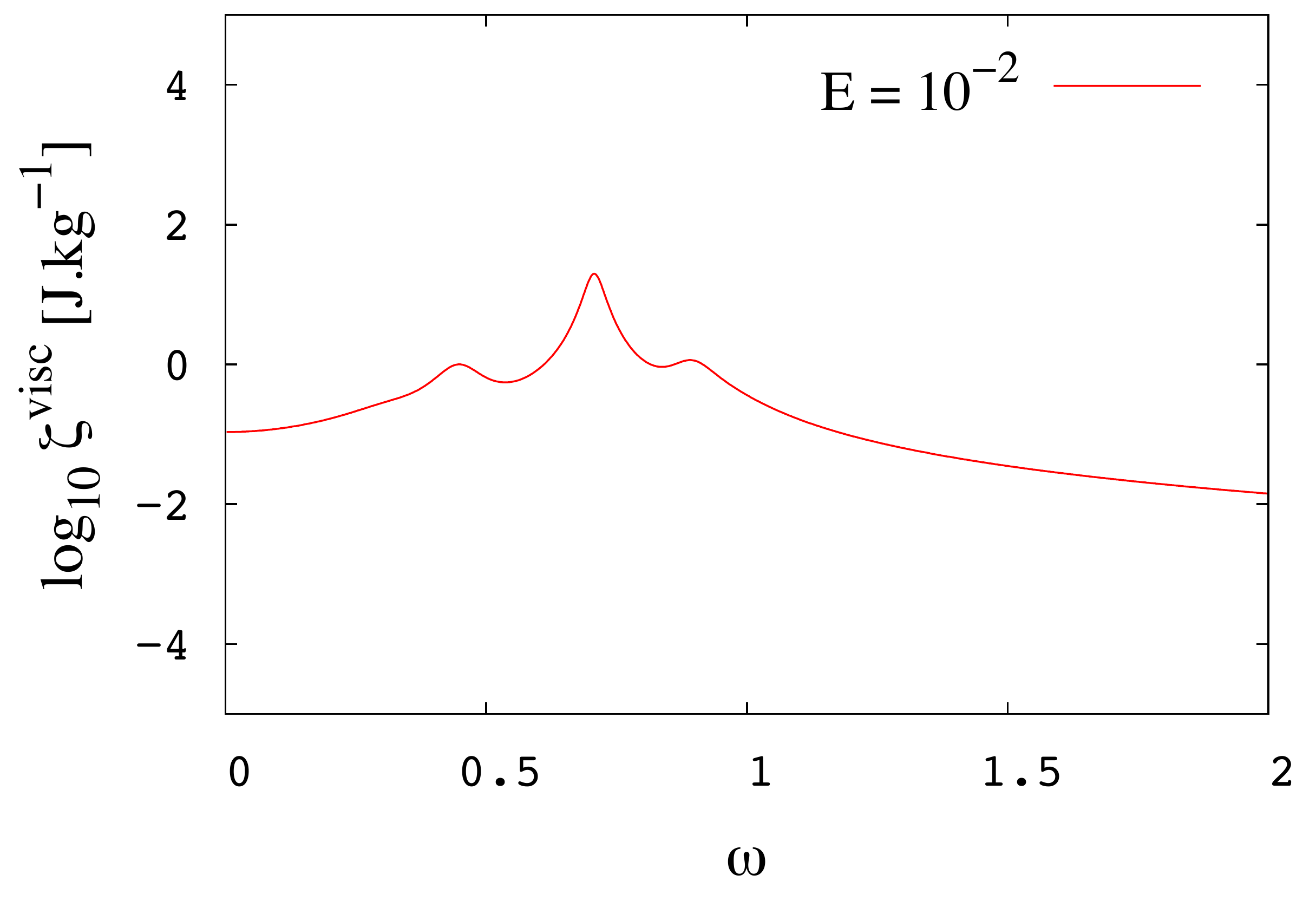}%
 \includegraphics[width=0.48\textwidth,clip]{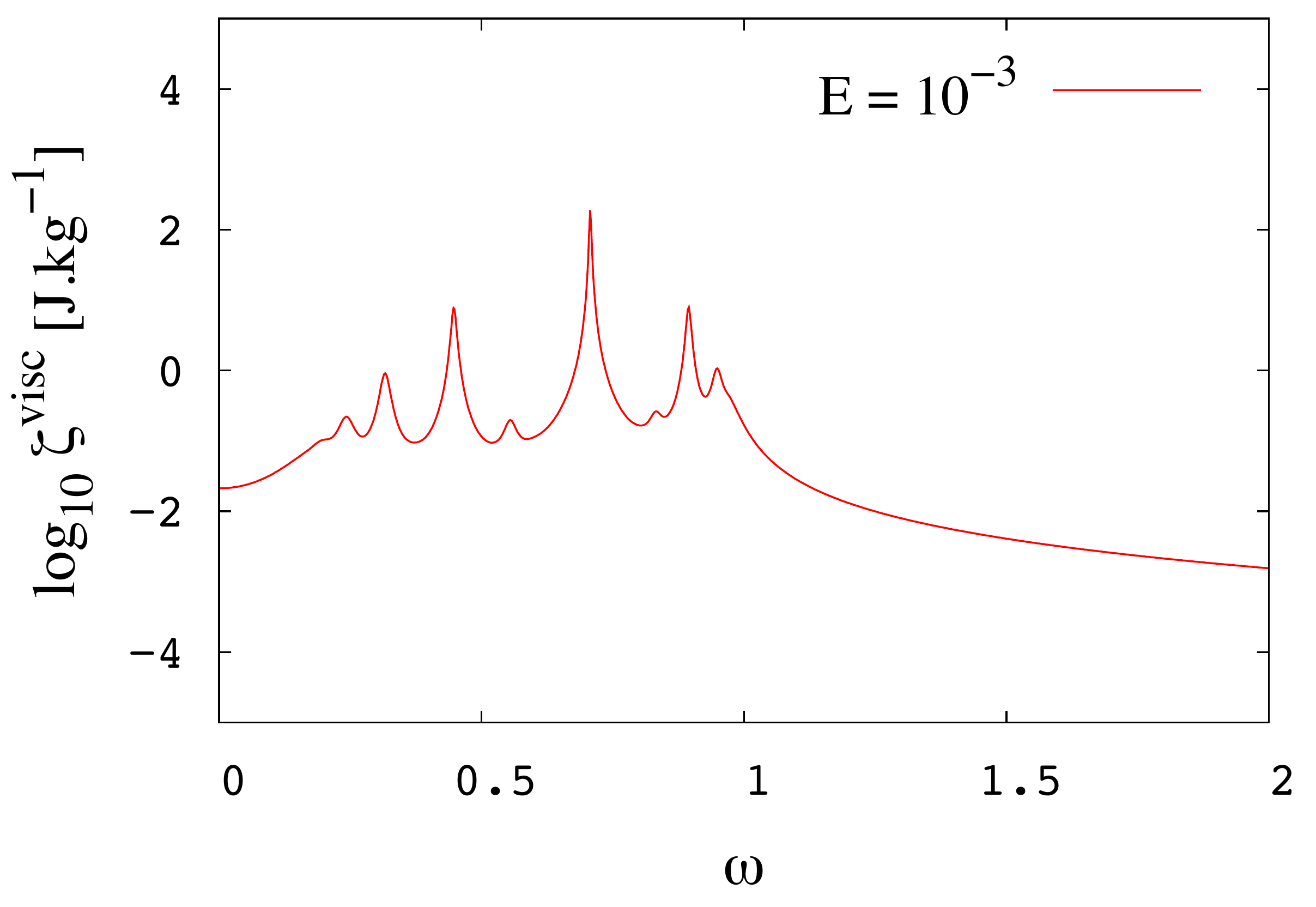} 
 \includegraphics[width=0.48\textwidth,clip]{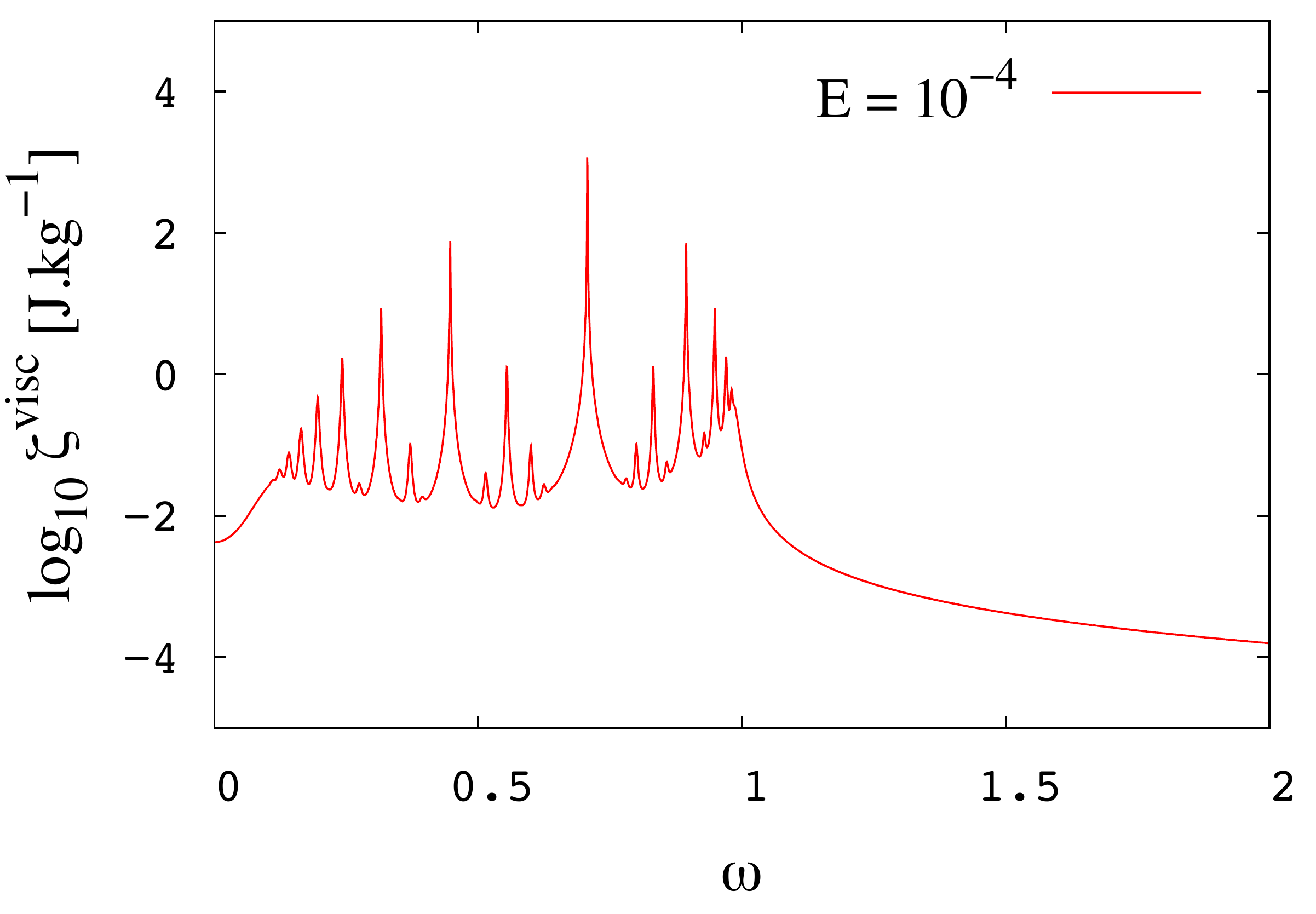}
 \includegraphics[width=0.48\textwidth,clip]{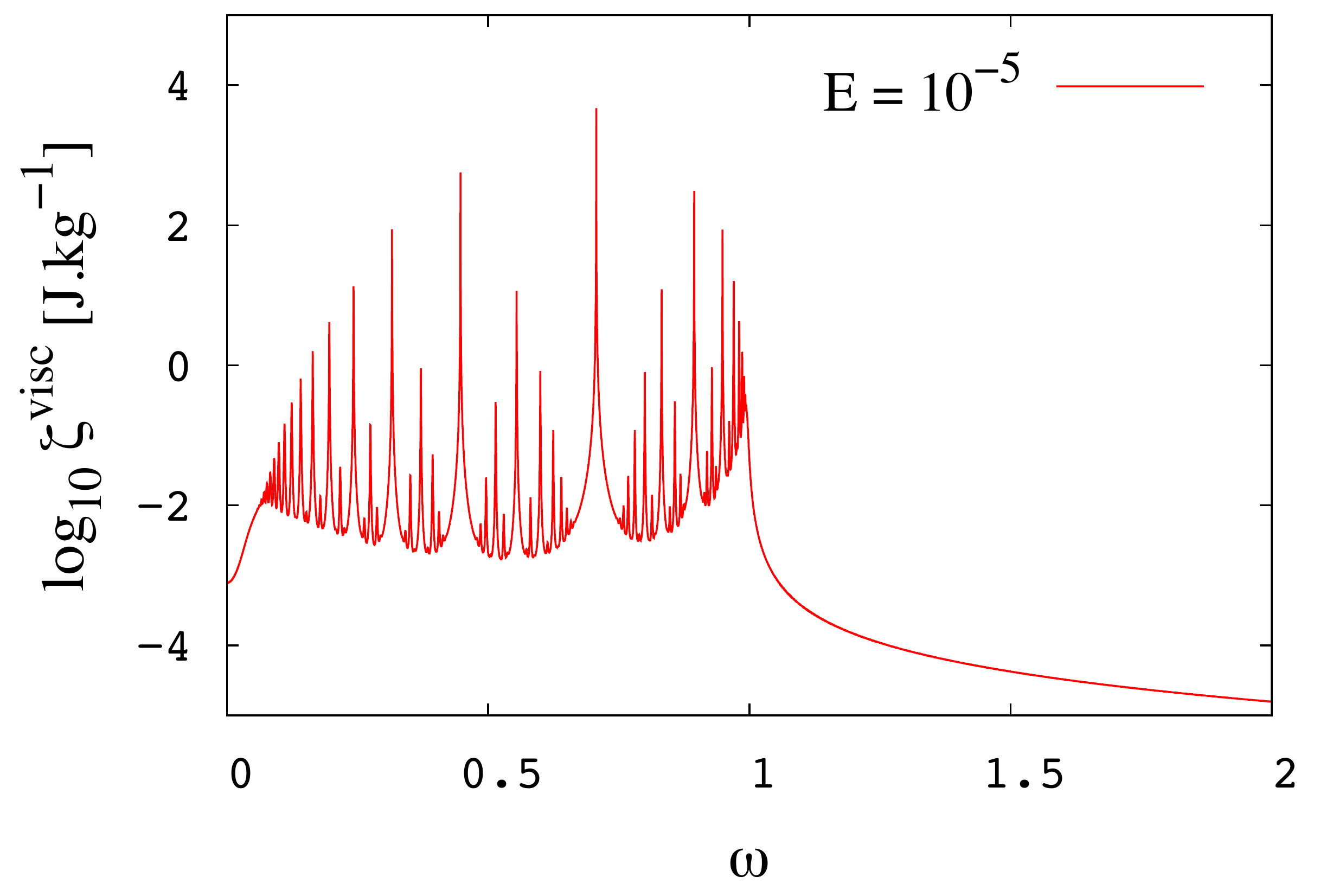}     
  \textsf{ \caption{\label{fig:spectre_inertiel_1} Frequency spectra of the energy per mass unit locally dissipated by viscous friction $ \zeta^{\rm visc} $ for inertial waves ($ A = 0 $) at the position $ \theta = 0 $ for $ K = 0 $ and different Ekman numbers. In abscissa,  the normalized frequency $ \omega = \chi / 2 \Omega $. {\bf Top left:} $ E = 10^{-2} $. {\bf Top right:} $ E = 10^{-3} $. {\bf Bottom left:} $ E = 10^{-4} $. {\bf Bottom right:} $ E = 10^{-5} $. The results obtained by \cite{OL2004} are recovered. Note that the non vanishing viscous dissipation at $\omega=0$ is the one of the geostrophic equilibrium.}}
\end{figure*}

\begin{figure*}[ht!]
 \centering
 \includegraphics[width=0.48\textwidth,clip]{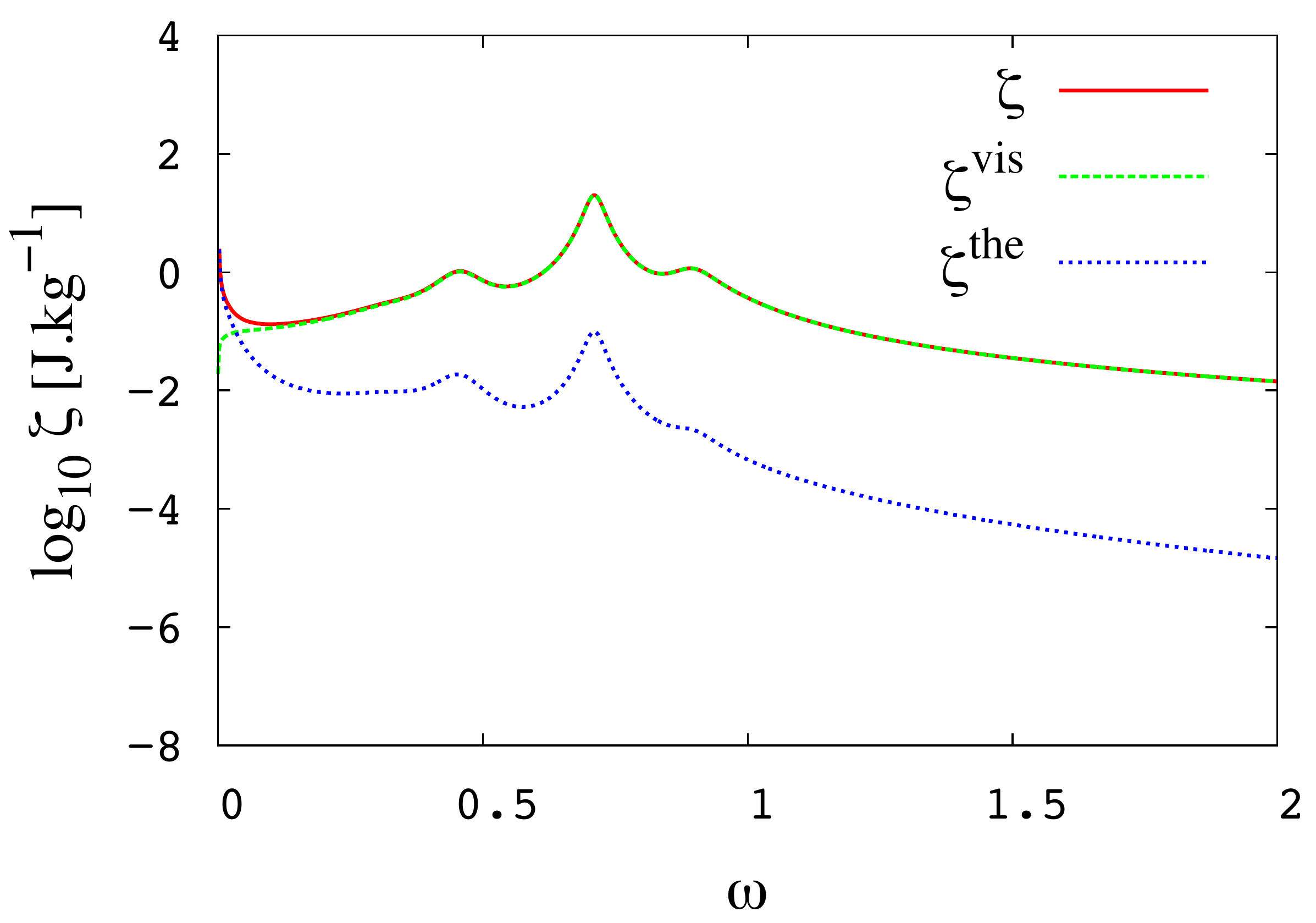}%
 \includegraphics[width=0.48\textwidth,clip]{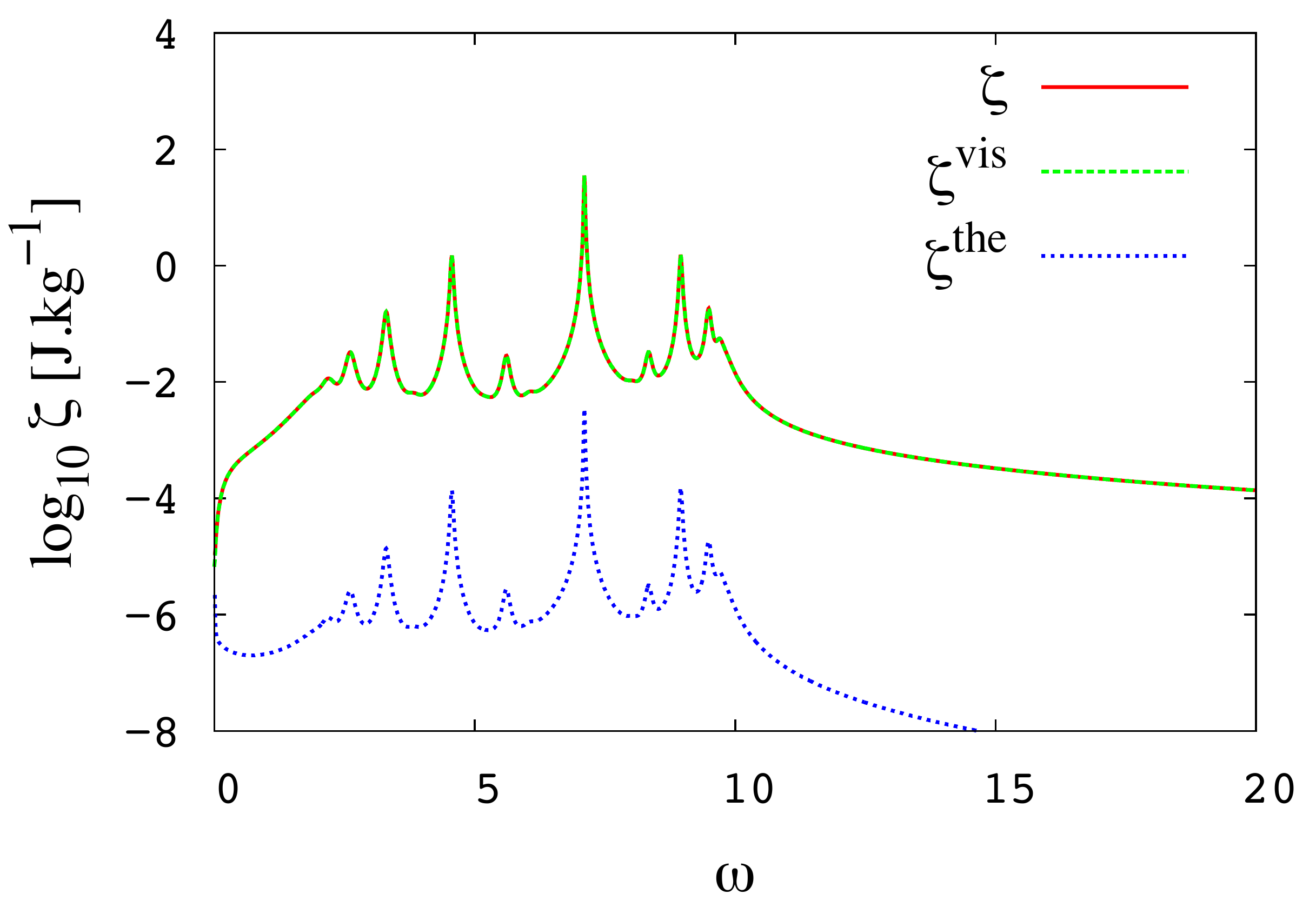}
 \includegraphics[width=0.48\textwidth,clip]{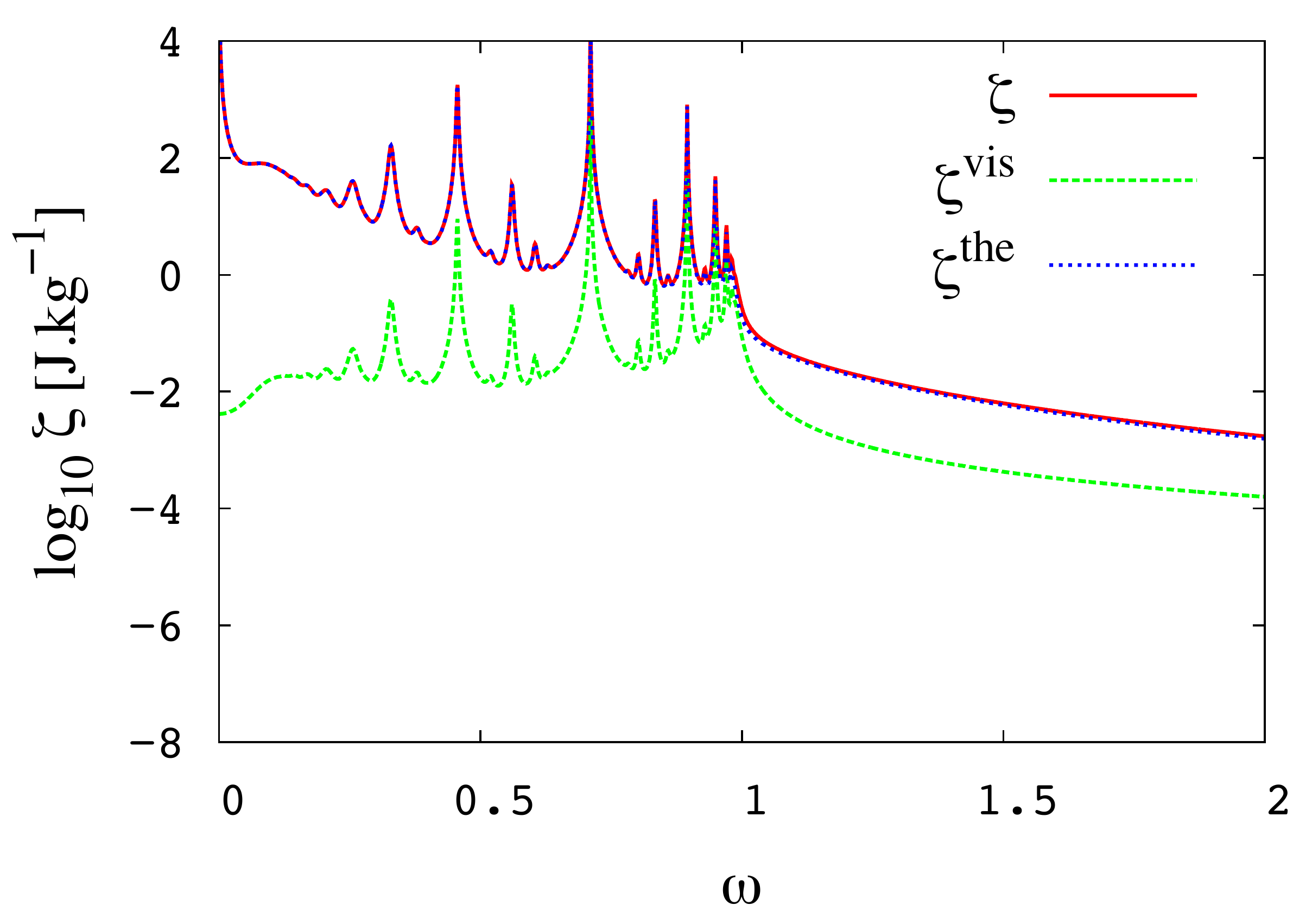}
 \includegraphics[width=0.48\textwidth,clip]{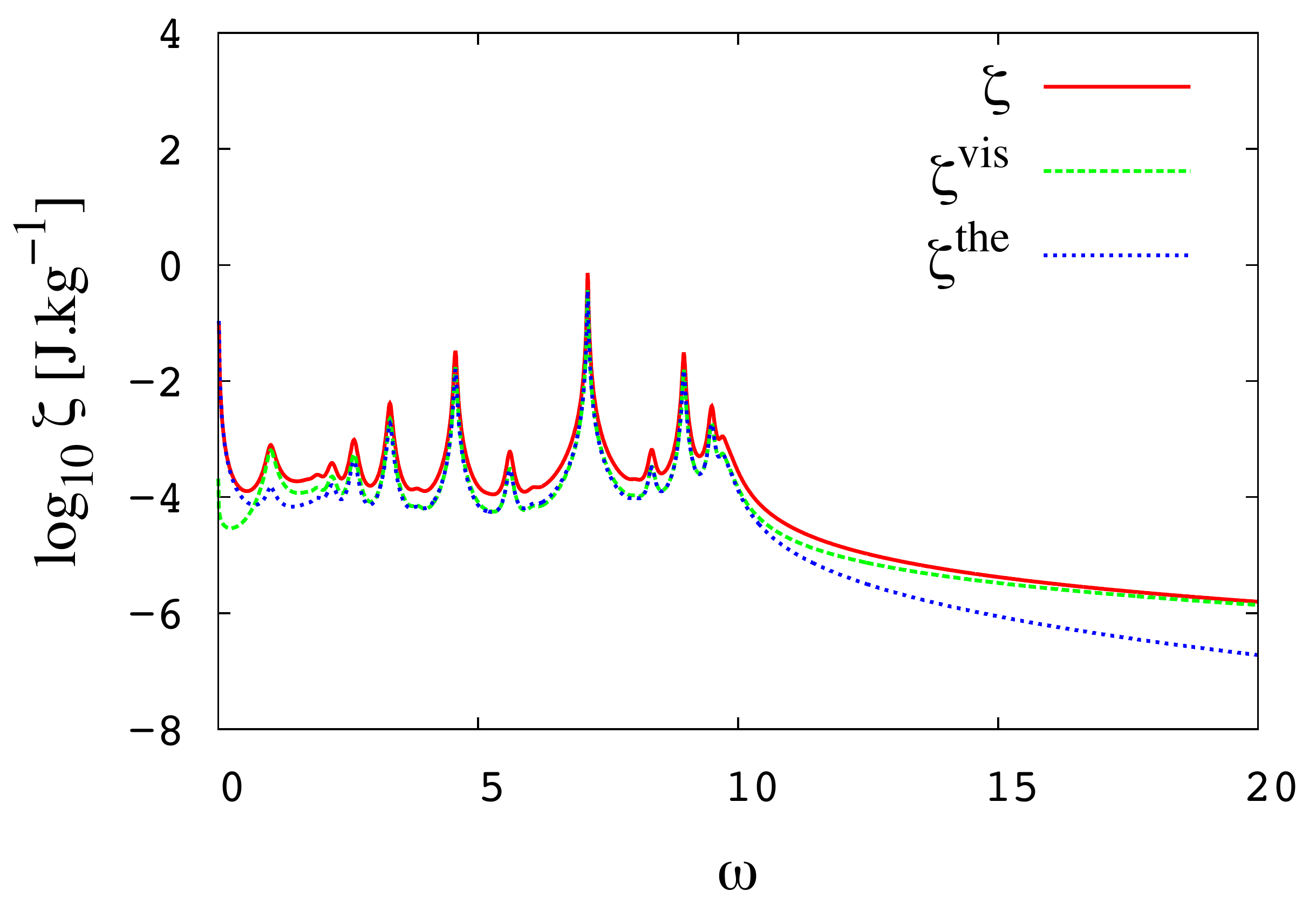}   
  \textsf{ \caption{\label{fig:spectre_inertiel_2} Frequency spectrum of the energy per mass unit dissipated in different regimes ($\zeta$) (continuous red line), and of its viscous and thermal contributions $ \zeta^{\rm visc} $ (dashed green line) and $ \zeta^{\rm therm} $ (dashed blue line). The latitude is fixed at $ \theta = 0 $. In abscissa,  the normalized frequency $ \omega = \chi / 2 \Omega $. {\bf Top left:} $ A = 1.0 \times 10^{-2} $; $ E = 1.0 \times 10^{-2} $; $ K = 1.0 \times 10^{-4} $. {\bf Top right:} $ A = 1.0 \times 10^{2} $; $ E = 1.0 \times 10^{-2} $; $ K = 1.0 \times 10^{-4} $. {\bf Bottom left:} $ A = 1.0 \times 10^{-2} $; $ E = 1.0 \times 10^{-4} $; $ K = 1.0 \times 10^{-2} $. {\bf Bottom right:} $ A = 1.0 \times 10^{2} $; $ E = 1.0 \times 10^{-4} $; $ K = 1.0 \times 10^{-2} $.}}
\end{figure*}

In Figs. \ref{fig:spectre_inertiel_1} and \ref{fig:spectre_inertiel_2}, we plot the frequency-spectra of $ \zeta^{\rm visc} $ for various sets of parameters. The abscissa measures the dimensionless frequency $ \omega = \chi / 2 \Omega $, and the vertical axis the local viscous dissipation per unit mass (in logarithmic scale). Figure~\ref{fig:spectre_inertiel_1} corresponds to pure inertial waves ($ A = 0 $) in a box located at the pole ($ \theta = 0 $). The plots illustrate most of the results of this section. First, the cutoff frequency for inertial modes, $ \omega_c = 1 $ (see the dispersion relation in Eq. \ref{dispersion}), clearly appears. Beyond $ \omega = 1 $, the dissipation decreases by several orders of magnitude and corresponds to the non-resonant background. Second, we can observe the sharp dependence of the dissipation to the tidal frequency $ \chi $. $ \zeta $ also varies with the Ekman number $ E $. The spectrum is regular for high values of $ E $, that correspond to waves damped by a strong viscous diffusion.  At the opposite, the number of resonances increases when $ E $ decreases.\\

Figure~\ref{fig:spectre_inertiel_2} gives an overview of the dependence of $ \zeta $, $ \zeta^{\rm visc} $ and $ \zeta^{\rm therm} $ on the control parameters. Each plot corresponds to a particular regime of tidal dissipation. The number of peaks is obviously correlated to the Ekman number and we retrieve the spectra of Fig~\ref{fig:spectre_inertiel_1} in the top-left and top-right plots where dissipation by viscous friction predominates. The ratio of the contributions, $ \zeta^{\rm visc} / \zeta^{\rm therm} $, depends on the Prandtl number, the dissipation being mainly due to viscous friction for high values of $ P_r $ and to heat diffusion for low ones. The cases of the bottom-left and bottom-right plots show that the critical Prandtl number, where both contributions are comparable, varies with $ A $. \\


 After this first qualitative approach, it becomes necessary to develop a quantitative physical description of the spectra taking into account both dissipation mechanisms. In a previous work, we had pointed out the importance of the role played by resonances in the evolution of planetary systems. Indeed, the variations of the distance between a star or a fluid planet and its planet or satellite are directly linked to the height and width of the peaks \citep[see][]{ADLPM2014}. It is now possible to understand how the latter depend on the physical fluid parameters.

\section{Properties of the resonances}

Numerical results show that the fluid behaves like a bandpass filter with clearly identified cutoff frequencies, $ \omega_{\rm inf} $ and $ \omega_{\rm sup} $, commented further (see Eq. \ref{omega_GS}). A typical dissipation spectrum is a batch of resonances presenting a large range of positions, widths and heights, and an apparent non-resonant background. Thanks to our analytical formulae of the viscous and thermal dissipations (Eqs. \ref{dissipation_visc} and \ref{dissipation_therm}), these properties can be established explicitly as functions of the colatitude $ \theta $, the ratio of squared frequencies $ A $, the Ekman number $ E $ and its analogue for thermal diffusion $ K $ (and the corresponding Prandtl number $P_r$). Thus, we will treat first the eigenfrequencies $ \omega_{mn} $ and the widths $ l_{mn} $ that do not depend on the forcing, and in a second step the heights $ H_{mn} $ and the non-resonant background $ H_{\rm bg} $. At the end, we will look at the number of resonances $ N_{\rm kc} $ and the sharpness ratio $ \Xi $, which is the ratio between the height of the main peak and the mean non-resonant background inside the frequency range $ \left[ \omega_{\rm inf} , \omega_{\rm sup} \right] $.

\subsection{Position and population of resonances}

The spectra of Fig. \ref{fig:spectre_inertiel_1} reveal a well-organized structure. Each peak is surrounded by two others of the nearest higher orders. The fractal pattern, accentuated by the particular forcing chosen, is apparent in the bottom-right plot ($ E = 10^{-5} $).  In this subsection, we retrieve this singular structure analytically, by studying the expression of $ \zeta $ (Eq. \ref{energy_visc}). The energy dissipated is written as an infinite sum of terms $ \zeta_{mn} $ associated to the degrees $ m $ and $ n $. As $ \zeta_{mn} \left( \omega \right) \sim \zeta_{-mn} \left( \omega \right) \sim \zeta_{m-n} \left( \omega \right) \sim \zeta_{-m-n} \left( \omega \right) $, $ m $ and $ n $ being natural integers, four terms of this series dominate all the others near a resonance. So, we assume:

\begin{equation}
\zeta^{\rm visc} \approx 4 \zeta_{mn}^{\rm visc} =  8 \pi E  \left( m^2 + n^2 \right) \left( \left| u_{mn}^2 \right| + \left| v_{mn}^2 \right| + \left| w_{mn}^2 \right| \right),
\label{zetamn_visc}
\end{equation}

\begin{equation}
\zeta^{\rm therm} \approx 4 \zeta_{mn}^{\rm therm} =  8 \pi A^{-2} E P_r^{-1}  \left( m^2 + n^2 \right) \left| b_{mn} \right|^2.
\label{zetamn_therm}
\end{equation}

$ \zeta^{\rm visc} $ and $ \zeta^{\rm therm} $ have the same denominator,

\begin{equation}
d_{mn} = \left| \left( m^2 + n^2 \right) \tilde{\omega}^2 - n^2 \cos^2 \theta - A m^2 \frac{\tilde{\omega}}{\hat{\omega}}  \right|^2.
\label{dmn}
\end{equation}

 Considering that the numerators varies smoothly compared to $ d_{mn} $, we deduce the eigenfrequencies $ \omega_{mn} $ (with $ \left( m,n \right) \in \mathbb{N}^{*2} $) from the dispersion relation (Eq. \ref{dispersion}). They correspond to the minima of the denominator $ d_{mn} $. Then, we factorize the expression given in Eq.~(\ref{dmn}) by $ 1/\hat{\omega} $, which varies smoothly like the numerator. It yields the polynomial $ P $:

\begin{equation}
P \left(  \omega \right) = \omega^6 + \left(  \beta^2 - 2 \alpha \right) \omega^4 + \left( \alpha^2 - 2\beta \gamma  \right) \omega^2 + \gamma^2,
\end{equation}

where $ \alpha $, $ \beta $ and $ \gamma $ are positive real coefficients:

\begin{equation}
\left\{
\begin{array}{rcl}
  \alpha & = & E \left( E + 2K \right) \left(m^2 + n^2 \right)^2 + \displaystyle \frac{n^2 \cos^2 \theta + m^2 A }{ m^2 + n^2 }\\
  \vspace*{0,1mm}\\
  \beta & = &  \left(K + 2 E \right) \left(m^2 + n^2 \right) \\
  \vspace*{0,1mm}\\
  \gamma & = & K  E^2 \left( m^2 + n^2 \right)^3 + n^2 \cos^2 \theta K + m^2 A E
\end{array}
.
\right.
\end{equation}

$ P \left( \omega \right) $ have the same minima as $ d_{mn} $. So, by differentiating $ P \left( \omega \right) $, we get:

\begin{equation}
P'\left( \omega \right) =  \omega Q \left( \omega \right),
\end{equation}

$ Q $ being an even fourth degree polynomial. Thus, solutions satisfy the equation:

\begin{equation}
Q \left(  \omega \right)  = 0,
\end{equation}

and exist only if the coefficients $ \alpha $, $ \beta $ and $ \gamma $ satisfy the condition:

\begin{equation}
\alpha^2 + 6 \beta \gamma + \beta^4 - 4 \alpha \beta^2 \geq 0.
\label{condition_positions}
\end{equation}

Given the form of the polynomials $ P $ and $ Q $, which involve only even terms, eigenfrequencies are symmetrical with respect to the vertical axis  ($ \omega = 0 $). Therefore, the domain studied can be restricted to the positive interval $ [0; + \infty [ $. There is a single physical solution if $ 2 \alpha > \beta^2 $ or $ \alpha^2 < 2 \beta \gamma $,

\begin{equation}
\omega_{mn} = \frac{1}{\sqrt{3}} \left[  2 \alpha - \beta^2 + \sqrt{ \beta^4 + \alpha^2 - 4 \alpha \beta^2 + 6 \beta \gamma }  \right]^{\frac{1}{2}},
\end{equation}

which corresponds to a minimum of $ P $, so a maximum of $ \zeta $. We have pointed out that $ \zeta_{mn}^{\rm visc} $ and $ \zeta_{mn}^{\rm therm} $ have the same denominator. Therefore, they also have the same eigenfrequencies. In a first time, consider the bi-parameter case of \cite{OL2004}: $ A = 0 $, $ K = 0 $ (the solution only depends on the position $ \theta $ and the Ekman number $ E $). The eigenfrequency $ \omega_{mn} $ becomes:

\begin{equation}
\omega_{mn} = \sqrt{ \frac{n^2 \cos^2 \theta}{m^2 + n^2} - E^2 \left( m^2 + n^2 \right)^2 }.
\label{frequency_inertial}
\end{equation}

Beyond a critical rank, there is no minimum any more. The first harmonics only are resonant because they verify the condition:

\begin{equation}
\frac{ n^2  }{  \left( m^2 + n^2 \right)^3 } > \left(  \frac{E}{\cos \theta} \right)^2.
\end{equation}

\begin{figure*}[ht!]
 \centering
 \includegraphics[width=0.48\textwidth,clip]{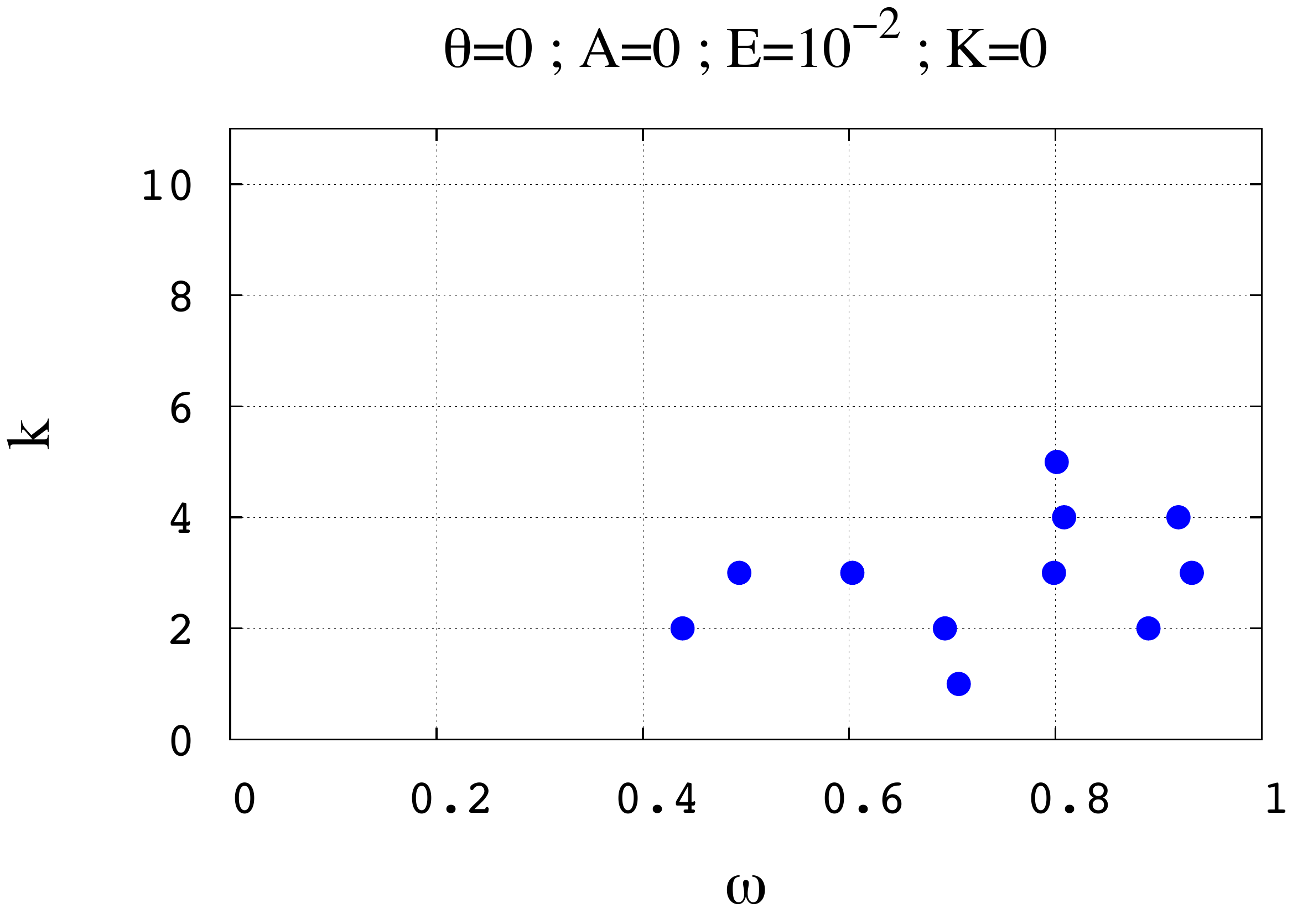}%
 \includegraphics[width=0.48\textwidth,clip]{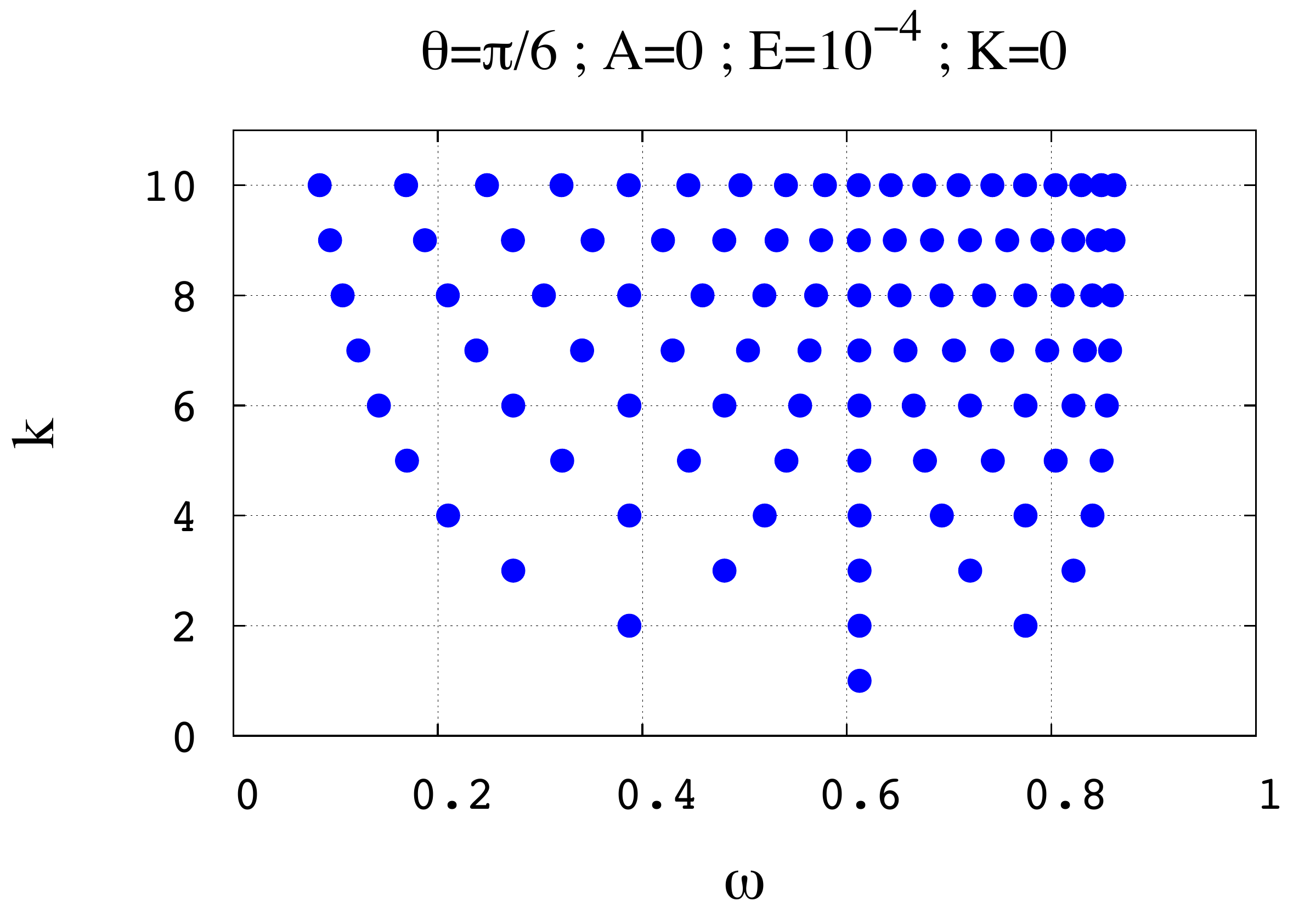} 
 \includegraphics[width=0.48\textwidth,clip]{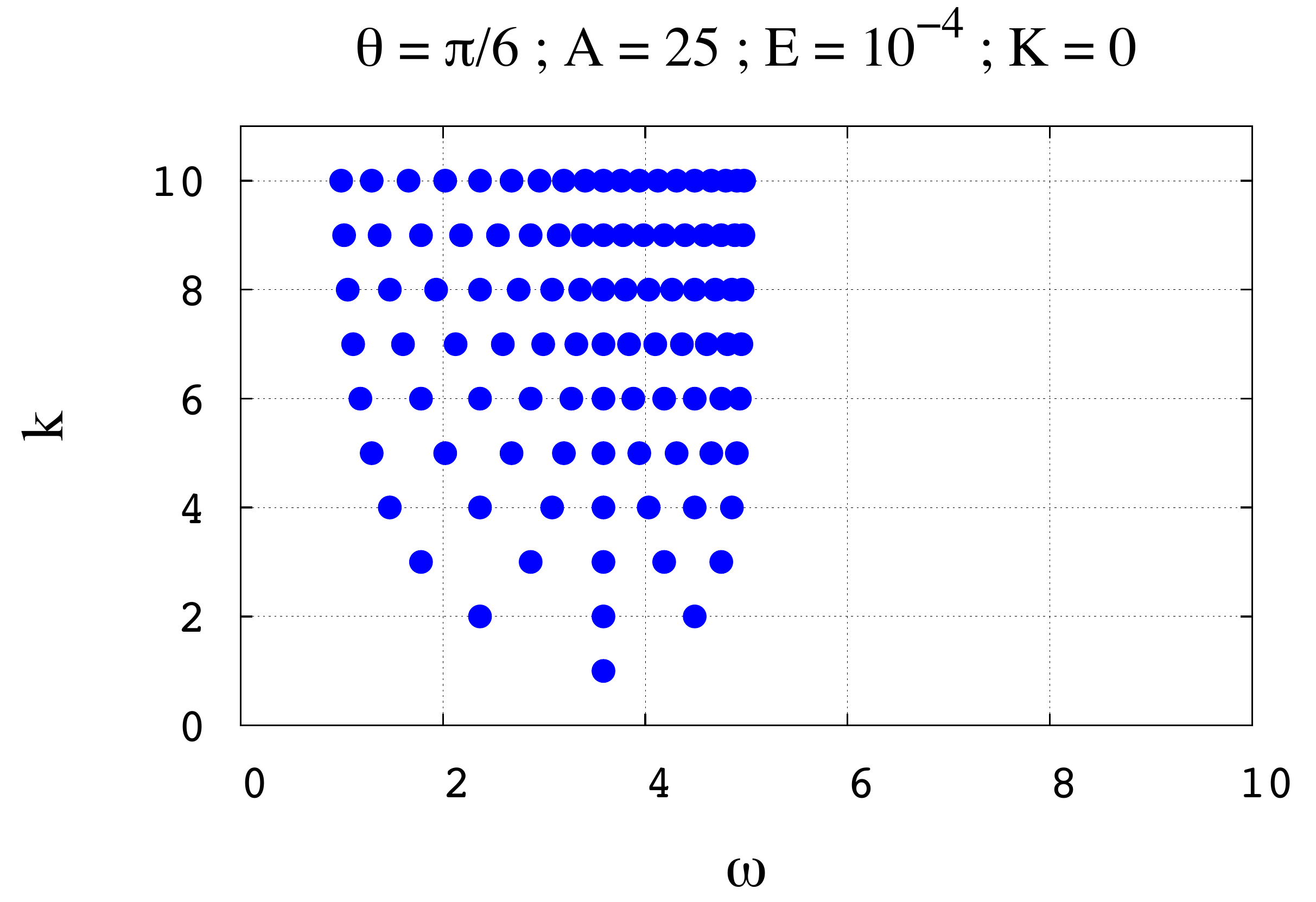}
 \includegraphics[width=0.48\textwidth,clip]{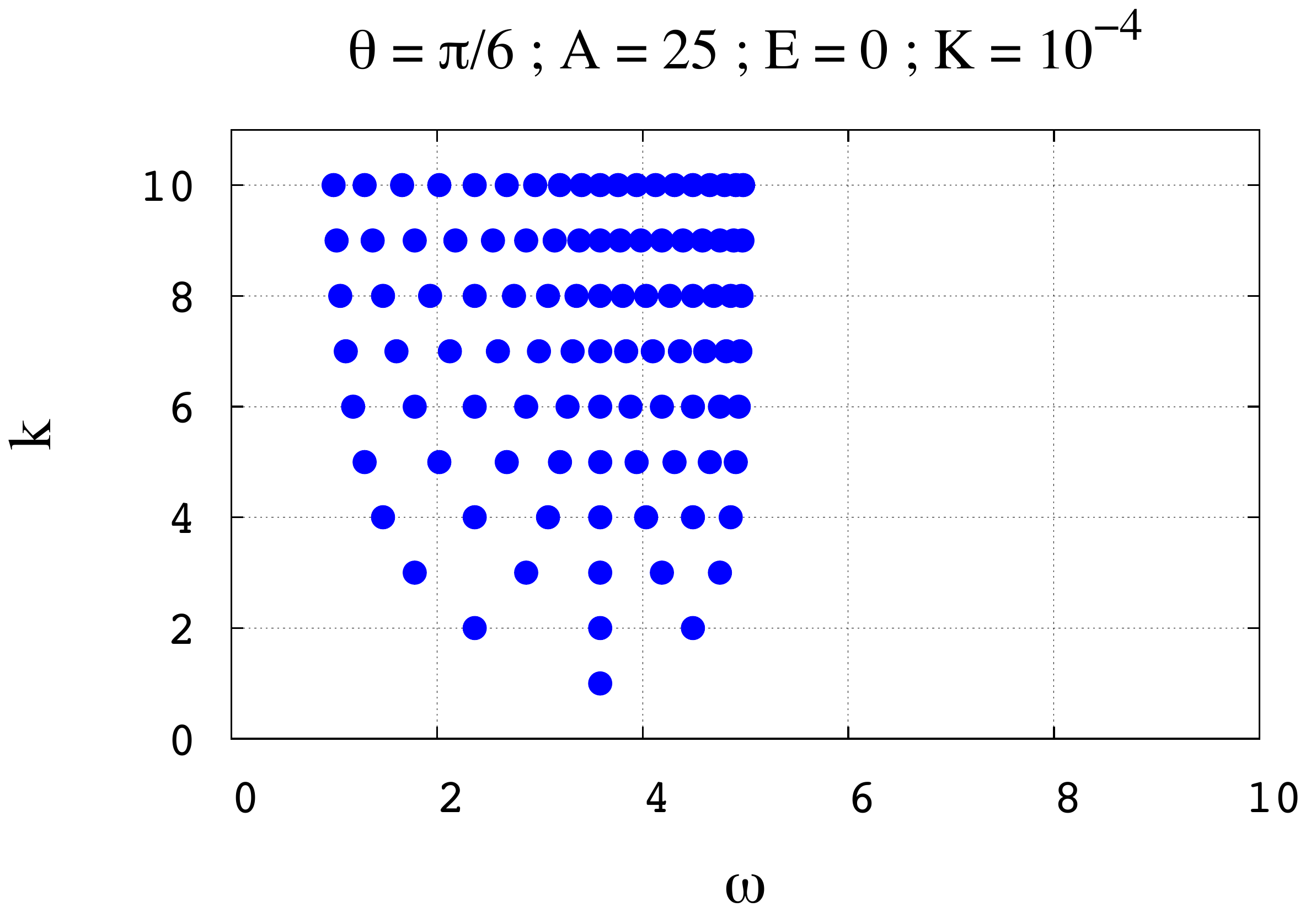}     
  \textsf{ \caption{\label{fig:positions} Structures of the frequency spectra of dissipation for inertial waves and gravito-inertial waves, with various sets of parameters. The positions of resonances (in abscissa, the normalized frequency $ \omega = \chi / 2 \Omega $) are indicated by blue points as functions of the characteristic rank $ k $ of the harmonics (ordinates). {\bf Top left:} $ \left( \theta ,A, E, K \right) = \left( 0 , 0, 10^{-2}, 0 \right) $. {\bf Top right:} $ \left( \theta ,A, E, K \right) = \left( \pi / 6 , 0, 10^{-4}, 0 \right) $. {\bf Bottom left:} $ \left( \theta ,A, E, K \right) = \left( \pi / 6 , 25, 10^{-4}, 0 \right) $. {\bf Bottom right:} $ \left( \theta ,A, E, K \right) = \left( \pi / 6 , 25, 0, 10^{-4} \right) $.}}
\end{figure*}

Thus, there is obviously a maximum number of resonances that decreases with $ E $ and $ \theta $ in the bi-parameter case. However, we will see that this maximum does not correspond to the effective number of resonances. This latter is actually strongly constrained by the level of the non-resonant background which is the cause of the variations observed in Figs. \ref{fig:spectre_inertiel_1} and \ref{fig:spectre_inertiel_2}. By plotting the analytical eigenfrequencies $ \omega_{\rm mn} $ (Fig.~\ref{fig:positions}), we retrieve the global structure of spectra. Each blue point corresponds to a mode $ \left( m,n \right) $. The characteristic rank of the harmonic $ k = \max \left\{  \vert m \vert , \vert  n \vert \right\} $ is introduced to reduce the doublet $ \left( m , n \right) $ to one index only. It represents an approximative degree of the resonance. The cutoff frequency of inertial waves $ \omega_{c} = \cos \theta $ distinctly appears on Fig. \ref{fig:positions} given that $ E \ll 1 $.  It can also be deduced from the simplified expression of $ \omega_{mn} $:

\begin{equation}
\omega_{mn} \approx \frac{n}{\sqrt{m^2 + n^2}} \cos \theta.
\end{equation}

The plots representing the eigenfrequencies (Fig.~\ref{fig:positions}) expose the symmetries of the dispersion relation in the quasi-adiabatic approximation, that we develop from now on.\\

The quasi-adiabatic approximation means that the diffusivities $ E $ and $ K $ are first-order infinitesimals with respect to $ \cos \theta $ and $ \sqrt{A} $. Such an assumption corresponds to low viscous and thermal diffusivities ($\nu$ and $\kappa$) relevant in stellar and planetary interiors. Thus, for the main resonances ($ k  < 10 $), the coefficients $ \alpha $, $ \beta $ and $ \gamma $ become:

\begin{equation}
\left\{
\begin{array}{rcl}
  \alpha & = & \displaystyle \frac{n^2 \cos^2 \theta + m^2 A }{ m^2 + n^2 }\\
  \vspace*{0,1mm}\\
  \beta & = &  \left(K + 2 E \right) \left(m^2 + n^2 \right) \\
  \vspace*{0,1mm}\\
  \gamma & = & n^2 \cos^2 \theta K + m^2 A E
\end{array}
\right.
\label{coefficients}
\end{equation}

and the frequency of the resonance $ \left( m,n \right) $,

\begin{equation}
\omega_{mn} = \sqrt{\alpha} \left( 1 + \varepsilon \right),
\end{equation}

where $ \varepsilon $ is a second order bias:

\begin{equation}
\varepsilon = \frac{\beta}{2 \alpha} \left( \frac{\gamma}{\alpha} - \beta  \right).
\end{equation}

Therefore, when $ E $ and $ K $ are small and can be neglected, resonances are located at the eigenfrequencies

\begin{equation}
\omega_{mn} = \sqrt{ \frac{n^2 \cos^2 \theta + A m^2}{ m^2 + n^2} }
\label{frequency}
\end{equation}

that correspond to the dispersion relation of gravito-inertial waves in the adiabatic case given in Eq. (\ref{dispersion2}). They give us the boundaries of the frequency range in the local model,

\begin{equation}
\begin{array}{ccc}
   \omega_{\rm inf} \approx \displaystyle \frac{N}{2 \Omega} & \mbox{and}  & \omega_{\rm sup} \approx \cos \theta \quad \mbox{if} \quad N<2\Omega \\
\end{array}
\end{equation}

or the contrary if $ N > 2 \Omega $. Note that the exact expressions of $ \omega_{\rm inf} $ and $ \omega_{\rm sup} $ (Eq. \ref{omega_GS}) are actually a bit more complex than the previous ones. In Figs.~\ref{fig:spectre_inertiel_2} and \ref{fig:positions}, the interval $ \left[ \omega_{\rm inf}, \omega_{\rm sup} \right] $ delimits peaks and blue points zones. As the impact of $ E $ and $ K $ is of second order, the structure of the batch of resonances only depends on $ A $ and the colatitude $ \theta $ under the quasi-adiabatic approximation. This is the reason why the plots of Fig. \ref{fig:positions} where $ \left\{E,K\right\} \ll \left\{\sqrt{A}, \cos{\theta}\right\} $ have the same appearance. By taking $ A = 0 $, we recover the expression that corresponds to pure inertial waves (see Eq. \ref{frequency_inertial}). \\

Moreover, Eq.~(\ref{frequency}) points out an hyper-resonant case related to quasi-inertial waves ($ A < 1 $) and characterized by the equality:

\begin{equation} 
A = \cos^2 \theta.
\end{equation}

When it is verified, all the peaks are superposed at the eigenfrequency:

\begin{equation}
\omega_p = \omega_c =  \cos \theta, 
\end{equation}

and form a huge single peak. So, if $ N< 2 \Omega $ (quasi-inertial waves), there is a critical colatitude $ \theta_0 = \arccos \left( N / \left( 2 \Omega \right) \right) $ at which the dissipation spectrum of tidal gravito-inertial waves reduces to a single resonance. The frequency range decreases when the latitude comes closer to $ \theta_0 $. As there is no such critical colatitude for $ A > 1 $, usual gravito-inertial waves are not affected by this effect. Their frequency range $ \left[ \omega_{\rm inf} , \omega_{\rm sup} \right] $ is only larger at the equator than at the pole. \\

The number of resonances constituting a spectrum can be estimated from the expressions of the eigenfrequencies $ \omega_{mn} $ (Eq. \ref{frequency}). Indeed, due to the definition of $ k $, there are $ 2k -1 $ points per line on Fig. \ref{fig:positions}. But the symmetries of $ \omega_{mn} $ make that some of these points have the same positions. Indeed, note that $ \omega_{m'n'} = \omega_{mn} $ if $ m' = p m $ and $ n' = pn $ with $ p \in \mathbb{Z}^* $. This explains why resonances are not as numerous as modes. We denote $ k_c $  the maximal rank of the harmonics dominating the background and $ N_{kc} $ the number of effective resonant peaks. The layer of harmonics $ k $ brings $ p_k $ new peaks:

\begin{equation}
p_k = 2k -1 - \displaystyle \sum_{ i \  | \ (k / k_i \in \mathbb{N} \backslash \{ 0,1 \} , \atop k_i \  {\rm prime \  number})} p_i\,.
\end{equation}

So, the effective number of resonances can be computed using the following recurrence series:

\begin{equation}
N_{kc} = \displaystyle \sum_{k=1}^{k_c} p_k.
\label{Nkc_exact}
\end{equation} 

\begin{figure}[htb]
\centering
{\includegraphics[width=0.475\textwidth]
{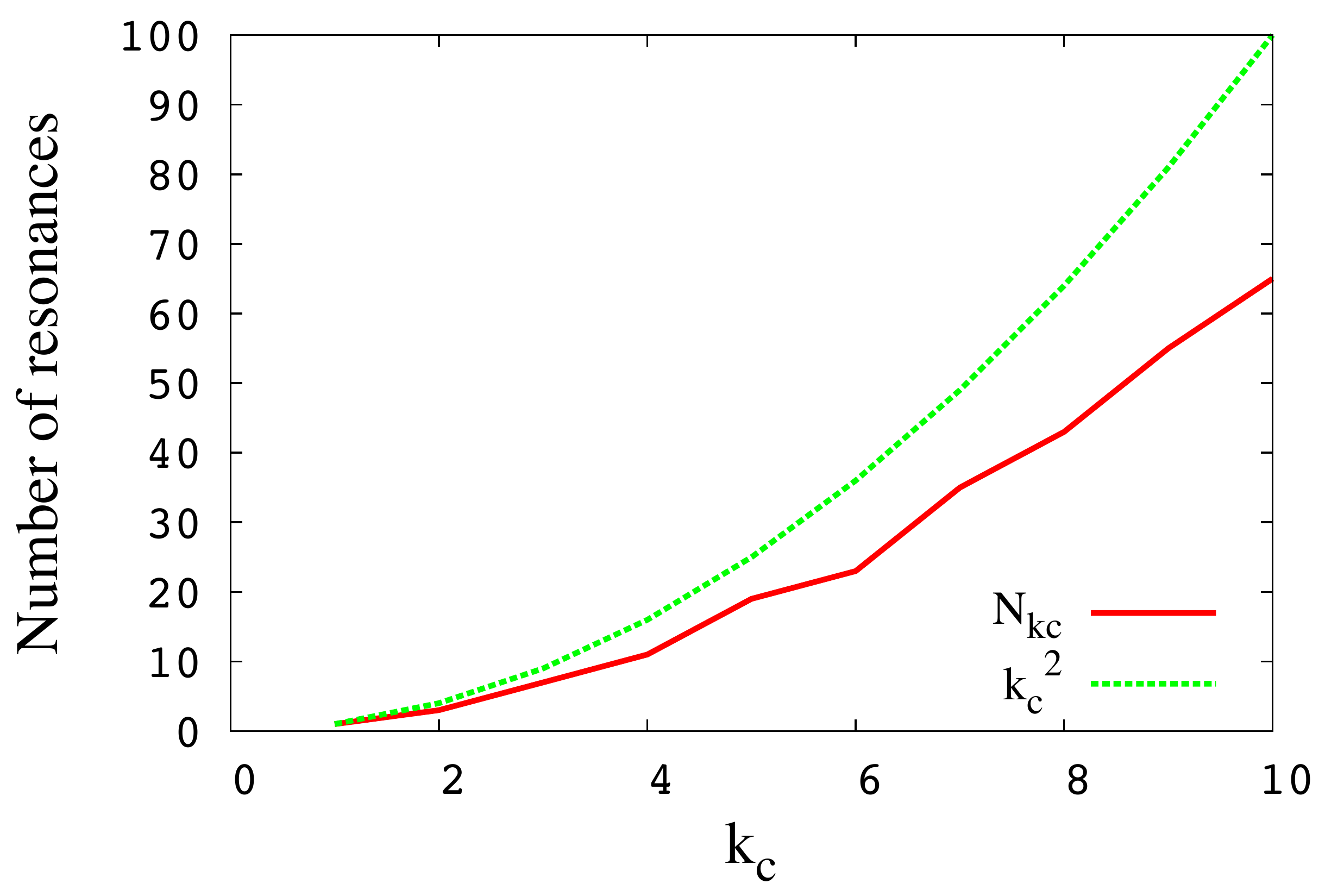}
\textsf{\caption{\label{fig:number_peaks} The real number of resonances $ N_{kc} $ (continuous red line) and its first order approximation $ k_{c}^2 $ (dashed green line) as a function of the rank of the highest harmonics $ k_c $ for the main resonances ($ 1 \leq k \leq 10 $).  } }}
\end{figure}

The values of $ N_{kc} $ and $ k_c^2 $ for the first orders are given in Table~\ref{number_peaks_values} and plotted in Fig.~\ref{fig:number_peaks}. We notice a slight difference between the two curves, the effective number of peaks growing slower than the number of modes. As a first approximation, we can consider that $ N_{kc} \propto k_{c}^2 $. This assumption will be used to compute the scaling laws of $ N_{\rm kc} $.

\begin{table}[h]
\centering
    \begin{tabular}{ c | cccccccccc }
      \hline
      \hline
      $ k_c $ & $ 1 $ & $ 2 $ & $ 3 $ & $ 4 $ & $ 5 $ & $ 6 $ & $ 7 $ & $ 8 $ & $ 9 $ & $ 10 $\\
      $ k_c^2 $ & $ 1 $ & $ 4 $ & $ 9 $ & $ 16 $ & $ 25 $ & $ 36 $ & $ 49 $ & $ 64 $ & $ 81 $ & $ 100 $\\
      $ N_{kc} $ & $ 1 $ & $ 3 $ & $ 7 $ & $ 11 $ & $ 19 $ & $ 23 $ & $ 35 $ & $ 43 $ & $ 55 $ & $ 65 $\\
      \hline
    \end{tabular}
    \textsf{\caption{\label{number_peaks_values} Numerical comparison between the number of peaks $ N_{kc} $ and the number of modes $ k_c^2 $, $ k_c $ being the rank of the highest harmonics, for the main resonances ($ 1 \leq k \leq 10 $).}}
 \end{table}

\subsection{Width of resonances}

The width of a resonance is a characteristic of great interest. Indeed, in a previous work \citep{ADLPM2014}, we studied a coplanar two-body system of semi-major axis $ a $ and showed that $ a $ was submitted to rapid changes due to resonances \citep[see also][]{WS1999,WS2001,WS2002}. In this work, the system is constituted by a fluid rotating planet and a pointlike satellite. The variation $ \Delta a $ of the semi-major axis $ a $ is related to the width at mid-height $ l_p $ and sharpness ratio $ \Xi_p = H_p / H_{\rm bg} $ of the $p$-th resonance encountered. Indeed, $ \Delta a / a \propto l_p \Xi_p^{1/4} $, i.e. the amplitude $ \Delta a $ linearly varies as a function of the width of the peak that causes it, this width being the signature of the dissipation. In this section, we compute an analytical formula of the width at mid-height $ l_{mn} $ as a function of the control parameters of the system. Similarly to the eigenfrequencies, $ l_{mn} $ are fully determined by the left-hand side of Eq. (\ref{NavierStokes}). We suppose that $ \xi_{mn} $, the numerator of $ \zeta_{mn} $ in Eqs. (\ref{zetamn_visc}) or (\ref{zetamn_therm}), varies smoothly compared to its denominator $ d_{mn} $ (see Eq. \ref{dmn}). Then, the width at mid-height is computed from the relation:

\begin{equation}
\zeta_{mn} \left( \omega_{mn} + \frac{l_{mn}}{2} \right) =  \frac{1}{2} \zeta_{mn} \left( \omega_{mn}  \right),   
\end{equation}

that can also be expressed:

\begin{equation}
\frac{\xi_{mn}}{P \left(  \omega_{mn} + \displaystyle \frac{l_{mn}}{2} \right) } = \frac{1}{2} \frac{\xi_{mn}}{P \left(  \omega_{mn} \right) }.
\end{equation}

Therefore, we solve the equation:

\begin{equation}
P \left( \omega_{mn} + \frac{l_{mn}}{2} \right) = 2 P \left(  \omega_{mn} \right).
\end{equation}

So, the $ l_{mn} $ are the same for $ \zeta^{\rm visc} $ and $ \zeta^{\rm therm} $, like the eigenfrequencies. In the context of the quasi-adiabatic approximation, we obtain:

\begin{equation}
l_{mn} = \left( m^2 + n^2 \right) \frac{ A m^2 K + \left( 2 n^2 \cos^2 \theta +A m^2 \right) E }{n^2 \cos^2 \theta + A m^2}.
\label{lmn}
\end{equation}

Looking at the form of this expression, we introduce two critical numbers proper to the mode $ \left( m,n \right) $,

\begin{equation}
\begin{array}{ccc}
A_{mn} \left( \theta \right) = \displaystyle \frac{2 n^2}{m^2} \cos^2 \theta & \mbox{and} & P_{r;mn} \left(  \theta, A \right) = \displaystyle \frac{ A }{A + A_{mn} \left(\theta \right)},
\end{array}
\label{Apr}
\end{equation}

that determine asymptotical behaviors. The expression of $ l_{mn} $ thus becomes:

\begin{equation}
l_{mn} = \left( m^2 + n^2 \right) \frac{ E \left( A + A_{mn} \right) \left( P_{r} + P_{r;mn} \right) }{P_{r} \left( A + \displaystyle{\frac{1}{2}} A_{mn} \right)}.
\label{lmn_2}
\end{equation}

\begin{figure}[htb]
\centering
{\includegraphics[width=0.475\textwidth]
{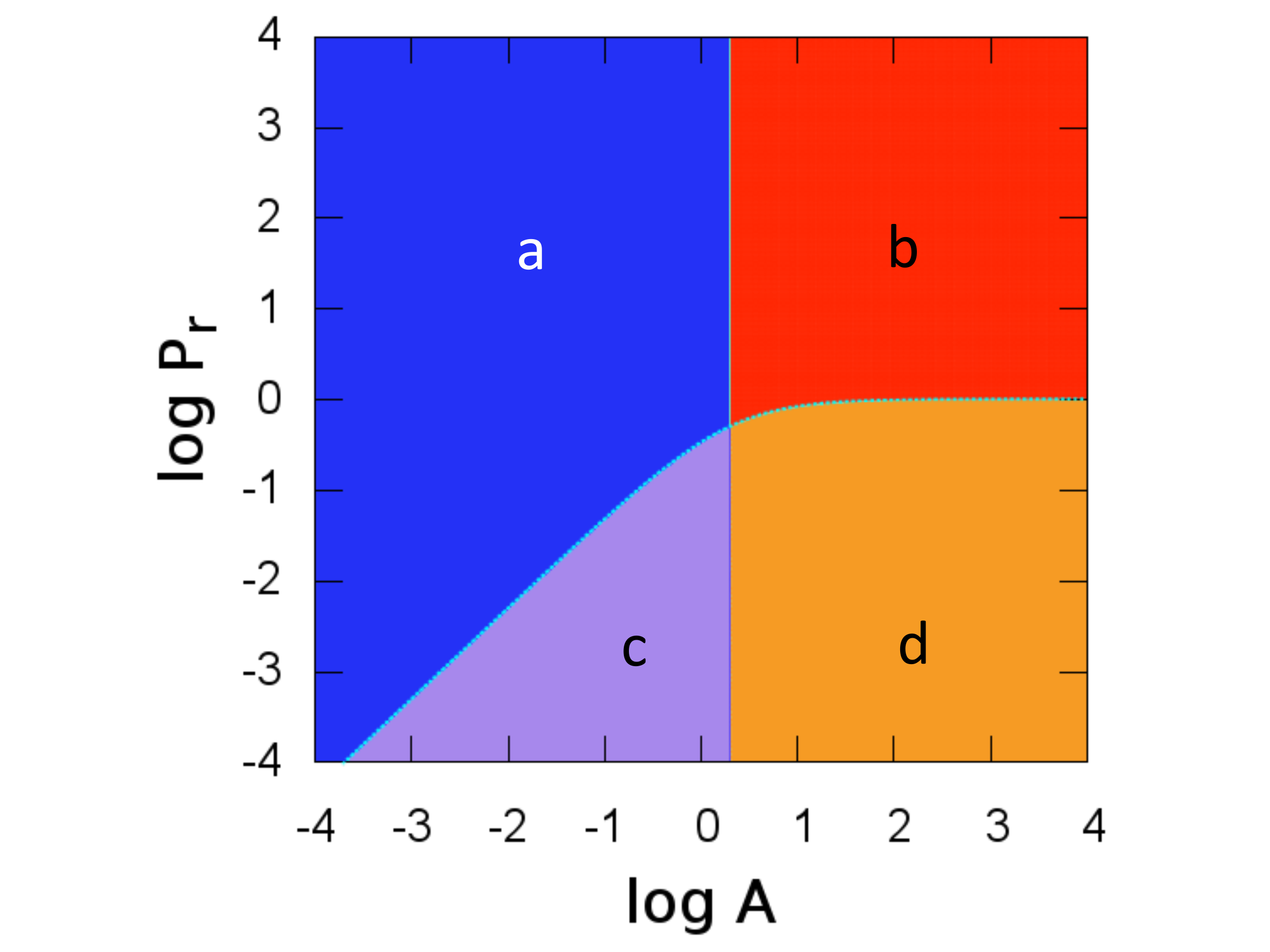}
\textsf{\caption{\label{fig:domaines} Asymptotic domains. The areas on the left (\textbf{a} and \textbf{c}) correspond to inertial waves, the ones on the right (\textbf{b} and \textbf{d}) to gravito-inertial waves. Resonances are damped by viscosity in areas \textbf{a} and \textbf{b}, and by thermal diffusivity in areas \textbf{c} and \textbf{d}. The corresponding domains of parameters are: \textbf{a:} $ A \ll A_{mn} $ and $ P_{r} \gg P_{r;mn} $; \textbf{b:} $ A \gg A_{mn} $ and $ P_{r} \gg P_{r;mn} $; \textbf{c:} $ A \ll A_{mn} $ and $ P_{r} \ll P_{r;mn} $; \textbf{d:} $ A \gg A_{mn} $ and $ P_{r} \ll P_{r;mn} $. } }}
\end{figure}

$ A \ll A_{mn} $ characterizes inertial waves and $ A \gg A_{mn} $ gravito-inertial waves. In the same way, if $ P_{r} \ll P_{r;mn} $, the resonance is damped by thermal diffusion and if $ P_{r} \gg P_{r;mn} $, it is damped by viscous friction. We thus identify four possible regimes, summarized in Fig. \ref{fig:domaines}:
\begin{itemize}
  \item[1.] $ A \ll A_{mn} $ and $ P_{r} \gg P_{r;mn} $: inertial waves damped by viscous diffusion;\\
  \item[2.] $ A \gg A_{mn} $ and $ P_{r} \gg P_{r;mn} $: gravity waves damped by viscous diffusion;\\
  \item[3.] $ A \ll A_{mn} $ and $ P_{r} \ll P_{r;mn} $: inertial waves damped by thermal diffusion;\\
  \item[4.] $ A \gg A_{mn} $ and $ P_{r} \ll P_{r;mn} $: gravity waves damped by thermal diffusion.
\end{itemize}

The scaling laws of Tab.~\ref{largeurs} are directly deduced from Eq.~(\ref{lmn}). They illustrate the differences between the four identified regimes. First, we focus on inertial waves. If the viscous term overpowers the term of heat diffusion, then we are in the case studied by \cite{OL2004}. The width at mid-height of resonances linearly varies with the Ekman number. For $ E = 10^{-2} $, peaks are larger than for $ E = 10^{-5} $ (see Fig. \ref{fig:spectre_inertiel_1}).  Elsewhere, the width is proportional to $ K $: for a given $ A > 0 $, the resonances would widen with $ K $ as they do with $ E $ in the previous case. Now, if we examine gravito-inertial waves, they behave similarly as inertial waves, linearly widening with $ E $ and $ K $ in the corresponding regimes defined above. Finally, note that $ l_{mn} $ always depends on a single parameter except in the case of inertial waves damped by thermal diffusion, for which the square frequencies ratio ($A$) has also a linear impact.\\

\begin{table}[h]
\centering
    \begin{tabular}{ c | c c }
      \hline
      \hline
      \textsc{Domain} & $ A \ll A_{mn} $ & $ A \gg A_{mn} $ \\
      \hline
      \vspace{0.1mm}\\
      $ P_{r} \gg P_{r;mn} $ & $ 2 \left( m^2 + n^2 \right)E  $ & $  \left( m^2 + n^2 \right) E  $ \\
      \vspace{0.1mm}\\
      $ P_{r} \ll P_{r;mn} $ & $ \displaystyle  \frac{m^2  \left( m^2 + n^2 \right) A E P_{r}^{-1} }{n^2 \cos^2 \theta} $ & $  \left( m^2 + n^2 \right) E P_{r}^{-1} $\\
      \vspace{0.1mm}\\
       \hline
    \end{tabular}
    \textsf{\caption{\label{largeurs} Scaling laws of the width at mid-height $ l_{mn} $ of the resonance associated to the doublet $ \left( m , n \right) $. $ l_{mn} $ is written as a function of the control parameters $ A $, $ E $ and $ K\equiv E P_{r}^{-1}$.}}
 \end{table}

\begin{figure*}[ht!]
 \centering
 \includegraphics[width=0.48\textwidth,clip]{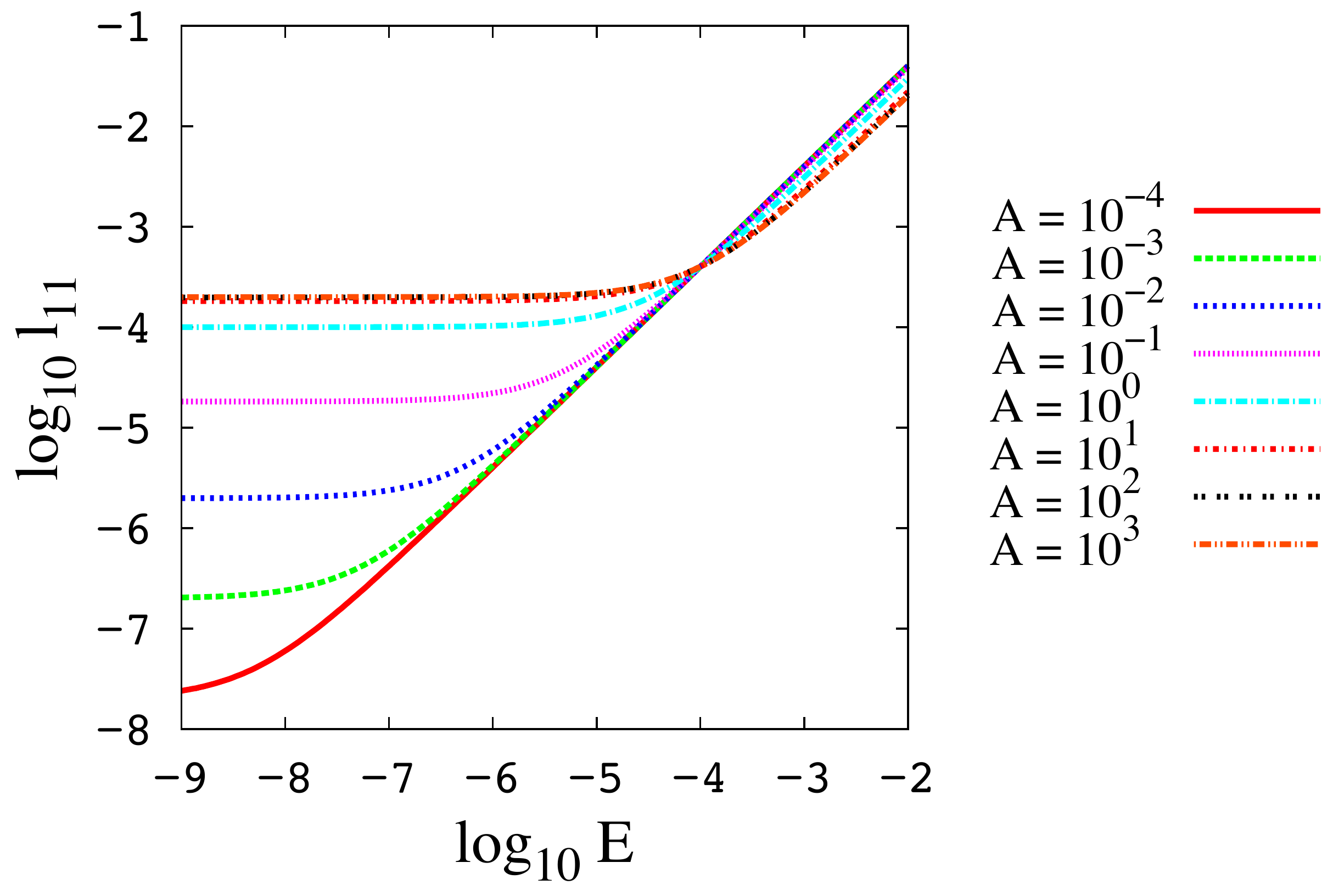}%
 \includegraphics[width=0.48\textwidth,clip]{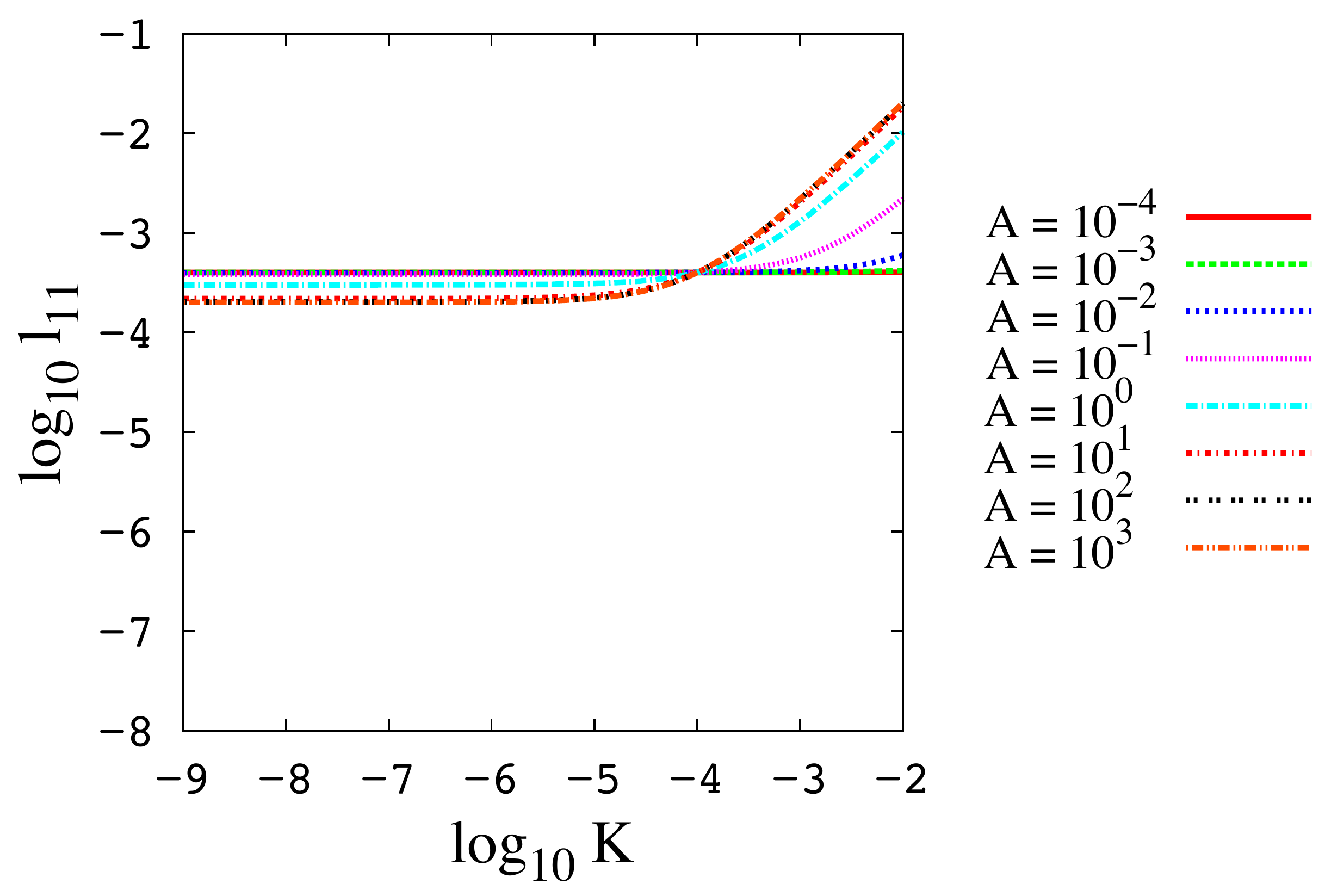} 
 \includegraphics[width=0.48\textwidth,clip]{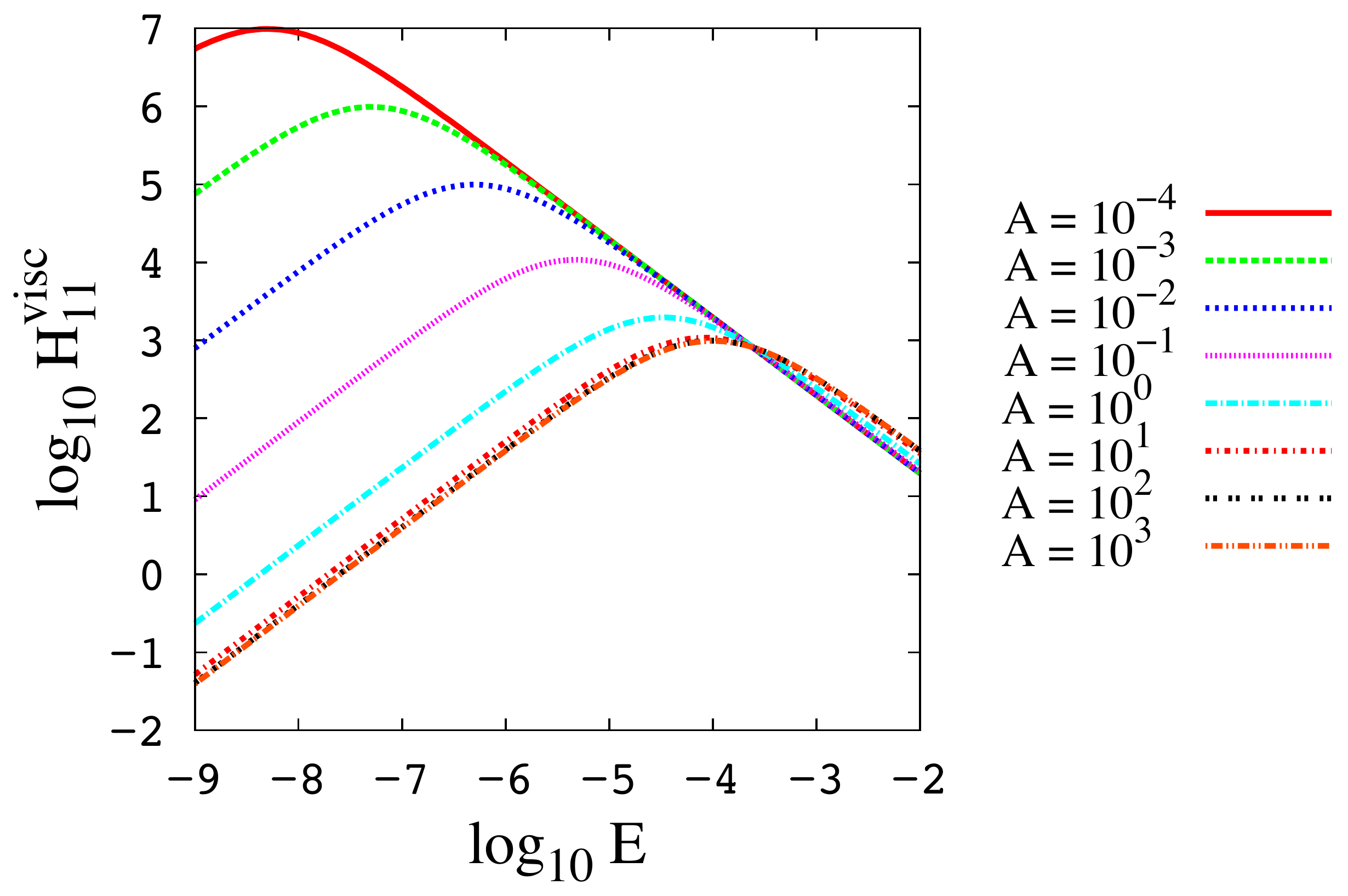}
 \includegraphics[width=0.48\textwidth,clip]{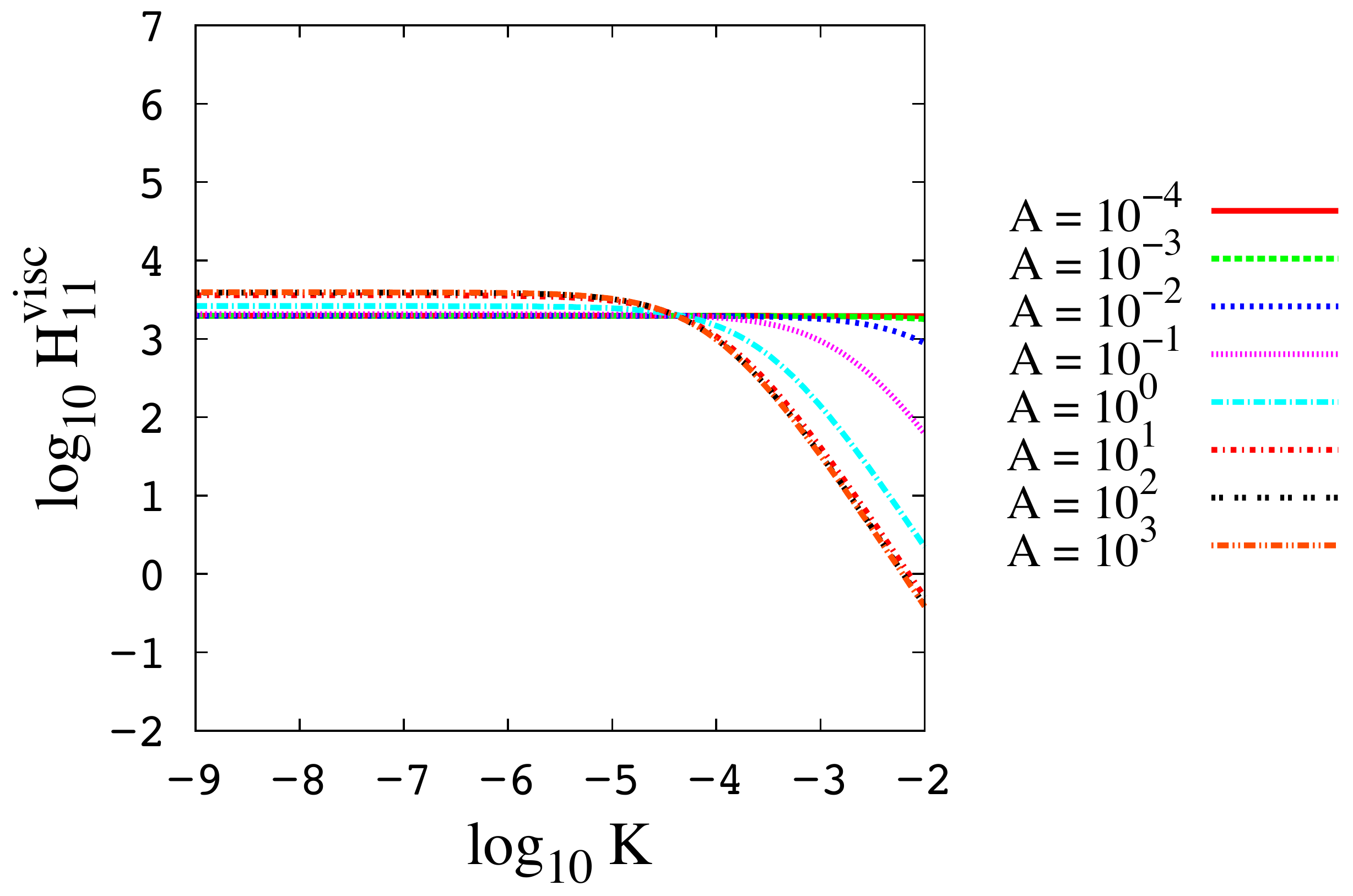}     
  \textsf{ \caption{\label{fig:lH_EA} Width at mid-height $ l_{11} $ and height $ H_{11}^{\rm visc} $ of the main resonance of $ \zeta^{\rm visc} $ as a function of the Ekman number $ E $ and its analogue for heat diffusion $ K $  for different values of $ A $ (in logarithmic scales). {\bf Top left:} $ l_{11} - E $. {\bf Top right:} $ l_{11} - K $. {\bf Bottom left:} $ H_{11}^{\rm visc} - E $. {\bf Bottom right:} $ H_{11}^{\rm visc} - K $.}}
\end{figure*}

Assuming that the resonances all have the same qualitative behavior, we plot the width at mid-height $ l_{11} $ of the main one from Eq. (\ref{lmn}). We thus retrieve in the plots the regimes predicted analytically as illustrated in Fig.~\ref{fig:lH_EA}, in particular transition zones. Concerning the nature of the damping, the transition zones are indicated by corners. They correspond to the critical Prandtl number $ P_{r;11} $. $ l_{11} $ also varies with the nature of the waves. If $ A \ll 1 $, the Coriolis acceleration drives the dynamics and the perturbation generates quasi-inertial waves.  If $A\!>\!\!>\!1$, it generates quasi-gravity waves. This is the reason why we can observe two asymptotical curves on the plots. If we come back to the case of inertial waves damped by viscous diffusion \citep{OL2004}, $ l_{11} $ is proportional to $ E $ and does not varies with $ K $. Thus, the peaks become thinner when $ E $ decreases that explains the behaviors observed in Figs.~\ref{fig:spectre_inertiel_1} and \ref{fig:spectre_inertiel_2}. In Appendix B, the reader will find color maps of the relative difference between the prediction of the analytical expression (Eq. \ref{lmn}), and the real theoretical result computed from the complete formula of $ \zeta^{\rm visc} $ (Eq. \ref{energy_visc}). They validate the expression of $ l_{11} $ in asymptotical regimes and highlight transition zones where it cannot be used.\\

Thanks to the scaling laws of $ l_{mn} $, it becomes possible to constrain the variations of orbital parameters caused by a resonance of the internal dissipation. For instance, the amplitude $ \Delta a $ for the semi-major axis $ a $ of our two-body system \citep{ADLPM2014} verify the law $ \Delta a /a \propto l_{p} \Xi_p^{1/4} \propto E /\Xi_p^{1/4} $ in the regime of inertial waves damped by viscous friction.


\subsection{Amplitude of resonances}

The height of a resonant peak also has an influence on the secular evolution of orbital parameters. Thus, we establish an analytical formula of the height $ H_{mn} $ of the peaks. Of course, the height of resonances depends on the tidal forcing $ \textbf{f} $, which requires that we choose a form for it. Following \cite{OL2004}, we take perturbation coefficients of the form:

\begin{equation}
\begin{array}{ccc}
f_{mn} = i \displaystyle \frac{F}{ \left| m  \right|  n^2}, &  g_{mn} = 0, & h_{mn} = 0,  
\end{array}
\end{equation}

and assuming $ \left\{E,K\right\} \ll \left\{\sqrt{A},\cos \theta\right\}$, we note that:

\begin{equation}
P \left( \omega_{mn}  \right) \approx \left[  Am^2 K + \left( 2n^2 \cos^2 \theta + A m^2  \right)E \right]^2, 
\end{equation} 

so that we can compute simple expressions for the heights of the peaks of $ \zeta^{\rm visc} $ and $ \zeta^{\rm therm} $.

\subsubsection{Dissipation by viscous friction}

First, we look at the resonances of $ \zeta^{\rm visc} $. Naming $ H_{mn}^{\rm visc} $ their heights, we obtain the relation:

\begin{equation}
H_{mn}^{\rm visc} = \frac{8 \pi F^2 E}{m^2 n^2 \left( m^2 + n^2  \right)^2} \frac{ \left(  2 n^2 \cos^2 \theta + A m^2 \right) \left(  n^2 \cos^2 \theta + A m^2  \right)  }{ \left[ Am^2 K + \left( 2n^2 \cos^2 \theta + A m^2  \right)E\right]^2 },
\label{H_visc}
\end{equation}

where we recognize the critical numbers $ A_{mn} $ and $ P_{r;mn} $ introduced in the previous section:

\begin{equation}
H_{mn}^{\rm visc} = \frac{8 \pi F^2 P_{r;mn}^2}{m^2 n^2 \left( m^2 + n^2  \right)^2} \frac{ P_{r}^2   \left( A + {\displaystyle\frac{1}{2}} A_{mn}   \right) }{E \left( A + A_{mn} \right) \left(  P_{r} + P_{r;mn} \right)^2 }.
\label{Hmn_visc}
\end{equation}

Thus, the height is characterized by the same asymptotic domains as the width. The frontiers between them depend on $ A_{mn} $ and $ P_{r;mn} $ (Table \ref{table_H_visc}). But it has different scaling laws. For example, when dissipation is led by viscosity, $ H_{mn}^{\rm visc} $ is inversely proportional to the Ekman number contrary to $ l_{mn} $. Therefore, the heights of the resonances increase if the Ekman number $ E $ decreases, which corroborates the plots of Fig. \ref{fig:spectre_inertiel_1}. If dissipation is led by heat diffusion, then $ H_{mn}^{\rm visc} $ depend on both $ E $ and $ K $. Of course, the scaling laws of $ H_{mn}^{\rm visc} $ are related to the spectral decomposition of the forcing. This latter partly determines the decay law of the heights with the wave number $ k $:  $ H_{mn}^{\rm visc} \propto 1 / k^8 $. Here, $ f_{mn} \propto 1 / k^3 $. But if $ f_{mn} $ did not depend on  $ m $ and $ n $, then the decay law would be $ H_{mn}^{\rm visc} \propto 1 / k^2 $. Consequently, with the forcing chosen for this study, the height of the resonant peaks becomes comparable with the non-resonant background rapidly. Therefore, few terms $ \zeta_{mn}^{\rm visc} $ of the lowest orders are sufficient to plot the spectrum and there is no need to take into account a large number of harmonics to describe the dissipation properly.\\ 

In the planet-satellite system studied by \cite{ADLPM2014}, we can thus write $ \Delta a /a $ with these new scaling laws:

\begin{equation}
 \frac{\Delta a}{a} \propto l_p \left( \frac{H_p}{H_{\rm bg}} \right)^{1/4} \propto E^{3/4} H_{\rm bg}^{-1/4} \left(A,E,K \right)
\end{equation}

in the regime of inertial waves damped by viscous friction.

\begin{table}[htb]
\centering
    \begin{tabular}{ c | c c }
      \hline
      \hline
      \textsc{Domain} & $ A \ll A_{mn} $ & $ A \gg A_{mn} $ \\
      \hline
      \vspace{0.1mm}\\
      $ P_{r} \gg P_{r;mn} $ & $ \displaystyle \frac{4 \pi F^2 E^{-1}}{m^2 n^2 \left( m^2 + n^2 \right)^2  } $ & $ \displaystyle \frac{8 \pi F^2 E^{-1} }{m^2 n^2 \left( m^2 + n^2 \right)^2  } $ \\
      \vspace{0.1mm}\\
      $ P_{r} \ll P_{r;mn} $ & $ \displaystyle \frac{16 \pi F^2 n^2 \cos^4 \theta A^{-2} E^{-1} P_{r}^{2}}{m^6 \left( m^2 + n^2 \right)^2 } $ & $ \displaystyle \frac{8 \pi F^2 E^{-1} P_{r}^2}{m^2 n^2 \left( m^2 + n^2 \right)^2  }  $\\
      \vspace{0.1mm}\\
       \hline
    \end{tabular}
    \textsf{\caption{\label{table_H_visc} Asymptotic behaviors of the height $ H_{mn}^{\rm visc} $ of the resonance of $ \zeta^{\rm visc} $ associated to the doublet $ \left( m , n \right) $.}}
 \end{table}

We plot the height of the main resonance $ H_{11}^{\rm visc} $ (Fig. \ref{fig:lH_EA}) as a function of $ A $, $ E $ and $ K $, using the analytical scaling laws (Eq. \ref{H_visc} and Tab. \ref{table_H_visc}). The maximum of $ H_{11}^{\rm visc} $ (Fig. \ref{fig:lH_EA}, bottom-left panel) results from the inversion of the dependence on $ E $, for $ P_{r} \sim P_{r;11} $. So, it highlights the transition zone between viscous friction and thermal damping. As expected, the position of this maximum varies with $ A $ until $ A \gg 1 $, where $ P_{r;11} \approx 1 $ asymptotically. We recall that Eq. (\ref{H_visc}) is relevant for small values of $ E $ and $ K $ (see Fig.~\ref{fig:ecarts} in Appendix B).

\subsubsection{Dissipation by thermal diffusion}

An expression of $ H_{mn}^{\rm therm} $ of the same form as Eq.~(\ref{H_visc}) can be deduced from Eq.~(\ref{energy_therm}),

\begin{equation}
H_{mn}^{\rm therm} = \frac{8 \pi K F^2}{n^2 \left( m^2 + n^2 \right)^2} \frac{n^2 \cos^2 \theta + m^2 A}{\left[ A m^2 K + \left( 2 n^2 \cos^2 \theta + A m^2 \right)E \right]^2},
\end{equation}

which can also be written as a function of the transition parameters $ A_{mn} $ and $ P_{r;mn} $,

\begin{equation}
H_{mn}^{\rm therm} = \frac{8 \pi F^2 }{ m^2 \left( m^2 + n^2 \right)^2} \frac{ P_{r} \left(  A + {\displaystyle\frac{1}{2}} A_{mn} \right) }{E  \left( A + A_{mn} \right)^2 \left( P_{r} + P_{r;mn} \right)^2  }.
\label{Hmn_therm}
\end{equation}

Thus, the asymptotic regimes of $ H_{mn}^{\rm therm} $ correspond to the four zones identified for $ l_{mn} $ and $ H_{mn}^{\rm visc} $ (see Fig.~\ref{fig:domaines}).

\begin{table}[htb]
\centering
    \begin{tabular}{ c | c c }
      \hline
      \hline
      \textsc{Domain} & $ A \ll A_{mn} $ & $ A \gg A_{mn} $ \\
      \hline
      \vspace{0.1mm}\\
      $ P_{r} \gg P_{r;mn} $ & $ \displaystyle \frac{2 \pi F^2  E^{-1} P_{r}^{-1}}{n^4 \left( m^2 + n^2 \right)^2 \cos^2 \theta}   $ & $ \displaystyle \frac{8 \pi F^2 A^{-1} E^{-1} P_{r}^{-1} }{m^2 n^2 \left( m^2 + n^2 \right)^2  } $ \\
      \vspace{0.1mm}\\
      $ P_{r} \ll P_{r;mn} $ & $ \displaystyle \frac{8 \pi F^2 \cos^2 \theta A^{-2} E^{-1} P_{r}}{m^4 \left( m^2 + n^2 \right)^2 }  $ & $ \displaystyle \frac{8 \pi F^2 A^{-1} E^{-1} P_{r}}{m^2 n^2 \left( m^2 + n^2 \right)^2  } $\\
      \vspace{0.1mm}\\
       \hline
    \end{tabular}
    \textsf{\caption{\label{table_H_therm} Asymptotic behaviors of the height $ H_{mn}^{\rm therm} $ of the resonance of $ \zeta^{\rm therm} $ associated to the doublet $ \left( m , n \right) $.}}
 \end{table}

\subsubsection{Total dissipation}

As $ \zeta^{\rm visc} $ and $ \zeta^{\rm therm} $ have the same resonance eigenfrequencies, their peaks are superposed. Therefore, when one of the dissipation mechanisms predominates at resonances, the observed peaks are all scaled by the law corresponding to this mechanism. To establish the scaling laws of the total energy, we have to delimit the domains of dissipation by viscous friction, where $ H_{mn} \sim H_{mn}^{\rm visc} $, and of dissipation by heat diffusion, where $ H_{mn} \sim H_{mn}^{\rm therm} $. So, we introduce the ratio $ \varsigma_{mn} $,

\begin{equation}
\varsigma_{mn} = \frac{H_{mn}^{\rm visc}}{H_{mn}^{\rm therm}} = P_{r} \left( A + A_{mn} \right).
\end{equation}

\begin{figure*}[htb]
 \centering
 \includegraphics[width=0.48\textwidth,clip]{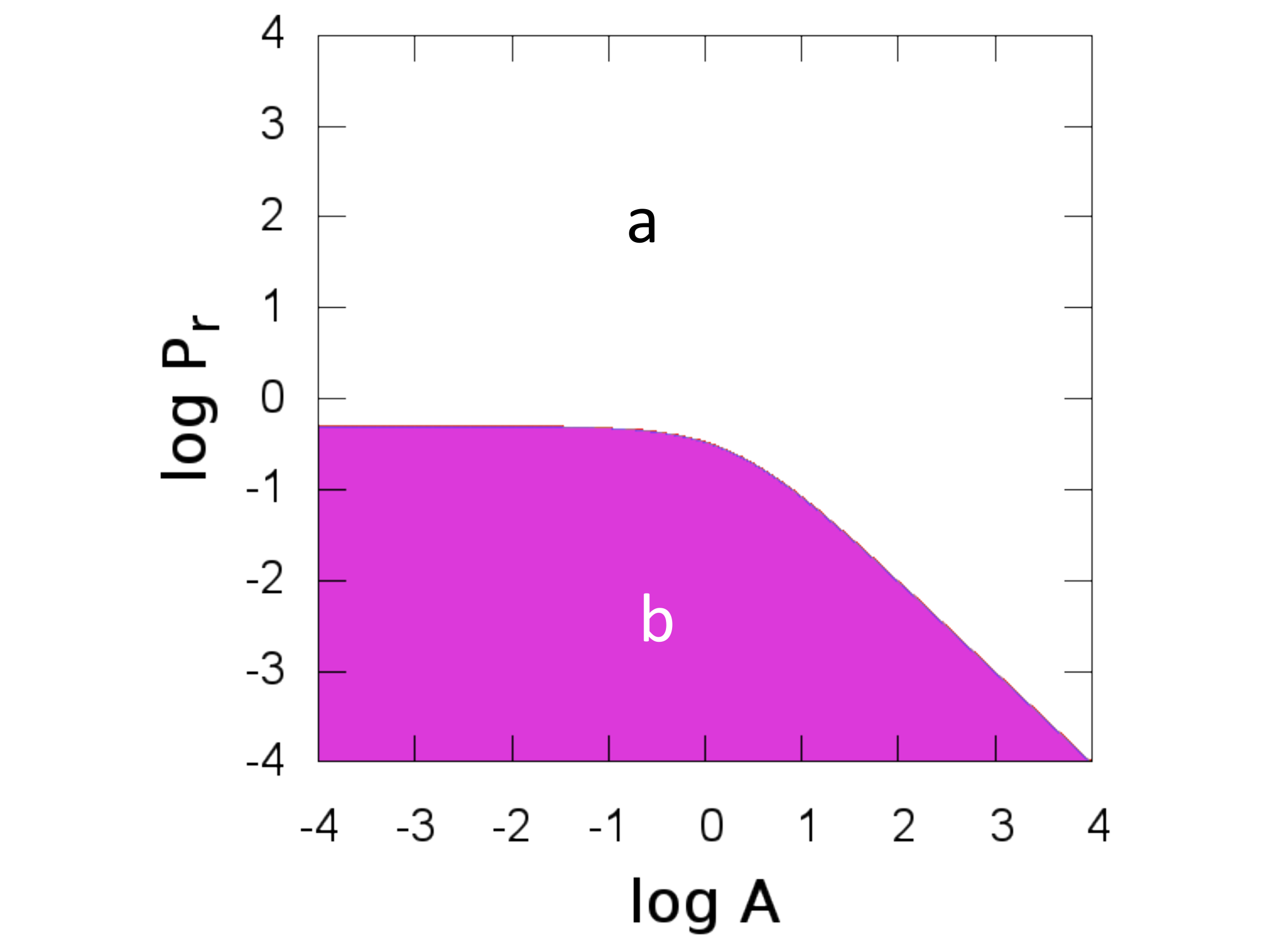}%
 \includegraphics[width=0.48\textwidth,clip]{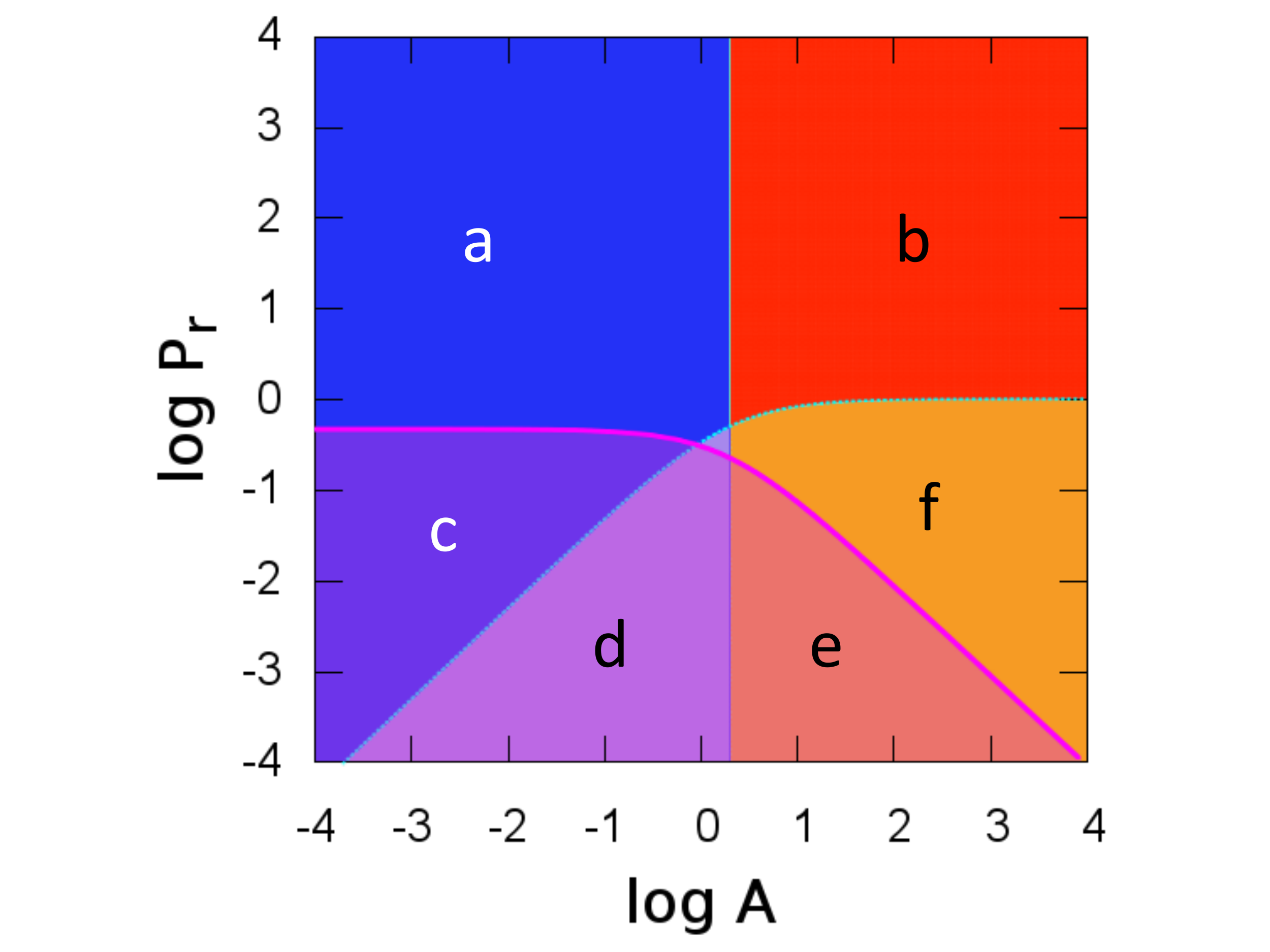}  
 
\textsf{\caption{\label{fig:map_zones_2} \textbf{Left:} Zones of predominances for dissipative mechanisms. In the pink area (\textbf{b}), $ \zeta^{\rm therm} \gg \zeta^{\rm visc} $: the dissipation is mainly due to thermal diffusion. In the white area (\textbf{a}), it is led by viscous friction. The transition zone corresponds to $ P_{r} \approx P_{r;11}^{\rm diss} $, where $ \zeta^{\rm visc} \sim \zeta^{\rm therm} $. \textbf{Right:} Asymptotic domains with the predominance zones. Low Prandtl-number areas of Fig.~\ref{fig:domaines} are divided in sub-areas corresponding to the locally predominating dissipation mechanism. In \textbf{a}, \textbf{b} and \textbf{f}, $ \zeta^{\rm therm} \ll \zeta^{\rm visc} $ while in \textbf{c}, \textbf{d} and \textbf{e}, $ \zeta^{\rm therm} \gg \zeta^{\rm visc} $.}}
\end{figure*}

The transition zone is characterized by $ \varsigma_{mn} \sim 1 $, or $ P_{r} \sim P_{r;mn}^{\rm diss} $, where 
\begin{equation}
P_{r;mn}^{\rm diss} \left( \theta, A \right)  = \frac{1}{A + A_{mn} \left( \theta \right) }
\end{equation}
is the Prandtl number indicating this transition zone. If $ P_{r} \ll P_{r;mn}^{\rm diss} $, then $ H_{mn} \sim H_{mn}^{\rm therm} $ (Fig.~\ref{fig:map_zones_2}, left panel); otherwise, $ H_{mn} \sim H_{mn}^{\rm visc} $. Taking into account the new zones appearing with this condition on $ P_{r} $ (Fig.~\ref{fig:map_zones_2}, right panel), we obtain the scaling laws for $ H_{mn} $, the heights of resonances of the total dissipation given in Tab.~\ref{table_H}. For the sake of simplicity we introduce a new Prandtl number,

\begin{equation}
P_{r;mn}^{\rm reg} \left( A \right) = \max \left(P_{r;mn}^{\rm diss} \left( A \right), P_{r;mn} \left( A \right)  \right),
\end{equation}

that separates the regimes where $P_{r}$ intervenes in scaling laws from others.

\begin{table*}[htb]
\centering
\begin{tabular}{| c | c | c | c | c |} 
   \hline
   \hline
   \textsc{Domain} & \multicolumn{2}{c|}{$ A \ll A_{mn} $} & \multicolumn{2}{c|}{$ A \gg A_{mn} $} \\
   \hline
   \vspace{0.1mm} & \multicolumn{2}{c|}{} & \multicolumn{2}{c|}{} \\
    $ P_{r} \gg P_{r;mn}^{\rm reg} $ & \multicolumn{2}{c|}{ $ \displaystyle \frac{4 \pi F^2 E^{-1}}{m^2 n^2 \left( m^2 + n^2 \right)^2} $  } & \multicolumn{2}{c|}{$ \displaystyle \frac{8 \pi F^2 E^{-1}}{m^2 n^2 \left( m^2 + n^2 \right)^2} $} \\
   \vspace{0.1mm} & \multicolumn{2}{c|}{} & \multicolumn{2}{c|}{} \\
    \hline
   \vspace{0.1mm} &  & & &\\
    \multirow{5}{*}{$ P_{r} \ll P_{r;mn}^{\rm reg} $} & $ P_{r} \gg P_{r;mn} $ & $ \displaystyle \frac{2 \pi F^2 E^{-1} P_{r}^{-1}}{n^4 \left( m^2 + n^2 \right)^2 \cos^2\theta} $ & $ P_{r} \gg P_{r;mn}^{\rm diss} $ & $ \displaystyle \frac{8 \pi F^2 E^{-1} P_{r}^2}{m^2 n^2 \left( m^2 + n^2 \right)^2} $ \\
    \vspace{0.1mm} &  & & &\\
    \cline{2-5} 
    \vspace{0.1mm} &  & & &\\
           & $ P_{r} \ll P_{r;mn} $ & $ \displaystyle \frac{8 \pi F^2 \cos^2 \theta A^{-2} E^{-1} P_{r}}{m^4 \left( m^2 + n^2  \right)^2} $ & $ P_{r} \ll P_{r;mn}^{\rm diss} $ & $ \displaystyle \frac{8 \pi F^2 A^{-1} E^{-1} P_{r}}{m^2 n^2 \left( m^2 + n^2 \right)^2} $ \\
    \vspace{0.1mm} &  & & &\\
    \hline
\end{tabular}
\textsf{\caption{\label{table_H} Asymptotic behaviors of the height $ H_{mn} $ of the resonance of $ \zeta $ associated to the doublet $ \left( m , n \right) $. Scaling laws correspond to the areas of Fig.~\ref{fig:map_zones_2}.}}
\end{table*}

\subsection{The non-resonant background}

The height of the non-resonant background $ H_{\rm bg} $ is the magnitude of the energy dissipated inside the frequency range $ \left[  \omega_{\rm inf} ,\omega_{\rm sup} \right] $, in non-resonant areas. It only depends on the first term of the sum in the expression of $ \zeta $ (Eq.~\ref{energy_visc}). Indeed, each mode is related to a typical wavelength that increases when the wave numbers $ m $ and $ n $ decay. Thus, at the eigenfrequencies $ \omega = \omega_{mn} $ ($ \left( m,n \right) \in \mathbb{N}^2 $), the perturbation presents a characteristic pattern of wavelengths $ \lambda_h = L / m $ in the $ x $ direction and $ \lambda_v = L/n $ in the $ z $ direction. But it is dominated by the lowest-order pattern $ m = n = 1 $ in the absence of resonance. If we had $ L \sim R $, then our first mode would correspond to the large-scale hydrostatic adjustment of the flow in phase with the perturber, the equilibrium tide or non wave-like displacement introduced above \citep[see e.g.][]{RMZ2012,Ogilvie2013}.\\

In this framework, the height of the non-resonant background gives us information about the mean dissipation and the smooth component of the tidal quality factor ($ Q $). Therefore, it plays an important role in the secular evolution of planetary systems. Indeed, it is also necessary to compute the sharpness ratio $ \Xi\equiv H_{11}/H_{\rm bg}  $ intervening in the expression of $ \Delta a $. Thus, in this subsection, we estimate $ H_{\rm bg} $ by computing the term of the main resonance, $ \zeta_{11} $, at the frequency $ \omega_{\rm bg} = \left( \omega_{11} + \omega_{21} \right)/2 $, that can be written:

\begin{equation}
\omega_{ \rm bg} = \omega_{11} \left( 1 + \varepsilon_{12}  \right),
\end{equation}

 with the relative distance between the two peaks,
 
 \begin{equation}
  \varepsilon_{12} = \displaystyle \frac{1}{2} \frac{ \omega_{21} - \omega_{11} }{ \omega_{11}}.
  \label{eps12}
 \end{equation}
 
Note that if $ A = \cos^2 \theta $ (critical hyper-resonant case), the characteristic level of the non-resonant background is not defined and $ \varepsilon_{12} = 0 $. Considering that the contributions of the main peaks are approximately the same, we write the characteristic height of the background: 
 
 \begin{equation}
 H_{\rm bg} \approx 4 \zeta_{11} \left( \omega_{\rm bg} \right).
 \end{equation}

\subsubsection{Dissipation by viscous friction}

We first focus on the contribution of viscous friction,

\begin{equation}
H_{\rm bg}^{\rm visc} \approx 4 \zeta_{11}^{\rm visc} \left( \omega_{\rm bg} \right).
\end{equation}

Assuming $\left\{E, K\right\} \ll \left\{ \sqrt{A},\cos \theta \right\}$, the previous expression becomes

\begin{equation}
H_{\rm bg}^{\rm visc} = 4 \pi F^2 E \varepsilon_{12}^2 \frac{  \left( A + \cos^2 \theta \right)  \left[ C_{\rm grav} \left( \varepsilon_{12} \right)   A  +C_{\rm in} \left( \varepsilon_{12} \right) \cos^2 \theta  \right]  }{   \left( A + \cos^2 \theta \right)^3 \varepsilon_{12}^2  + \xi \left( \theta, A, E,K  \right) }
\label{Hbg_visc_brut}
\end{equation}

with

\begin{equation}
\xi \left( \theta, A, E,K  \right) = 2 \alpha^2 \beta^2 \left( 1 + 4 \varepsilon_{12}  \right)  -  4 \alpha \beta \gamma \left( 1 + 2 \varepsilon_{12}  \right) + 2 \gamma^2.
\label{fonction_xi}
\end{equation}

The functions $ C_{\rm in} \left( \varepsilon_{12} \right)  $ and $ C_{\rm grav}  \left( \varepsilon_{12} \right) $ are expressed by

\begin{equation}
\begin{array}{lcl}
C_{\rm in}^{\rm visc} \left( \varepsilon_{12} \right) & = &  \displaystyle \frac{ \left( 2+2 \varepsilon_{12} + \varepsilon_{12}^2 \right) \left( 1 +\varepsilon_{12} \right)^2 }{\varepsilon_{12}^2} 
,\\
\vspace{0.1mm}\\
C_{\rm grav}^{\rm visc} \left( \varepsilon_{12} \right) & = & \displaystyle \frac{\left( 1 + \varepsilon_{12} \right)^4  }{\varepsilon_{12}^2} 
.\\
\end{array}
\label{Cin_Cgrav}
\end{equation}

The coefficients $ \beta $ and $ \gamma $ depend on $ E $ and $ K $ linearly (Eq. \ref{coefficients}). On the contrary, neither the Ekman number nor the dimensionless thermal diffusivity have an influence on $ \omega_{11} $ and $ \varepsilon_{12} $. Therefore, the background linearly varies with the Ekman number if the condition 

\begin{equation}
  \left|  \xi \left( \theta, A, E,K  \right) \right|  \ll  \left( A + \cos^2 \theta \right)^3 \varepsilon_{12}^2 
  \label{condition_xi}
\end{equation}

is satisfied. To unravel what this implies on the physical parameters, we need to study two asymptotic cases separately: 
\begin{itemize}
   \item[1.] $ A \ll \cos^2 \theta $ (quasi-inertial waves),\\
   \item[2.] $ A \gg \cos^2 \theta $ (gravito-inertial waves).\\
\end{itemize}

By deducing the main eigenfrequency $ \omega_{11} = \sqrt{\left( A + \cos^2 \theta \right)/2} $ from the dispersion relation (Eq.~\ref{dispersion}), we compute the relative distance between the two main resonances,

\begin{equation}
\begin{array}{lcl}
\varepsilon_{12} \approx  \displaystyle \frac{ \sqrt{2} - \sqrt{5} }{ \sqrt{20} }  =  \varepsilon_{\rm in} & \mbox{if} &  A \ll \cos^2 \theta, \\
\vspace{0.1mm}\\
\varepsilon_{12} \approx  \displaystyle \frac{ 2 \sqrt{2} - \sqrt{5} }{\sqrt{20} }  =  \varepsilon_{\rm grav} & \mbox{if} &  A \gg \cos^2 \theta.  \\
\end{array}
\end{equation}

Thus, the relative distance between $ \omega_{11} $ and $ \omega_{\rm bg} $ belongs to the interval $ \left[ \varepsilon_{\rm in},\varepsilon_{\rm grav} \right]  $, $ \varepsilon_{\rm in} $ and $ \varepsilon_{\rm grav} $ being the boundaries corresponding to the asymptotic cases $ A \ll \cos^2 \theta $ (i.e. the quasi-inertial waves) and $ A \gg \cos^2 \theta $ (i.e. the gravito-inertial waves) respectively. Numerically, $ \varepsilon_{\rm in} \approx -0.183 $ and $ \varepsilon_{\rm grav} \approx 0.132 $ and the asymptotic values of the coefficients $ C_{\rm in}^{\rm visc} $ and $ C_{\rm grav}^{\rm visc} $ are:

\begin{equation}
\begin{array}{lclcl}
    \mathcal{C}_{\rm in}^{\rm visc} & = & C_{\rm in}^{\rm visc} \left(  \varepsilon_{\rm in} \right) & = & 32.87\\
    \mathcal{C}_{\rm grav}^{\rm visc} & = & C_{\rm grav}^{\rm visc} \left(  \varepsilon_{\rm grav} \right) & = & 93.74. 
\end{array}
\end{equation}

From Eq.~(\ref{eps12}), we note that the dependence of the non-resonant background on $ E $ is linear only if:

\begin{equation}
\min \left\{ \sqrt{A}, \cos \theta  \right\} \gg  \max \left\{ E, K \right\},
\label{condition_Hbg}
\end{equation}

which corresponds to the quasi-adiabatic approximation met before and assumed in the whole study. We thus finally obtain the expression of $ H_{\rm bg}^{\rm visc} $ for inertial waves and gravity waves (Table \ref{background_visc}), 

\begin{equation}
H_{\rm bg}^{\rm visc} = 4 \pi F^2 E  \frac{  \mathcal{C}_{\rm grav}^{\rm visc}  A  + \mathcal{C}_{\rm in}^{\rm visc}  \cos^2 \theta  }{   \left( A + \cos^2 \theta \right)^2 }.
\label{Hbg_visc}
\end{equation}

Note that the background does not depend on the Prandtl number under the quasi-adiabatic approximation. Its level is only defined by the ratio $ A/\cos^2 \theta \sim A / A_{11} $. So, the level of the non-resonant background is completely independent on the thermal diffusivity $ K $ in both asymptotic regimes. Besides, $ H_{\rm bg}^{\rm visc} $ increases with $ E $ whereas $ H_{mn}^{\rm visc} $ decays.\\ 

Note that thanks to our result for $ H_{\rm bg}^{\rm visc} $, we now have a complete scaling law for the magnitude of the variation of the semi-major axis caused by a resonance of the dissipation spectra in the regime of inertial waves damped by viscous friction in a convective planetary or stellar region \citep{ADLPM2014}: 
\begin{equation}
\frac{\Delta a}{a} \propto l_p \left( \frac{H_p}{H_{\rm bg}} \right)^{1/4} \propto E^{1/2}.
\label{eq:semi}
\end{equation}

\begin{table}[htb]
\centering
    \begin{tabular}{  c c }
      \hline
      \hline
      $ A \ll A_{11} $ & $ A \gg A_{11} $ \\
      \hline
      \vspace{0.1mm}\\
     $ \displaystyle \frac{4 \pi \mathcal{C}_{\rm in}^{\rm visc} F^2 E}{\cos^2 \theta} $ & $ 4 \pi  \mathcal{C}_{\rm grav}^{\rm visc} F^2 A^{-1} E $ \\
      \vspace{0.1mm}\\
       \hline
    \end{tabular}
    \textsf{\caption{\label{background_visc} Asymptotic behaviors of the non-resonant background level $ H_{\rm bg}^{\rm visc} $ of the spectrum. The case where $ A \ll \cos^2 \theta $ ($ A \gg \cos^2 \theta $) corresponds to quasi- (gravito-)inertial waves.}}
 \end{table}

\begin{figure*}[ht!]
 \centering
 \includegraphics[width=0.48\textwidth,clip]{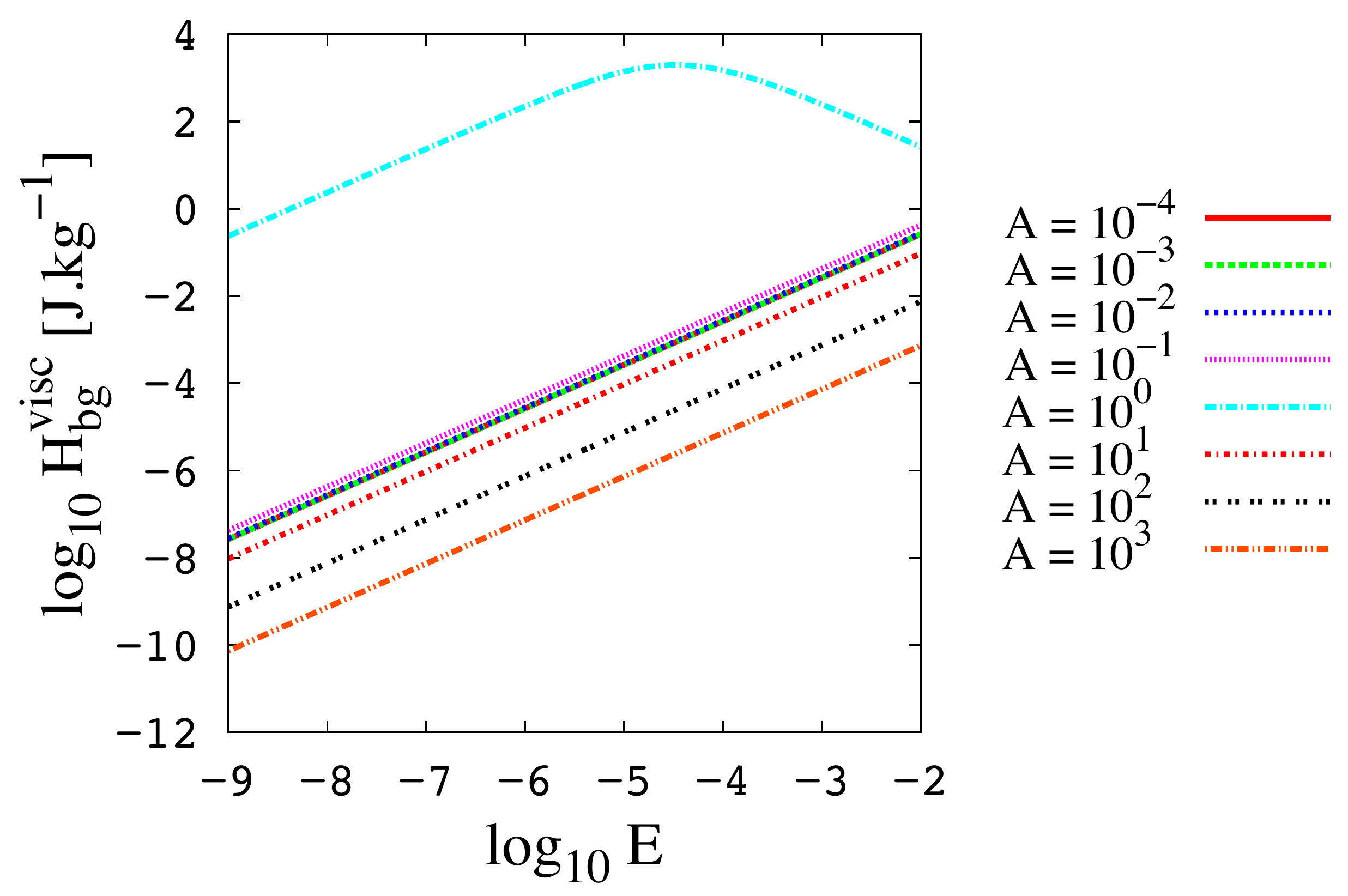}%
 \includegraphics[width=0.48\textwidth,clip]{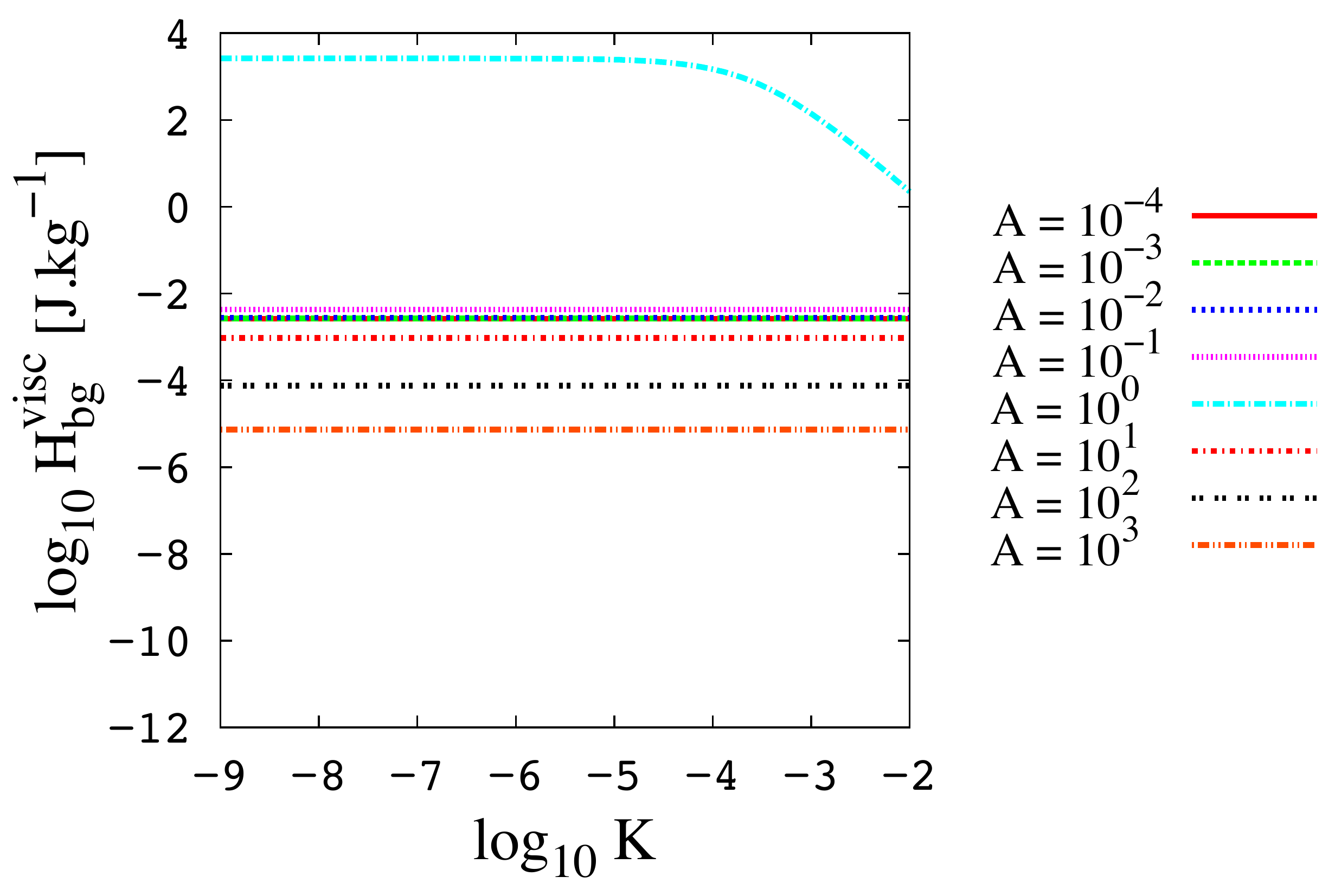} 
 \includegraphics[width=0.48\textwidth,clip]{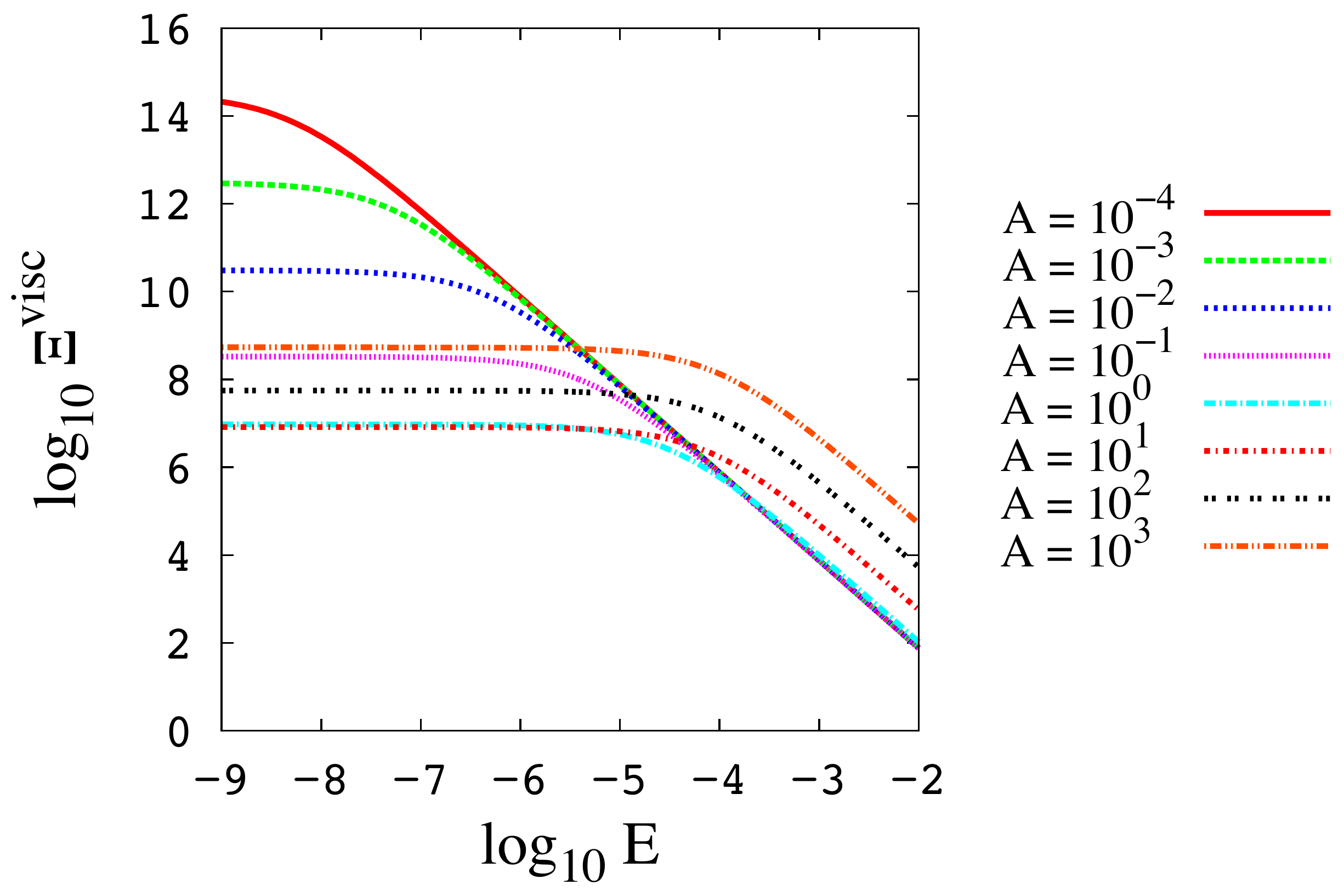}
 \includegraphics[width=0.48\textwidth,clip]{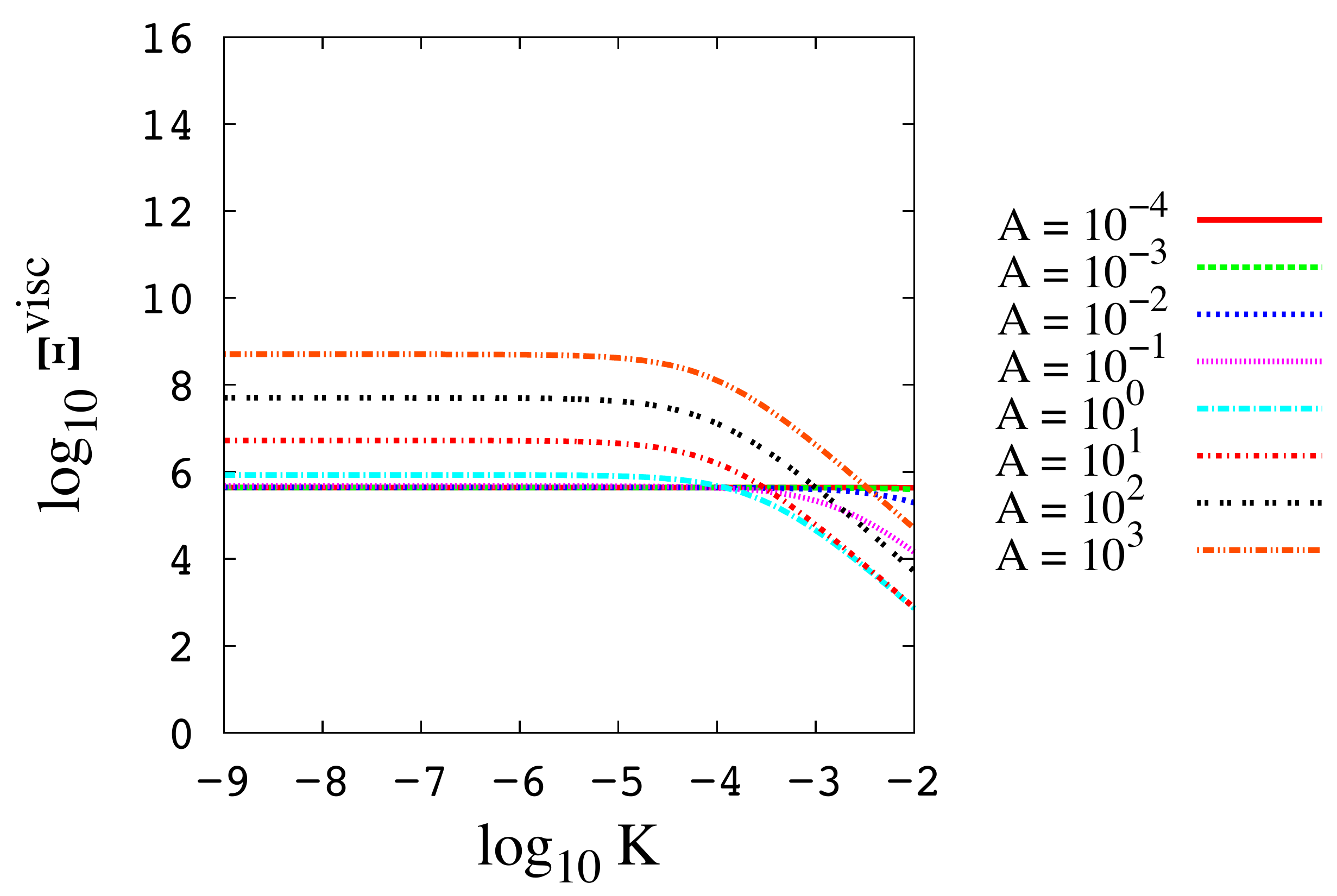}     
  \textsf{ \caption{\label{fig:HXi_EA} Non-resonant background level $ H_{\rm bg}^{\rm visc} $ and sharpness ratio $ \Xi^{\rm visc}$ as a function of $ E $ (with $ K = 10^{-4} $) and $ K $ (with $ E=10^{-4} $) for different values of $ A $ (in logarithmic scales). {\bf Top left:} $ H_{\rm bg}^{\rm visc} - E $. {\bf Top right:} $ H_{\rm bg}^{\rm visc} - K $. {\bf Bottom left:} $ \Xi^{\rm visc} - E $. {\bf Bottom right:} $ \Xi^{\rm visc} - K $.}}
\end{figure*}


It is interesting here to introduce a sharpness ratio $ \Xi $ corresponding to the sensitivity of dissipation to the tidal frequency $ \omega $. $ \Xi $ is defined as the ratio between the height of the resonance and the background level:

\begin{equation}
\Xi = \frac{H_{11}}{H_{\rm bg}}.
\label{sharpness}
\end{equation}

Therefore, the sharpness ratio estimates the relative magnitude of the variations of the dissipation due to the main resonant peak and compares the strengths of the dynamical and the equilibrium/non-wave like tides. A high sharpness ratio  characterizes a peaked spectrum, and a low one a smooth dependence of the dissipation on $ \omega $. As demonstrated in \cite{ADLPM2014}, this parameter is important for the dynamical evolution of planet-moon and star-planet systems. Indeed, the variation of orbital parameters is correlated to $ \Xi $, the evolution of these laters being smooth in absence of resonances and erratic if $ \Xi \gg 1 $ \citep[][]{ADLPM2014}.\\

We can define $ \Xi $ for each of both contributions:

\begin{equation}
\begin{array}{ccc}
    \displaystyle \Xi^{\rm visc} = \frac{H_{11}^{\rm visc}}{H_{\rm bg}^{\rm visc}} & \mbox{and} & \displaystyle \Xi^{\rm therm} = \frac{H_{11}^{\rm therm}}{H_{\rm bg}^{\rm therm}}.
\end{array}
\end{equation}

The transition zones of $ H_{11}^{\rm visc} $ and $ H_{\rm bg}^{\rm visc} $ are the same. Assuming the quasi-adiabatic approximation (Eq. \ref{condition_Hbg}), and considering the analytical expressions of $ H_{11}^{\rm visc} $ (Eq. \ref{H_visc}) and $ H_{\rm bg}^{\rm visc} $ (Eq. \ref{Hbg_visc_brut}), we write the sharpness ratio:

\begin{equation}
\Xi^{\rm visc} = \frac{1}{2} \frac{ \left(  2 \cos^2  + A \right) \left[  \left(  A + \cos^2 \theta  \right)^3 \varepsilon_{12}^2 +  \xi \left( \theta, A,E, K \right)  \right] }{ \varepsilon_{12}^2  \left[  A K +  \left( 2 \cos^2 + A \right) E \right]^2  \left[  C_{\rm in} \cos^2 \theta + C_{\rm grav} A \right]   }.
\label{sharpness1}
\end{equation}

This expression may be simplified in asymptotic domains (see Fig. \ref{fig:domaines}), where:
\begin{itemize}
  \item[$ \bullet $]   $ \left|  \xi \left( \theta, A, E,K  \right) \right|  \ll  \left( A + \cos^2 \theta \right)^3 \varepsilon_{12}^2\quad$ (see Eq. \ref{condition_xi}),\\
  \item[$ \bullet $] $ C_{\rm in}^{\rm visc} \left( \varepsilon_{12} \right) = C_{\rm in}^{\rm visc} \left( \varepsilon_{\rm in} \right) = \mathcal{C}_{\rm in}^{\rm visc}\quad$ (see Eq. \ref{Cin_Cgrav}),\\
  \item[$ \bullet $] $ C_{\rm grav}^{\rm visc}  \left( \varepsilon_{12} \right) = C_{\rm grav}^{\rm visc} \left( \varepsilon_{\rm grav} \right) = \mathcal{C}_{\rm grav}^{\rm visc} $.
\end{itemize}

Thus, the sharpness ratio becomes:

\begin{equation}
\Xi^{\rm visc} = \frac{1}{2} \frac{ \left(  2 \cos^2  + A \right)  \left(  A + \cos^2 \theta  \right)^3 }{ \left[  A K +  \left( 2 \cos^2 + A \right) E \right]^2  \left[  \mathcal{C}_{\rm in}^{\rm visc} \cos^2 \theta + \mathcal{C}_{\rm grav}^{\rm visc} A \right]   }.
\label{sharpness2}
\end{equation}

Like the height of the resonances ($ H_{mn}^{\rm visc} $), $ \Xi^{\rm visc} $ obviously presents symmetrical behaviors for gravito-inertial waves (Table \ref{sharpness_visc}). It is inversely proportional to the square of the dominating dimensionless diffusivity, $ E $ (for $ P_r \gg P_{r;11} $) or $ K $ (for $ P_r \ll P_{r;11} $). So, the sensitivity of tidal dissipation to the frequency $ \omega $ increases quadratically when the diffusivities decrease (Fig.~\ref{fig:HXi_EA}, bottom-left and bottom-right panels). Likewise, $ \Xi $ increases with $ A $ quadratically when $ A \gg A_{11} $. If $ A \ll A_{11} $, then it is related to the colatitude $ \theta $. One will notice that $ \Xi_{\rm visc} $ does not depends on the forcing at all. Neither the amplitude $ F $ of the force, nor the form of the perturbation coefficients $ f_{mn} \propto 1/ \left| m \right| n^2 $ intervene in the expressions of Eq. (\ref{sharpness1}) and (\ref{sharpness2}). Given the link existing between the sharpness ratio and the tidal quality factor $ Q $ modeling dissipation discussed above, we can deduce the qualitative evolution of orbital parameters (Eq. \ref{eq:semi}). Indeed, if a planet is composed by a fluid with a large viscosity, the variation of the semi-major axis of its companion will be smooth. In the opposite case of a low viscosity planet, it becomes strongly erratic, with intermittent rapid changes of the orbit due to resonances \citep[e.g][]{ADLPM2014}.

\begin{table}[htb]
\centering
    \begin{tabular}{ c | c c }
      \hline
      \hline
      \textsc{Domain} & $ A \ll A_{11}  $ & $ A \gg A_{11} $ \\
      \hline
      \vspace{0.1mm}\\
      $ P_{r} \gg P_{r;11} $ & $  \displaystyle \frac{\cos^2 \theta E^{-2}}{4 \mathcal{C}_{\rm in}^{\rm visc} }  $ & $ \displaystyle \frac{A E^{-2}}{2 \mathcal{C}_{\rm grav}^{\rm visc}}   $ \\
      \vspace{0.1mm}\\
      $ P_{r} \ll P_{r;11} $ & $ \displaystyle \frac{\cos^6 \theta A^{-2} E^{-2} P_{r}^2}{\mathcal{C}_{\rm in}^{\rm visc}} $ & $ \displaystyle \frac{A E^{-2} P_{r}^2}{2 \mathcal{C}_{\rm grav}^{\rm visc} }  $\\
      \vspace{0.1mm}\\
       \hline
    \end{tabular}
    \textsf{\caption{\label{sharpness_visc} Asymptotic expressions of the sharpness ratio $ \Xi^{\rm visc} $ characterizing the spectrum of dissipation by viscous friction.}}
 \end{table}


\subsubsection{Dissipation by thermal diffusion}

Similarly as $ H_{\rm bg}^{\rm visc} $, the height of the non-resonant background of dissipation by heat diffusion is assumed to be approximatively:

\begin{equation}
H_{\rm bg}^{\rm therm} \approx 4 \zeta_{11}^{\rm therm} \left( \omega_{\rm bg} \right).
\end{equation}

Assuming the quasi-adiabatic approximation, we can write it: 

\begin{equation}
H_{\rm bg}^{\rm therm} = 4 \pi F^2 E P_{r}^{-1} \frac{ \left( A + \cos^2 \theta \right) \left( 1 + \varepsilon_{12}  \right)^2  }{ \left( A + \cos^2 \theta \right)^3 \varepsilon_{12}^2 + \xi \left( \theta, A, E, K  \right) },
\end{equation}

where $ \xi $ has been defined above (see Eq.~\ref{fonction_xi}). We introduce the function $ C^{\rm therm} $ of $ \varepsilon_{12} $,

\begin{equation}
C^{\rm therm} \left( \varepsilon_{12} \right) = \left(  1 + \frac{1}{\varepsilon_{12}} \right)^2,
\end{equation}

that will be a constant in asymptotical regimes,

\begin{equation}
\begin{array}{lclcl}
    \mathcal{C}_{\rm in}^{\rm therm} & = & C^{\rm therm} \left(  \varepsilon_{\rm in} \right) & = & 19.73\\
    \mathcal{C}_{\rm grav}^{\rm therm} & = & C^{\rm therm} \left(  \varepsilon_{\rm grav} \right) & = & 73.10. 
\end{array}
\end{equation}

When $ \left| \log \left( A / A_{11} \right) \right| \gg 1 $, the height of the peaks can simply be expressed:

\begin{equation}
H_{\rm bg}^{\rm therm} \approx \frac{4 \pi F^2 E P_{r}^{-1}}{ \left( A + \cos^2 \theta \right)^2 } C^{\rm therm} \left( \varepsilon_{12} \right),
\end{equation}

or by the interpolating function:

\begin{equation}
H_{\rm bg}^{\rm therm} \approx \frac{4 \pi F^2 E P_{r}^{-1}}{ \left( A + \cos^2 \theta \right)^3 } \left(  \mathcal{C}_{\rm grav}^{\rm therm} A + \mathcal{C}_{\rm in}^{\rm therm}  \cos^2 \theta  \right). 
\label{Hbg_therm}
\end{equation}

Considering the asymptotic regimes, we obtain the scaling laws given in Tab.~\ref{background_therm}. \\

\begin{table}[htb]
\centering
    \begin{tabular}{  c c }
      \hline
      \hline
      $ A \ll A_{11} $ & $ A \gg A_{11} $ \\
      \hline
      \vspace{0.1mm}\\
     $ \displaystyle \frac{4 \pi \mathcal{C}_{\rm in}^{\rm therm} F^2}{\cos^4 \theta} E P_{r}^{-1} $ & $ 4 \pi  \mathcal{C}_{\rm grav}^{\rm therm} F^2 \displaystyle A^{-2} E P_{r}^{-1} $ \\
      \vspace{0.1mm}\\
       \hline
    \end{tabular}
    \textsf{\caption{\label{background_therm} Asymptotic behaviors of the non-resonant background level $ H_{\rm bg}^{\rm therm} $ of the dissipation by heat diffusion. The case where $ A \ll A_{11} $ ($ A \gg A_{11} $) corresponds to quasi-(gravito-)inertial waves.}}
 \end{table}

The sharpness ratio of dissipation by heat diffusion can be written:

\begin{equation}
\Xi^{\rm therm} = \frac{1}{4} \frac{  P_{r}^2   \left[  \left( A + \displaystyle{\frac{1}{2}} A_{11} \right)^3 \varepsilon_{12} + \xi \left( \theta , A ,E,K \right)  \right] }{ E^2 \left( A + A_{11} \right)^2 \left( P_{r} + P_{r;11} \right)^2  \left( 1 + \varepsilon_{12} \right)^2 },
\end{equation}

expression which can be simplified in

\begin{equation}
\Xi^{\rm therm} = \frac{1}{4} \frac{P_r^2 \left( A + \displaystyle{\frac{1}{2}} A_{11} \right)^3  }{E^2 \left( A + A_{11} \right)^2 \left( P_{r} + P_{r;11} \right)^2 C^{\rm therm} \left( \varepsilon_{12} \right) }
\end{equation}

in asymptotic regions where $ \left| \log \left( A / A_{11} \right)  \right| \gg 1 $, or by

\begin{equation}
\Xi^{\rm therm} = \frac{1}{4} \frac{P_{r}^2 \left( A + \displaystyle\frac{1}{2} A_{11} \right)^3  }{ E^2 \left( A + A_{11} \right) \left( P_{r} + P_{r;11} \right)^2  \left(  \mathcal{C}_{\rm grav}^{\rm therm} A + \mathcal{C}_{\rm in}^{\rm therm}  A_{11}  \right) },
\label{Xi_therm}
\end{equation}

if we use the asymptotic interpolation of the function $ C^{\rm therm} $ 

\begin{equation}
C^{\rm therm} \left( \varepsilon_{12} \left( A \right) \right)  \sim \frac{ \mathcal{C}_{\rm grav}^{\rm therm} A + \mathcal{C}_{\rm in}^{\rm therm}  A_{11}  }{ A + A_{11}}.
\end{equation}

Using Eq.~(\ref{Xi_therm}), we establish the asymptotic scaling laws given in Tab.~\ref{sharpness_therm}. These laws are the same as the ones of Tab.~\ref{sharpness_visc}, which means that the variations of the energy with the tidal frequency do not depend on the dissipation mechanism in our local model. Therefore, the number of peaks, which is deduced from the sharpness ratio, will have the same scaling laws in both zones of Fig.~\ref{fig:map_zones_2} (left panel).

\begin{table}[htb]
\centering
    \begin{tabular}{ c | c c }
      \hline
      \hline
      \textsc{Domain} & $ A \ll A_{11}  $ & $ A \gg A_{11} $ \\
      \hline
      \vspace{0.1mm}\\
      $ P_{r} \gg P_{r;11} $ & $  \displaystyle \frac{\cos^2 \theta E^{-2} }{8 \mathcal{C}_{\rm in}^{\rm therm} }  $ & $ \displaystyle \frac{A E^{-2}}{2 \mathcal{C}_{\rm grav}^{\rm therm}}   $ \\
      \vspace{0.1mm}\\
      $ P_{r} \ll P_{r;11} $ & $ \displaystyle \frac{\cos^6 \theta A^{-2} E^{-2} P_{r}^2}{2 \mathcal{C}_{\rm in}^{\rm therm}} $ & $ \displaystyle \frac{A E^{-2} P_{r}^2}{2 \mathcal{C}_{\rm grav}^{\rm therm}}  $\\
      \vspace{0.1mm}\\
       \hline
    \end{tabular}
    \textsf{\caption{\label{sharpness_therm} Asymptotic expressions of the sharpness ratio $ \Xi^{\rm therm} $ characterizing the spectrum of dissipation by heat diffusion.}}
 \end{table}
 
\subsubsection{Total dissipation}

The background level of $ \zeta $, denoted $ H_{\rm bg} $, is the sum of $ H_{\rm bg}^{\rm visc} $ and $ H_{\rm bg}^{\rm therm} $. Given that $ \mathcal{C}_{\rm in}^{\rm therm} \sim \mathcal{C}_{\rm in}^{\rm visc}  $ and $ \mathcal{C}_{\rm grav}^{\rm therm} \sim \mathcal{C}_{\rm grav}^{\rm visc}  $, the ratio of the backgrounds is equivalent to the one of the heights, 

\begin{equation}
\frac{H_{\rm bg}^{\rm visc}}{H_{\rm bg}^{\rm therm}} \sim \varsigma_{11},
\end{equation}

and we express scaling laws of $ H_{\rm bg} $ in Tab.~\ref{background} for the regimes identified in Fig.~\ref{fig:map_zones_2} (left panel).\\

The asymptotic behaviors of the global sharpness ratio given in Tab.~\ref{table_Xi} are directly deduced from Tabs. \ref{sharpness_visc} and \ref{sharpness_therm}. We notice that $ \Xi^{\rm visc} \sim \Xi^{\rm therm} $ for all regimes. So, $ \Xi \sim \Xi^{\rm visc} $ and Tab.~\ref{table_Xi} can be approximated by Tab.~\ref{sharpness_visc}.

\begin{table}[htb]
\centering
    \begin{tabular}{c |  c c }
      \hline
      \hline
      \textsf{Domain} & $ A \ll A_{11} $ & $ A \gg A_{11} $ \\
      \hline
      \vspace{0.1mm}\\
     $ P_{r} \gg P_{r;11}^{\rm diss} $ & $ \displaystyle \frac{4 \pi \mathcal{C}_{\rm in}^{\rm visc} F^2 E}{\cos^2 \theta} $ & $ 4 \pi  \mathcal{C}_{\rm grav}^{\rm visc} F^2 A^{-1} E $ \\
      \vspace{0.1mm}\\
     $ P_{r} \ll P_{r;11}^{\rm diss} $ & $ \displaystyle \frac{4 \pi \mathcal{C}_{\rm in}^{\rm therm} F^2}{\cos^4 \theta} E P_{r}^{-1} $ & $ 4 \pi  \mathcal{C}_{\rm grav}^{\rm therm} F^2 \displaystyle A^{-2} E P_{r}^{-1} $ \\
      \vspace{0.1mm}\\
       \hline
    \end{tabular}
    \textsf{\caption{\label{background} Asymptotic behaviors of the non-resonant background level $ H_{\rm bg} $ of the total dissipation. The case where $ A \ll A_{11} $ ($ A \gg A_{11} $) corresponds to quasi-(gravito-)inertial waves. }}
 \end{table}

 \begin{table*}[htb]
\centering
\begin{tabular}{| c | c | c | c | c |} 
   \hline
   \hline
   \textsc{Domain} & \multicolumn{2}{c|}{$ A \ll A_{11} $} & \multicolumn{2}{c|}{$ A \gg A_{11} $} \\
   \hline
   \vspace{0.1mm} & \multicolumn{2}{c|}{} & \multicolumn{2}{c|}{} \\
    $ P_{r} \gg P_{r;11}^{\rm reg} $ & \multicolumn{2}{c|}{ $  \displaystyle \frac{\cos^2 \theta E^{-2}}{4 \mathcal{C}_{\rm in}^{\rm visc} } $  } & \multicolumn{2}{c|}{$\displaystyle \frac{A E^{-2}}{2 \mathcal{C}_{\rm grav}^{\rm visc}} $} \\
   \vspace{0.1mm} & \multicolumn{2}{c|}{} & \multicolumn{2}{c|}{} \\
    \hline
   \vspace{0.1mm} &  & & &\\
    \multirow{5}{*}{$ P_{r} \ll P_{r;11}^{\rm reg} $} & $ P_{r} \gg P_{r;11} $ & $ \displaystyle \frac{\cos^2 \theta E^{-2} }{8 \mathcal{C}_{\rm in}^{\rm therm} } $ & $ P_{r} \gg P_{r;11}^{\rm diss} $ & $ \displaystyle \frac{A E^{-2} P_{r}^2}{2 \mathcal{C}_{\rm grav}^{\rm visc} } $ \\
    \vspace{0.1mm} &  & & &\\
    \cline{2-5} 
    \vspace{0.1mm} &  & & &\\
           & $ P_{r} \ll P_{r;11} $ & $ \displaystyle \frac{\cos^6 \theta A^{-2} E^{-2} P_{r}^2}{2 \mathcal{C}_{\rm in}^{\rm therm}} $ & $ P_{r} \ll P_{r;11}^{\rm diss} $ & $ \displaystyle \frac{A E^{-2} P_{r}^2}{2 \mathcal{C}_{\rm grav}^{\rm therm}} $ \\
    \vspace{0.1mm} &  & & &\\
    \hline
\end{tabular}
\textsf{\caption{\label{table_Xi} Scaling laws of the sharpness ratio of $ \zeta $. They correspond to the areas of Fig.~\ref{fig:map_zones_2}.}}
\end{table*}

\begin{figure*}[ht!]
 \centering
 \includegraphics[width=0.48\textwidth,clip]{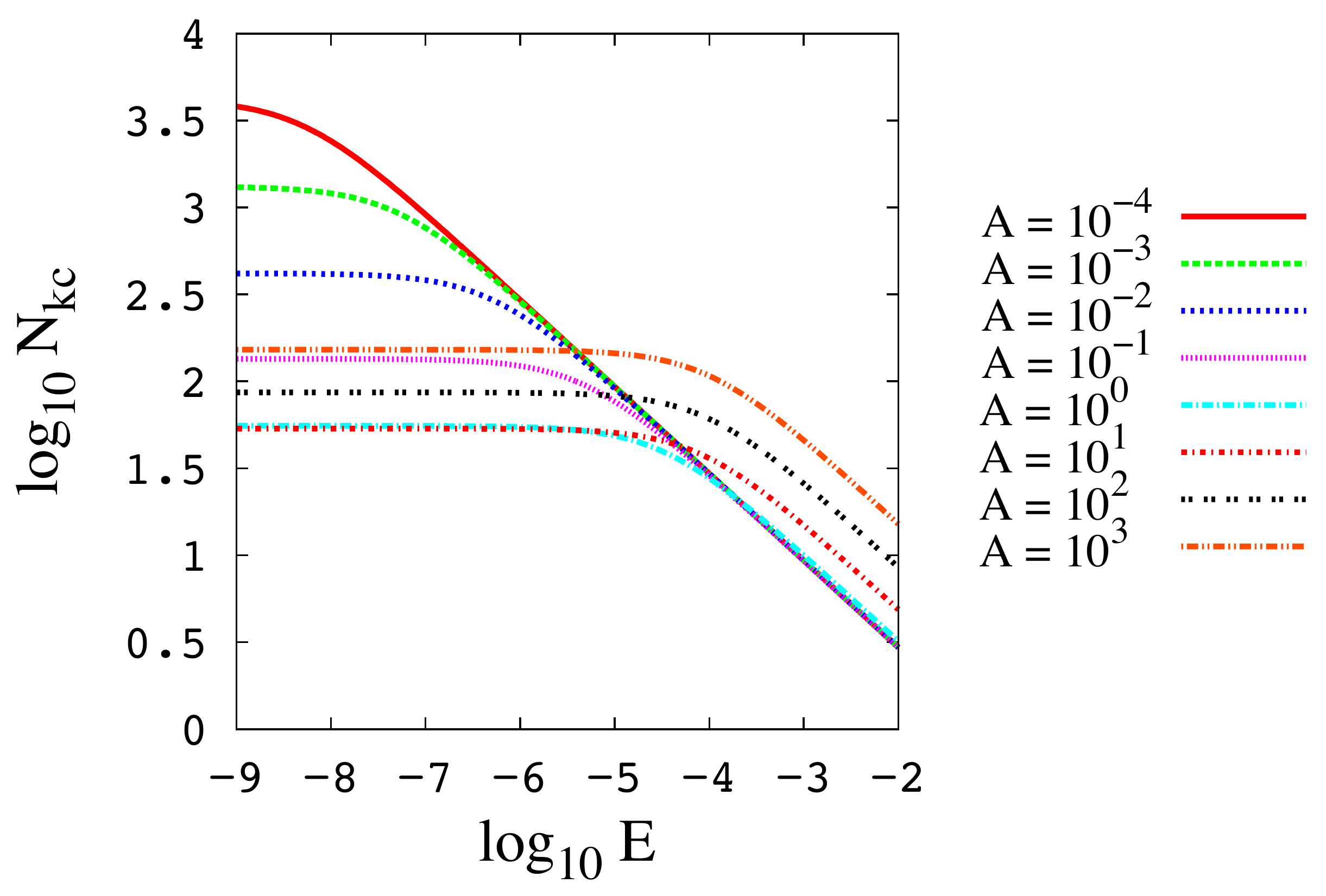}
 \includegraphics[width=0.48\textwidth,clip]{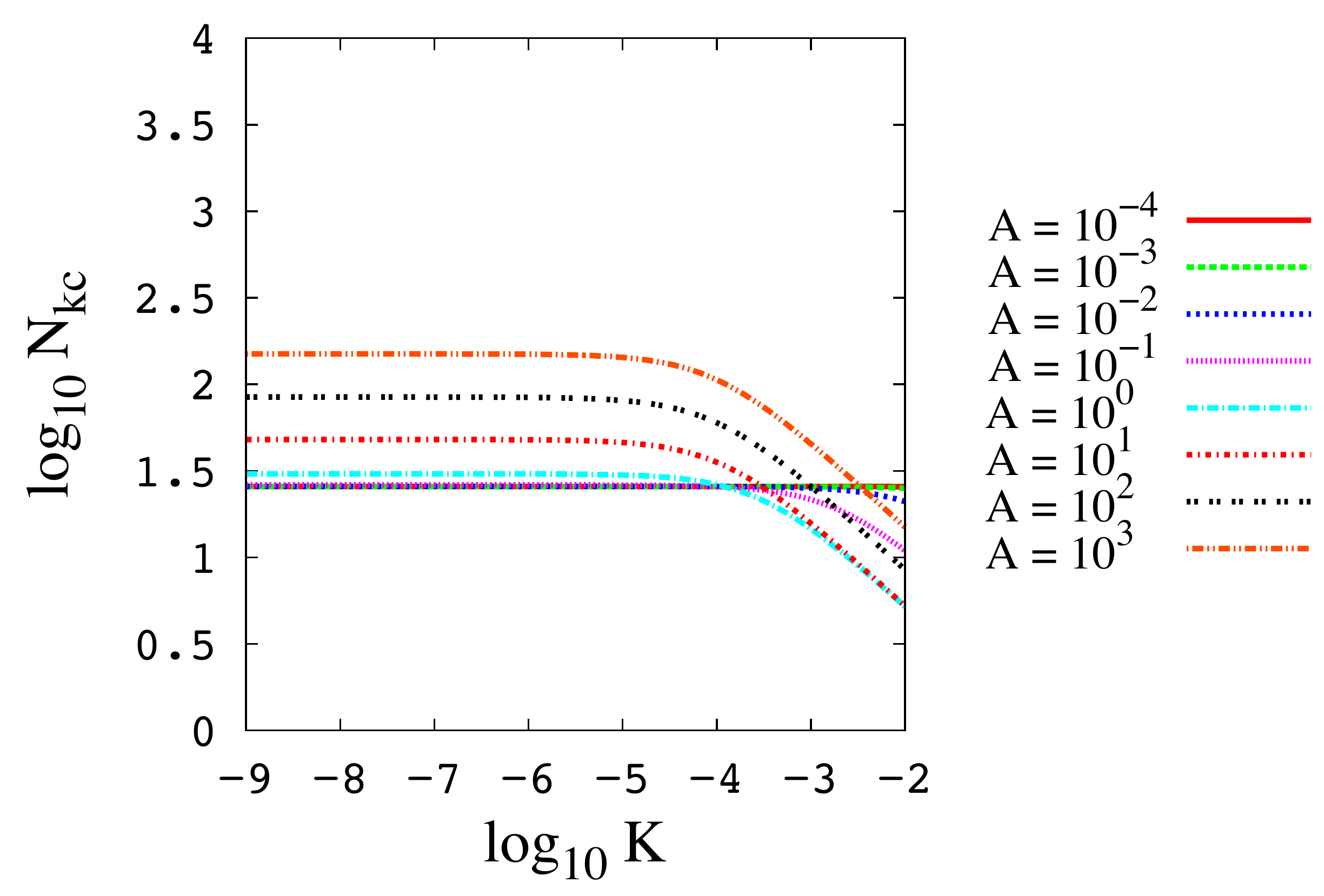}     
  \textsf{ \caption{\label{fig:Nkc_EA} Number of resonances $ N_{k_c} $ as a function of $ E $ (with $ K = 10^{-4} $) and $ K $ (with $ E=10^{-4} $) for different values of $ A $ (in logarithmic scales). {\bf Left:} $ N_{k_c} - E $. {\bf Right:} $ N_{k_c} - K $.}}
\end{figure*}


\subsection{Number of resonant peaks}

In spite of the theoretical great number of resonances allowed by the condition given in Eq. (\ref{condition_positions}) (see also Fig. \ref{fig:positions}), the only effective resonant peaks are the ones that exceed the level of the non-resonant background $ H_{\rm bg} $ (Figs. \ref{fig:spectre_inertiel_1} and \ref{fig:spectre_inertiel_2}) given by Eq. (\ref{Hbg_visc}) and (\ref{Hbg_therm}) \citep[see also the discussion in][]{RV2010}. This will tend to smooth the variations of the quality factor $ Q $ and the resulting evolution of the orbital parameters of a star-planet/planet-satellite system. Indeed, the resonances of highest orders are absorbed in the mean dissipation level of the resonant frequency range. The harmonics thus have to match a criterion to appear in the dissipation spectrum. This criterion simply corresponds to the inequality:

\begin{equation}
H_{mn} > H_{\rm bg},
\label{criterium}
\end{equation}

where $ H_{mn} $ is given by Eqs. (\ref{Hmn_visc}) and (\ref{Hmn_therm}). The approximate wave number $ k $ introduced before allows to compute the typical order of the smallest peaks $ k_c $. As noted before, $ \Xi \sim \Xi^{\rm visc} \sim \Xi^{\rm therm} $, which implies that $ k_c \sim k_c^{\rm visc} \sim k_c^{\rm therm} $. We replace the index $ m $ and $ n $ of the height $ H_{mn} $ by $ k $ and use Eq. (\ref{criterium}) to obtain:

\begin{equation}
k_c^{\rm visc} \sim \left\{ \frac{1}{2}  \frac{ \left( 2 \cos^2 \theta + A \right) \left[ \left( A + \cos^2 \theta \right)^3 \varepsilon_{12}^2 + \xi \left( \theta, A,E, K \right) \right] }{ \varepsilon_{12}^2 \left[ AK + \left( 2 \cos^2 + A \right) E  \right]^2 \left[ C_{\rm in}  \cos^2 \theta + C_{\rm grav}  A \right] }  \right\}^{ \displaystyle \frac{1}{8} }.
\label{kc1}
\end{equation}

This expression can be simplified in asymptotic cases. Assuming Eq. (\ref{condition_Hbg}), it becomes

\begin{equation}
k_c^{\rm visc} \sim \left\{ \frac{1}{2}  \frac{ \left( 2 \cos^2 \theta + A \right) \left( A + \cos^2 \theta \right)^3 }{ \left[ AK + \left( 2 \cos^2 + A \right) E  \right]^2 \left[ C_{\rm in}^{\infty} \cos^2 \theta + C_{\rm grav}^{\infty} A \right] }  \right\}^{ \displaystyle \frac{1}{8} }.
\label{kc2}
\end{equation}

In Tab.~\ref{ordre_max}, we give the scaling laws of $ k_c $ for the different asymptotic regimes. Taking into account resonances beyond this wave number does not change the global shape of the dissipation spectrum. In fact, in the situations corresponding to Figs~\ref{fig:spectre_inertiel_1} and \ref{fig:spectre_inertiel_2}, there is no need to go beyond $ k \sim 10 $ (Fig. \ref{fig:kc_EA}, top-left and top-right panels). This is amply sufficient to model the dissipation realistically. The exponent $ 1/8 $ of $ k_c $ in Eq. (\ref{kc1}) and (\ref{kc2}) is related to the spectral decomposition of the perturbing force.\\

 \begin{table}[htb]
\centering
    \begin{tabular}{ c | c c }
      \hline
      \hline
      \textsc{Domain} & $ A \ll A_{11} $ & $ A \gg A_{11}  $ \\
      \hline
      \vspace{0.1mm}\\
      $ P_{r} \gg P_{r;11} $ & $ k_c \sim  \left( \displaystyle \frac{\cos^2 \theta}{4 C_{\rm in}^{\infty} E^2}  \right)^{\displaystyle \frac{1}{8}}  $ & $ k_c \sim \left(  \displaystyle \frac{ A}{2 C_{\rm grav}^{\infty} E^2 } \right)^{\displaystyle  \frac{1}{8}} $ \\
      \vspace{0.1mm}\\
      $ P_{r} \ll P_{r;11} $ & $  k_c \sim \left( \displaystyle \frac{ \cos^6 \theta}{C_{\rm in}^{\infty} A^2 E^2 P_{r}^{-2}} \right)^{\displaystyle  \frac{1}{8}} $ & $ k_c \sim  \left( \displaystyle \frac{ A}{ 2 C_{\rm grav}^{\infty} E^2 P_{r}^{-2}} \right)^{\displaystyle  \frac{1}{8}}  $\\
      \vspace{0.1mm}\\
       \hline
    \end{tabular}
    \textsf{\caption{\label{ordre_max} Asymptotic behaviors of the maximal order of noticeable resonances $ k_c $.}}
 \end{table}

The wave number $ k_c $ gives us the number of peaks $ N_{k_c} $ (Eq.~\ref{Nkc_exact}) which is of great interest. As described above, we can assume $ N_{k_c}  \propto k_c^2 $ (Tab. \ref{number_peaks_values} and Fig. \ref{fig:number_peaks}). Thus, we immediately deduce the scaling laws of $ N_{\rm kc} $ (Tab. \ref{Nkc}) from those of the rank of the harmonics (Tab. \ref{ordre_max}) in the four asymptotic regimes (see Fig. \ref{fig:domaines}). In the same way as $ k_c $, $ N_{k_c} \sim N_{k_c}^{\rm visc} \sim N_{k_c}^{\rm therm} $. $ N_{k_c}^{\rm visc} $ is given by the analytical expression:
 
 \begin{equation}
N_{k_c}^{\rm visc} \sim \left\{ \frac{1}{2}  \frac{ \left( 2 \cos^2 \theta + A \right) \left[ \left( A + \cos^2 \theta \right)^3 \varepsilon_{12}^2 + \xi \left( \theta, A,E, K \right) \right] }{ \varepsilon_{12}^2 \left[ AK + \left( 2 \cos^2 + A \right) E  \right]^2 \left[ C_{\rm in} \cos^2 \theta + C_{\rm grav}  A \right] }  \right\}^{ \displaystyle \frac{1}{4} },
\end{equation}

which is asymptotically equivalent to: 

\begin{equation}
N_{k_c}^{\rm visc} \sim \left\{ \frac{1}{2}  \frac{ \left( 2 \cos^2 \theta + A \right) \left( A + \cos^2 \theta \right)^3 }{ \left[ AK + \left( 2 \cos^2 + A \right) E  \right]^2 \left[ C_{\rm in}^{\infty} \cos^2 \theta + C_{\rm grav}^{\infty} A \right] }  \right\}^{ \displaystyle \frac{1}{4} }.
\end{equation}
 
So, $ N_{k_c} \propto E^{-1/2} $ for inertial waves damped by viscous diffusion (see Fig.~\ref{fig:Nkc_EA}, left and right panels). In this regime, the number of peaks decays when the Ekman number increases as can be observed in Fig. \ref{fig:spectre_inertiel_1}. The critical degree $ k_c $, the number of resonances $ N_{k_c} $ and the sharpness ratio $ \Xi $ are related the ones to the others by the scaling equality:

\begin{equation}
N_{k_c} \sim k_c^2 \sim \Xi^{  \frac{1}{4} },
\end{equation}

in which, as pointed out above, the exponent $ 1/4 $ depends on the form of the coefficients of the perturbation. $ k_c $, $ N_{k_c} $, and $ \Xi $ all measure the smoothness of the dissipation spectrum. As the change rate of orbital parameters is proportional to the energy dissipated inside the bodies, the evolution of an orbital system will be regular if they are small, erratic otherwise.

 \begin{table}[htb]
\centering
    \begin{tabular}{ c | c c }
      \hline
      \hline
      \textsc{Domain} & $ A \ll A_{11} $ & $ A \gg A_{11} $ \\
      \hline
      \vspace{0.1mm}\\
      $ P_{r} \gg P_{r;11} $ & $ N_{k_c} \sim \left( \displaystyle \frac{\cos^2 \theta}{4 C_{\rm in}^{\infty} E^2}  \right)^{\displaystyle \frac{1}{4}} $ & $ N_{k_c} \sim\left(  \displaystyle \frac{ A}{2 C_{\rm grav}^{\infty} E^2 } \right)^{\displaystyle  \frac{1}{4}}  $ \\
      \vspace{0.1mm}\\
      $ P_{r} \ll P_{r;11} $ & $  N_{k_c} \sim  \left( \displaystyle \frac{ \cos^6 \theta}{C_{\rm in}^{\infty} A^2 E^2 P_{r}^{-2}} \right)^{\displaystyle  \frac{1}{4}} $ & $ N_{k_c} \sim  \left( \displaystyle \frac{ A}{2 C_{\rm grav}^{\infty} E^2 P_{r}^{-2}} \right)^{\displaystyle  \frac{1}{4}}  $\\
      \vspace{0.1mm}\\
       \hline
    \end{tabular}
    \textsf{\caption{\label{Nkc} Asymptotic behaviors of the number of peaks $ N_{k_c}^{\rm visc} $. The same scaling laws are obtained for $ N_{k_c} $.}}
 \end{table}

\subsection{Super-adiabaticity ($ N^2 < 0 $)}

\begin{figure*}[ht!]
 \centering
 \includegraphics[width=0.48\textwidth,clip]{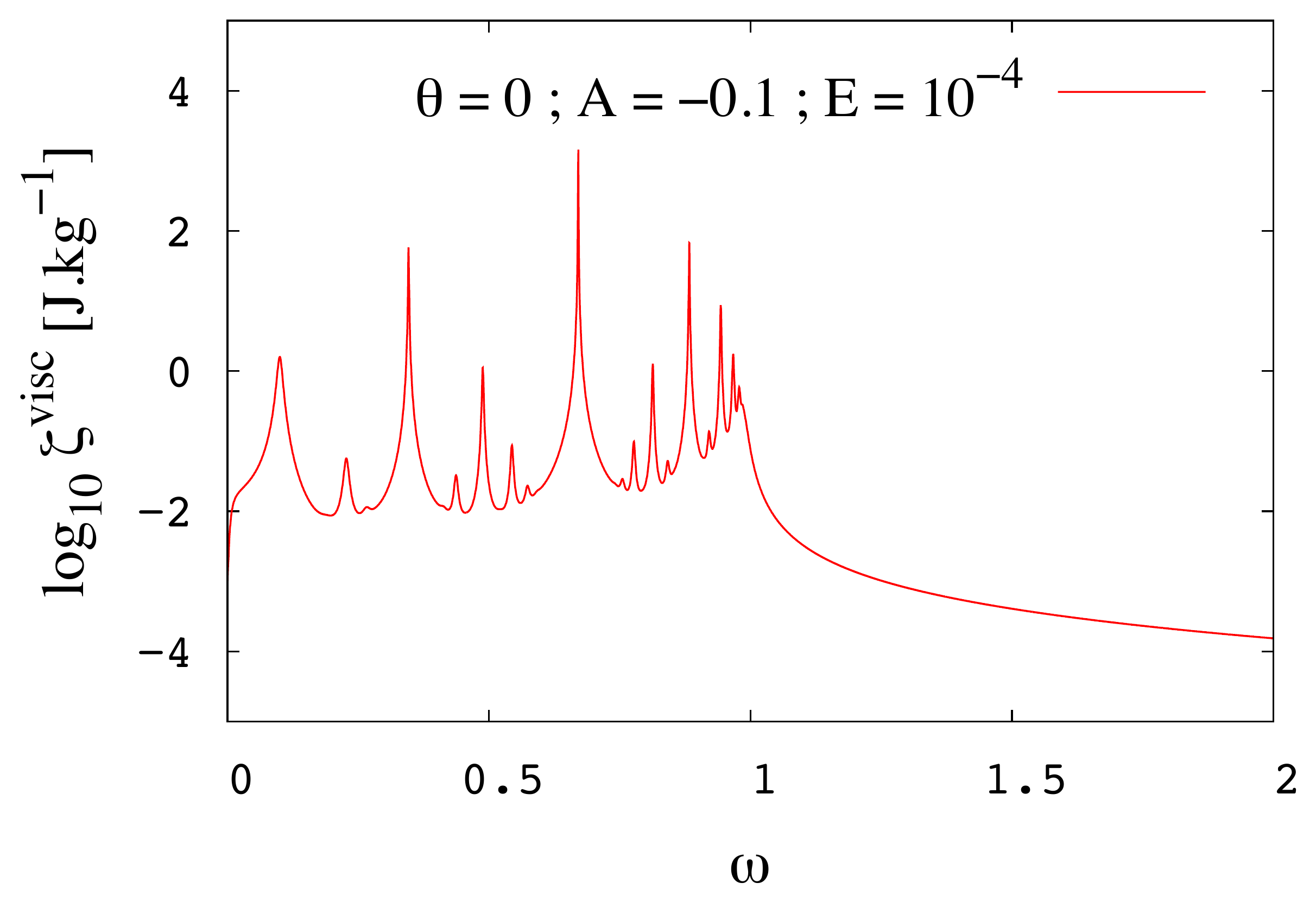}%
 \includegraphics[width=0.48\textwidth,clip]{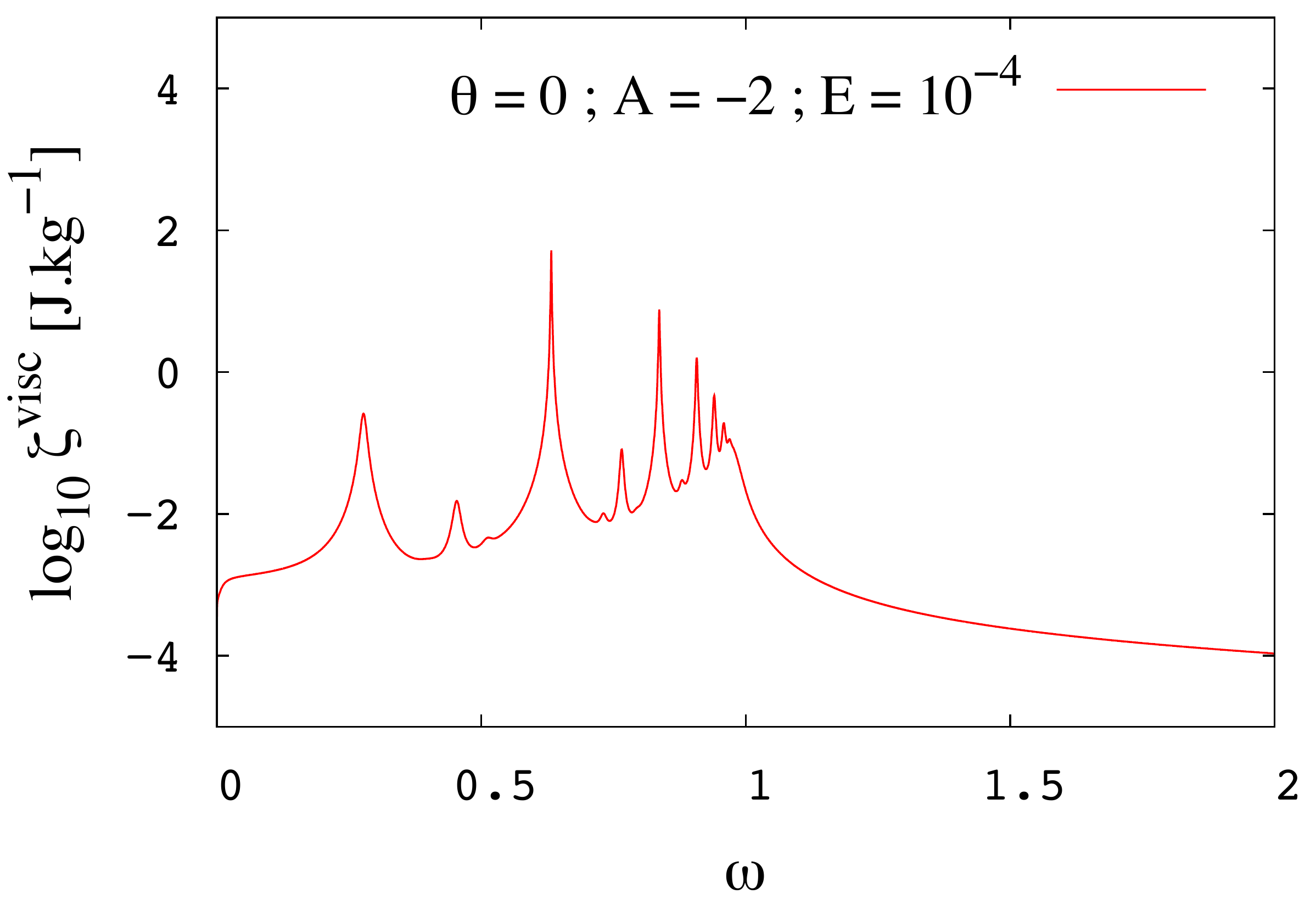} 
 \includegraphics[width=0.48\textwidth,clip]{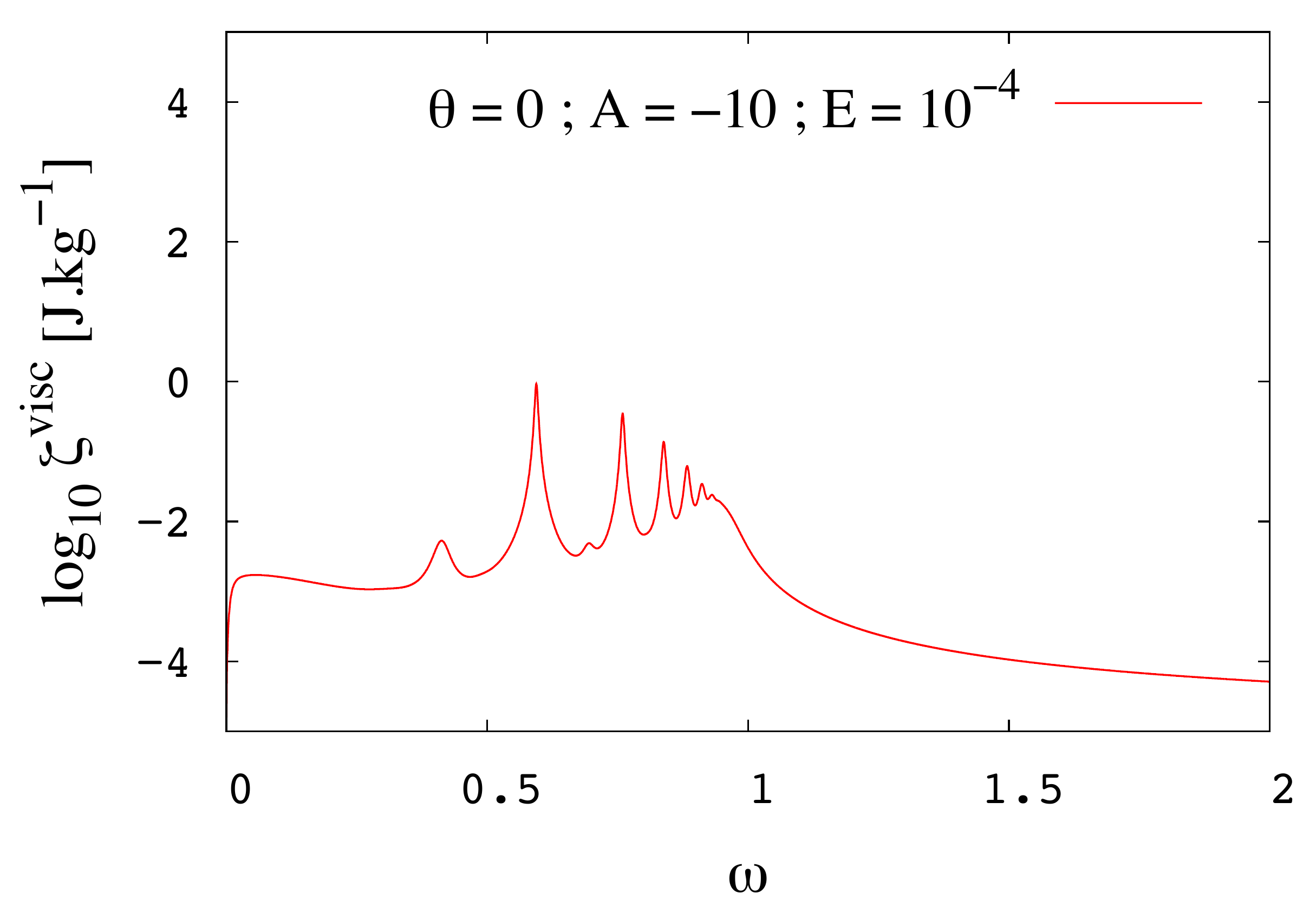}
 \includegraphics[width=0.48\textwidth,clip]{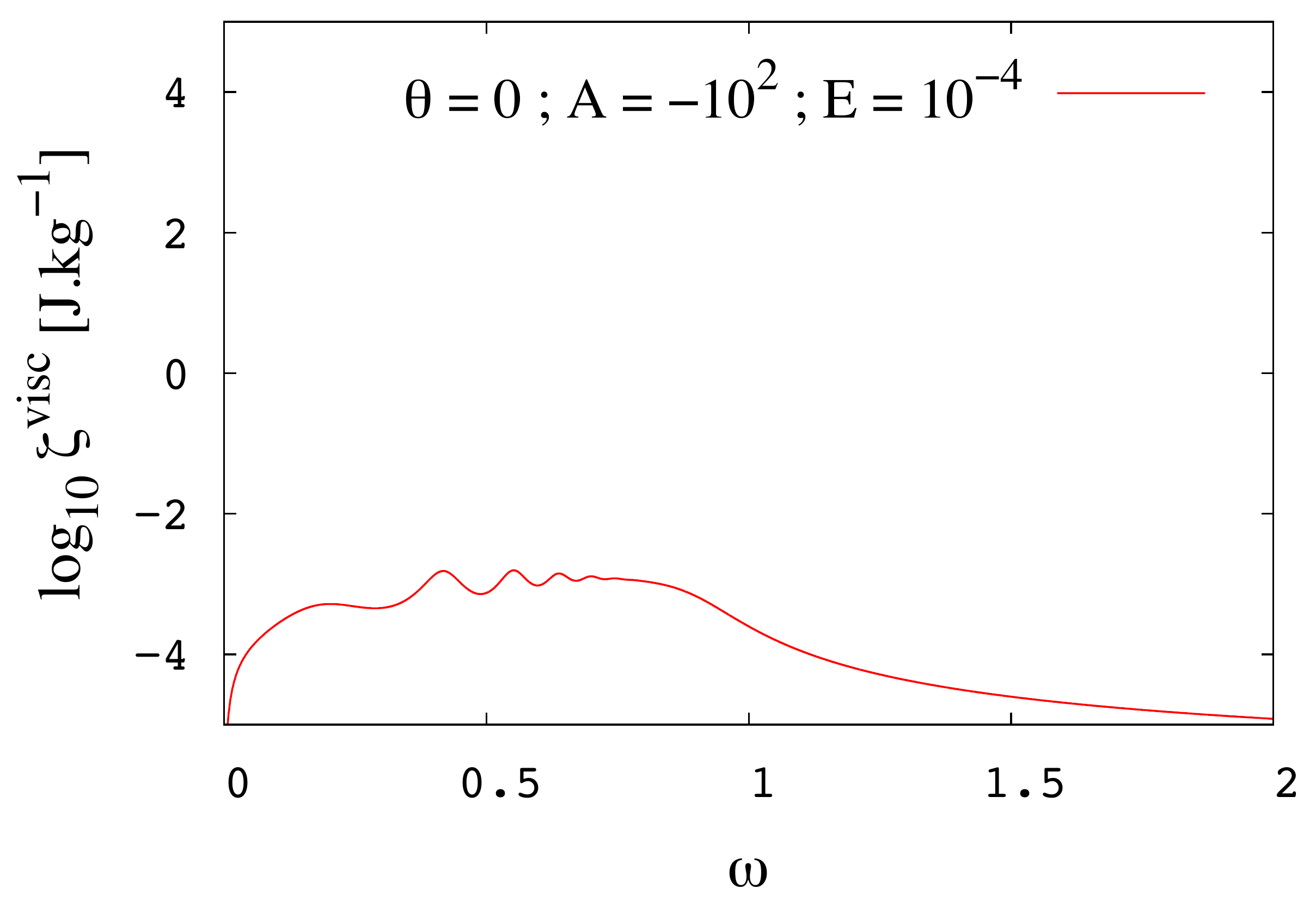}     
  \textsf{ \caption{\label{fig:N2neg} Spectra of $ \zeta^{\rm visc} $ in the case of pure inertial waves ($ E = 10^{-4} $ and $ K = 0 $) for $ A < 0 $. {\bf Top left:} $ A = - 0.1 $. {\bf Top right:} $ A = -2 $. {\bf Bottom left:} $ A = -10 $. {\bf Bottom right:} $ A = - 100 $.}}
\end{figure*}

The expression of $ \zeta^{\rm diss} $ (Eq.~\ref{energy_visc}) allows us to compute dissipation frequency-spectra if $ N^2 < 0 $ that corresponds to super-adiabaticity established by convection. Fig.~\ref{fig:N2neg} shows how the spectrum of the energy dissipated by viscous friction evolves when the magnitude of an imaginary Brunt-V\"ais\"al\"a frequency increases. The graphs plotted correspond to pure inertial waves. We will not study here the particular scaling laws describing their properties, which are obviously not the same as the ones found for classical configurations. Nevertheless, we can observe that strong values of $ \left| A \right| $ tend to flatten the spectrum. The height of the resonant peaks and the level of the background both decay. For exemple, the dissipation varies smoothly with the frequency when $ \left| A \right| = 10^2 $ contrary to the case of positive $ N^2 $. This situation will be studied in detail in a forthcoming work.

\section{Discussion}

Thanks to the analytical expression of the dissipated energy (Eq. \ref{energy_visc}), we have been able to compute the hydrodynamical scaling laws that link the properties of the dissipation to the dynamics of the fluid. They are summarized in Tab. \ref{scaling_laws} and show that the related dependence of the dissipation onto the tidal frequency ($ \omega $) and the dimensionless characteristic numbers of the fluid ($ A $, $ E $ and $ P_r $) is tightly bound to the regime of tidal waves. In particular, they allow to explain the tendencies observed on the spectra presented by \cite{OL2004}. For example, it was noticed that the number of resonances grew if the Ekman number decreases in the regime of inertial waves dissipated by viscous friction (see Fig. \ref{fig:spectre_inertiel_1}). Now, we derive laws for our local model that demonstrate that it scales as $ N_{k_c} \propto E^{-1/2} $. Likewise, the height of the peaks has been analytically proved to be inversely proportional to $ E $, contrary to the level of the non-resonant background for which we obtain $ H_{\rm bg} \propto E $. This explains why variations of the tidal frequency $ \omega $ strongly impact the dissipation when $ E $ tends to zero. Now, to understand how the regime of the waves determines the properties of the spectra, we have focused for instance on the sharpness ratio $ \Xi $. Table \ref{scaling_laws} shows that this number always increases when the dominant dimensionless diffusivity, $ E $ or $ K $, decreases. The difference between inertial waves and gravito-inertial waves resides in the linear dependence onto the stratification parameter $ A $. This kind of comparison can be done for any other property similarly. In the light of the scaling laws, we identify the parameters that will have a real impact on the viscous dissipation and therefore, through the quality factor $ Q $, on the long term evolution of planetary systems.


\begin{table*}[htb]
\centering
\begin{tabular}{ | c  |  c    l l  |  c    l l  |}
  \hline
  \hline
    \textsc{Domain} & \multicolumn{3}{c|}{$ A \ll A_{11} $}  & \multicolumn{3}{c|}{$ A \gg A_{11} $}   \\ \hline
     \vspace{0.01mm} & & & & & & \\
    \multirow{5}{*}{$ P_r \gg P_{r;11}^{\rm reg} $} & & $ l_{mn} \propto E $ & $ \omega_{mn} \propto \displaystyle \frac{n}{\sqrt{m^2 + n^2}} \cos \theta $  &  & $ l_{mn} \propto E $  & $ \omega_{mn} \propto \displaystyle \frac{m}{\sqrt{m^2 + n^2}}  \sqrt{A} $   \\
     \vspace{0.01mm} & & & & & & \\
    & & $ H_{mn} \propto E^{-1}  $  & $ N_{\rm kc} \propto E^{-1/2} $ &  & $ H_{mn} \propto E^{-1} $ &  $ N_{\rm kc} \propto A^{1/4} E^{-1/2} $    \\
     \vspace{0.01mm} & & & & & & \\
    &  & $ H_{\rm bg} \propto E $ & $ \Xi \propto E^{-2}  $ &  & $ H_{\rm bg} \propto A^{-1} E $ & $ \Xi \propto A E^{-2} $    \\
     \vspace{0.01mm} & & & & & & \\
     \hline
     \vspace{0.01mm} & & \multicolumn{1}{| l}{} & & & \multicolumn{1}{| l}{} & \\
   \multirow{12}{*}{$ P_r \ll P_{r;11}^{\rm reg} $} & \multirow{6}{*}{$ P_r \gg P_{r;11} $} & \multicolumn{1}{| l}{$ l_{mn} \propto  E  $} & $ \displaystyle \omega_{mn} \propto \frac{n}{\sqrt{m^2 + n^2}} \cos \theta $ & \multirow{6}{*}{$ P_r \gg P_{r;11}^{\rm diss} $}  & \multicolumn{1}{| l}{$ l_{mn} \propto E P_r^{-1} $}     & $ \displaystyle \omega_{mn} \propto \frac{m}{\sqrt{m^2 + n^2}}  \sqrt{A}  $  \\
     \vspace{0.01mm} & & \multicolumn{1}{| l}{} & & & \multicolumn{1}{| l}{} & \\
    & &  \multicolumn{1}{| l}{$ H_{mn} \propto E^{-1} P_r^{-1} $} & $ N_{\rm kc} \propto E^{-1/2}  $  & & \multicolumn{1}{| l}{$ H_{mn} \propto E^{-1} P_r^2 $} & $ N_{\rm kc} \propto A^{1/4} E^{-1/2} P_r^{1/2}  $   \\
     \vspace{0.01mm} & & \multicolumn{1}{| l}{} & & & \multicolumn{1}{| l}{} & \\
    & &  \multicolumn{1}{| l}{$ H_{\rm bg} \propto E P_r^{-1}  $} & $ \Xi \propto E^{-2} $ & & \multicolumn{1}{| l}{$ H_{\rm bg} \propto A^{-1} E $} & $ \Xi \propto A E^{-2} P_r^{2} $   \\
     \vspace{0.01mm} & & \multicolumn{1}{| l}{} & & & \multicolumn{1}{| l}{} & \\
         \cline{2-7} 
     \vspace{0.01mm} & & \multicolumn{1}{| l}{} & & & \multicolumn{1}{| l}{} & \\
      & \multirow{6}{*}{$ P_r \ll P_{r;11} $} & \multicolumn{1}{| l}{$ l_{mn} \propto A E P_r^{-1} $} & $ \displaystyle \omega_{mn} \propto \frac{n}{\sqrt{m^2 + n^2}} \cos \theta $ & \multirow{6}{*}{$ P_r \ll P_{r;11}^{\rm diss} $}  & \multicolumn{1}{| l}{$ l_{mn} \propto E P_r^{-1}  $}  & $ \displaystyle \omega_{mn} \propto \frac{m}{\sqrt{m^2 + n^2}}  \sqrt{A} $  \\
     \vspace{0.01mm} & & \multicolumn{1}{| l}{} & & & \multicolumn{1}{| l}{} & \\
    & &  \multicolumn{1}{| l}{$ H_{mn} \propto A^{-2} E^{-1} P_r $} & $ N_{\rm kc} \propto A^{-1/2} E^{-1/2} P_r^{1/2}   $ & & \multicolumn{1}{| l}{$ H_{mn} \propto A^{-1} E^{-1} P_r $} & $ N_{\rm kc} \propto A^{1/4} E^{-1/2} P_r^{1/2}  $   \\
     \vspace{0.01mm} & & \multicolumn{1}{| l}{} & & & \multicolumn{1}{| l}{} & \\
    & & \multicolumn{1}{| l}{$ H_{\rm bg} \propto E P_r^{-1} $} & $ \Xi \propto A^{-2} E^{-2} P_{r}^{2}  $ & & \multicolumn{1}{| l}{$ H_{\rm bg} \propto A^{-2} E P_r^{-1} $} & $ \Xi \propto A E^{-2} P_r^2 $   \\
     \vspace{0.01mm} & & \multicolumn{1}{| l}{} & & & \multicolumn{1}{| l}{} & \\
  \hline
  \hline
\end{tabular}
 \textsf{\caption{\label{scaling_laws} Scaling laws for the properties of the energy dissipated in the different asymptotic regimes. $ P_{r;11}^{\rm diss} $ indicates the transition zone between a dissipation led by viscous friction and a dissipation led by heat diffusion. $ A_{11} $ indicates the transition between tidal inertial and gravity waves.}}
 
 \end{table*}

\section{Conclusion and perspectives}

Through a local approach, we have studied the physics of tidal gravito-inertial waves in stars and in fluid planetary regions. Their viscous and thermal dissipation is one of the main contributor to the secular dynamics of spins and orbits. More precisely, it determines the so-called tidal quality factor $ Q\left(\chi\right) $ intervening in the dynamics \citep{GS1966,ADLPM2014}. This factor decreases when the dissipated tidal kinetic energy increases. Indeed, for an ideal elastic body, it takes very high values and the architecture of the orbital system is weakly affected by tides. At the opposite, if the body is the seat of a strong dissipation, the energy dissipated becomes important, the values of $Q$ are smaller and systems evolve because of tides. This strong link between the secular evolution of a planetary system and tidal dissipation in their components therefore motivates studies to understand its dependence on their internal structure and dynamics.\\

In the case of celestial bodies constituted of fluids, previous works highlighted a complex behavior \citep[e.g.][]{Zahn1975,Webb1980,OL2004,OL2007}. Indeed, liquid and gas layers do not dissipate energy like solid icy and rocky ones. The corresponding quality factor $ Q $ strongly varies as a function of forcing frequencies, fluid parameters (i.e. rotation, stratification and diffusivities), the geometry of the fluid container and boundary conditions and can vary over several orders of magnitude. This has important consequences for the evolution of spins and orbits \cite[e.g.][]{WS1999,ADLPM2014}. In this study, the dissipation caused by viscous friction and thermal diffusion has been estimated in a local Cartesian section of a rotating body in order to understand and quantify these dependences in the whole domain of parameters relevant in stars and planets; in this framework, the Cartesian fluid box is complementary with global models \citep{O2005,JO2014,Barker2013,Barker2014}. The equations of momentum, mass conservation and heat transport are solved analytically with solutions expanded in periodic Fourier series. It yields the following conclusions: \\
 
 \begin{itemize}
 
  \item[$ \bullet $] Tidal dissipation resulting from viscous friction and thermal diffusion is highly resonant. Gravito-inertial waves are excited by tides at identified frequencies belonging to the interval $ \left[ \chi_{\rm inf}  , \chi_{\rm sup}  \right] $, where $ \chi_{\rm inf} $ and $ \chi_{\rm sup} $ are in a first approximation the inertial ($ 2 \Omega $) and the Brunt-V\"ais\"al\"a ($N$) frequencies \citep[for more precise boundaries, see Eq. \ref{omega_GS} and][]{GS2005}. Thus, the fluid behaves like a bandpass filter. Inside the allowed frequency range, a batch of resonances can be excited by tides. Outside, the dissipation is weaker and varies smoothly with the forcing frequency that corresponds to the so-called equilibrium or non-wave like tide \citep{RMZ2012,Ogilvie2013}. In the resonant regime, a typical dissipation spectrum presents an organized structure of peaks which have different sizes and properties. These laters depend on the one hand on the spectral form of the forcing, in particular the amplitude of the harmonics which explains the fractal pattern of the studied case. On the other hand, they are narrowly bound to the fluid parameters: rotation ($ \Omega $), stratification ($ N $), thermal diffusivity ($ \kappa $) and viscosity ($ \nu $). \\
  
 \item[$ \bullet $] There are four asymptotic regimes for tidal waves, characterized by the frequency ratio $ A $ and the Prandtl number $ P_r $, and the positions of the transition borders are expressed as functions of the fluid parameters. The case when $ A \ll 1 $ corresponds to quasi-inertial waves. They are mainly driven by the Coriolis acceleration even when $ N^2 > 0 $. If $ A \gg 1 $, then quasi-gravity waves are excited, the Archimedean force predominating on the Coriolis term. Besides, depending on the value of the Prandtl number $ P_r $, the dissipation is driven either by viscous friction ($ P_r \gg P_{r;11} $) or by thermal diffusion ($ P_r \ll P_{r;11} $).\\
 
  \item[$ \bullet $] For each of these asymptotic behaviors, all the properties of the dissipation spectrum can be expressed as functions of the fluid parameters: the positions, widths at mid-height, heights and number of resonances, the level of the non-resonant background and the sharpness ratio. These scaling laws, derived analytically from the model, can be used as zero-order approaches to constrain the resulting rotational and orbital dynamics. In this context, they constitute a tool to explore the physics of dissipation and offer a first basis to validate the results obtained from direct numerical simulations and global models. For instance, they can be used to compute scaling laws for the turbulent convective dissipation of tidal inertial waves in the convective envelope of stars and giant planets as a function of their rotation \citep{MADGLP2014}.\\  
 \end{itemize}
 
In a near future, the method applied in this work will be extended to other dissipative systems with for example differential rotation \citep{BR2013} or stratified convection \citep{Leconte2012}. Moreover, we will include magnetic field and take into account the corresponding Ohmic dissipation in addition to viscous and thermal diffusions \citep[e.g.][]{Barker2014}. Tidal waves will become magneto-gravito-inertial waves \citep{MdB2011}. This will introduce new fluid parameters (the Alfv\'en frequency and the Elsasser and magnetic Prandtl numbers), new asymptotic behaviors and the corresponding scaling laws. Finally, non-linear interactions between tidal waves will be studied and characterized \cite[e.g.][]{Galtier2003,Senetal2012,Sen2013,Barker2013}.



\begin{acknowledgements}
The authors are grateful to the referee and the editor for their detailed review which has allowed to improve the paper. This work was supported by the French Programme National de Plan\'etologie (CNRS/INSU), the CNES-CoRoT grant at CEA-Saclay, the "Axe f\'ed\'erateur Etoile" of Paris Observatory Scientific Council, and the International Space Institute (ISSI; team ENCELADE 2.0).
\end{acknowledgements}

\bibliographystyle{aa}  
\bibliography{ADMLP2015} 

\begin{thebibliography}{81}
\expandafter\ifx\csname natexlab\endcsname\relax\def\natexlab#1{#1}\fi

\bibitem[{{Albrecht} {et~al.}(2012){Albrecht}, {Winn}, {Johnson}, {Howard},
  {Marcy}, {Butler}, {Arriagada}, {Crane}, {Shectman}, {Thompson}, {Hirano},
  {Bakos}, \& {Hartman}}]{Albrecht2012}
{Albrecht}, S., {Winn}, J.~N., {Johnson}, J.~A., {et~al.} 2012, \apj, 757, 18

\bibitem[{{Alexander}(1973)}]{Alexander1973}
{Alexander}, M.~E. 1973, \apss, 23, 459

\bibitem[{{Auclair-Desrotour} {et~al.}(2014){Auclair-Desrotour}, {Le
  Poncin-Lafitte}, \& {Mathis}}]{ADLPM2014}
{Auclair-Desrotour}, P., {Le Poncin-Lafitte}, C., \& {Mathis}, S. 2014, \aap,
  561, L7

\bibitem[{{Barker} \& {Lithwick}(2013)}]{Barker2013}
{Barker}, A.~J. \& {Lithwick}, Y. 2013, \mnras, 435, 3614

\bibitem[{{Barker} \& {Lithwick}(2014)}]{Barker2014}
{Barker}, A.~J. \& {Lithwick}, Y. 2014, \mnras, 437, 305

\bibitem[{{Baruteau} \& {Rieutord}(2013)}]{BR2013}
{Baruteau}, C. \& {Rieutord}, M. 2013, Journal of Fluid Mechanics, 719, 47

\bibitem[{{Berthomieu} {et~al.}(1978){Berthomieu}, {Gonczi}, {Graff},
  {Provost}, \& {Rocca}}]{Berthomieuetal1978}
{Berthomieu}, G., {Gonczi}, G., {Graff}, P., {Provost}, J., \& {Rocca}, A.
  1978, \aap, 70, 597

\bibitem[{{Bolmont} {et~al.}(2012){Bolmont}, {Raymond}, {Leconte}, \&
  {Matt}}]{Bolmont2012}
{Bolmont}, E., {Raymond}, S.~N., {Leconte}, J., \& {Matt}, S.~P. 2012, \aap,
  544, A124

\bibitem[{{Braviner} \& {Ogilvie}(2015)}]{BO2015}
{Braviner}, H.~J. \& {Ogilvie}, G.~I. 2015, \mnras, 447, 1141

\bibitem[{{C{\'e}bron} {et~al.}(2013){C{\'e}bron}, {Bars}, {Gal}, {Moutou},
  {Leconte}, \& {Sauret}}]{Cebron2013}
{C{\'e}bron}, D., {Bars}, M.~L., {Gal}, P.~L., {et~al.} 2013, \icarus, 226,
  1642

\bibitem[{{Cebron} {et~al.}(2012){Cebron}, {Le Bars}, {Moutou}, \& {Le
  Gal}}]{Cebron2012}
{Cebron}, D., {Le Bars}, M., {Moutou}, C., \& {Le Gal}, P. 2012, \aap, 539, A78

\bibitem[{{Correia} {et~al.}(2014){Correia}, {Bou{\'e}}, {Laskar}, \&
  {Rodr{\'{\i}}guez}}]{Correiaetal2014}
{Correia}, A.~C.~M., {Bou{\'e}}, G., {Laskar}, J., \& {Rodr{\'{\i}}guez}, A.
  2014, \aap, 571, A50

\bibitem[{{Correia} \& {Laskar}(2003)}]{Correia2003}
{Correia}, A.~C.~M. \& {Laskar}, J. 2003, Journal of Geophysical Research
  (Planets), 108, 5123

\bibitem[{{Correia} {et~al.}(2008){Correia}, {Levrard}, \& {Laskar}}]{CLL2008}
{Correia}, A.~C.~M., {Levrard}, B., \& {Laskar}, J. 2008, \aap, 488, L63

\bibitem[{{Cowling}(1941)}]{Cowling1941}
{Cowling}, T.~G. 1941, \mnras, 101, 367

\bibitem[{{Darwin}(1879)}]{Darwin1879}
{Darwin}, G.~H. 1879, Royal Society of London Proceedings Series I, 30, 1

\bibitem[{{Efroimsky}(2012)}]{Efroimsky2012}
{Efroimsky}, M. 2012, \apj, 746, 150

\bibitem[{{Efroimsky} \& {Lainey}(2007)}]{EL2007}
{Efroimsky}, M. \& {Lainey}, V. 2007, Journal of Geophysical Research
  (Planets), 112, 12003

\bibitem[{{Egbert} \& {Ray}(2000)}]{EgbertRay2000}
{Egbert}, G.~D. \& {Ray}, R.~D. 2000, \nat, 405, 775

\bibitem[{{Egbert} \& {Ray}(2001)}]{EgbertRay2001}
{Egbert}, G.~D. \& {Ray}, R.~D. 2001, \jgr, 106, 22475

\bibitem[{{Emelyanov} \& {Nikonchuk}(2013)}]{EN2013}
{Emelyanov}, N.~V. \& {Nikonchuk}, D.~V. 2013, \mnras, 436, 3668

\bibitem[{{Fabrycky} {et~al.}(2012){Fabrycky}, {Ford}, {Steffen}, {Rowe},
  {Carter}, {Moorhead}, {Batalha}, {Borucki}, {Bryson}, {Buchhave},
  {Christiansen}, {Ciardi}, {Cochran}, {Endl}, {Fanelli}, {Fischer}, {Fressin},
  {Geary}, {Haas}, {Hall}, {Holman}, {Jenkins}, {Koch}, {Latham}, {Li},
  {Lissauer}, {Lucas}, {Marcy}, {Mazeh}, {McCauliff}, {Quinn}, {Ragozzine},
  {Sasselov}, \& {Shporer}}]{Fabryckyb2012}
{Fabrycky}, D.~C., {Ford}, E.~B., {Steffen}, J.~H., {et~al.} 2012, \apj, 750,
  114

\bibitem[{{Ferraz-Mello}(2013)}]{FM2013}
{Ferraz-Mello}, S. 2013, Celestial Mechanics and Dynamical Astronomy, 116, 109

\bibitem[{{Galtier}(2003)}]{Galtier2003}
{Galtier}, S. 2003, \pre, 68, 015301

\bibitem[{{Gerkema} \& {Shrira}(2005{\natexlab{a}})}]{GS2005}
{Gerkema}, T. \& {Shrira}, V.~I. 2005{\natexlab{a}}, Journal of Fluid
  Mechanics, 529, 195

\bibitem[{{Gerkema} \& {Shrira}(2005{\natexlab{b}})}]{GS2005b}
{Gerkema}, T. \& {Shrira}, V.~I. 2005{\natexlab{b}}, Journal of Geophysical
  Research (Oceans), 110, 1003

\bibitem[{{Goldreich} \& {Keeley}(1977)}]{GK1977}
{Goldreich}, P. \& {Keeley}, D.~A. 1977, \apj, 212, 243

\bibitem[{{Goldreich} \& {Soter}(1966)}]{GS1966}
{Goldreich}, P. \& {Soter}, S. 1966, \icarus, 5, 375

\bibitem[{{Greenberg}(2009)}]{Greenberg2009}
{Greenberg}, R. 2009, \apjl, 698, L42

\bibitem[{{Henning} {et~al.}(2009){Henning}, {O'Connell}, \&
  {Sasselov}}]{Henning2009}
{Henning}, W.~G., {O'Connell}, R.~J., \& {Sasselov}, D.~D. 2009, \apj, 707,
  1000

\bibitem[{{Hut}(1981)}]{Hut1981}
{Hut}, P. 1981, \aap, 99, 126

\bibitem[{{Jacobson}(2010)}]{Jacobson2010}
{Jacobson}, R.~A. 2010, \aj, 139, 668

\bibitem[{{Jouve} \& {Ogilvie}(2014)}]{JO2014}
{Jouve}, L. \& {Ogilvie}, G.~I. 2014, Journal of Fluid Mechanics, 745, 223

\bibitem[{{Kaula}(1964)}]{Kaula1964}
{Kaula}, W.~M. 1964, Reviews of Geophysics and Space Physics, 2, 661

\bibitem[{Kelvin(1863)}]{Kelvin1863}
Kelvin, L. 1863, Phil. Trans. Roy. Soc. London, Treatise on Natural Philosophy,
  2, 837

\bibitem[{{Konopliv} {et~al.}(2011){Konopliv}, {Asmar}, {Folkner}, {Karatekin},
  {Nunes}, {Smrekar}, {Yoder}, \& {Zuber}}]{K2011}
{Konopliv}, A.~S., {Asmar}, S.~W., {Folkner}, W.~M., {et~al.} 2011, \icarus,
  211, 401

\bibitem[{{Konopliv} \& {Yoder}(1996)}]{KY1996}
{Konopliv}, A.~S. \& {Yoder}, C.~F. 1996, \grl, 23, 1857

\bibitem[{{Lainey} {et~al.}(2009){Lainey}, {Arlot}, {Karatekin}, \& {van
  Hoolst}}]{Lainey2009}
{Lainey}, V., {Arlot}, J.-E., {Karatekin}, {\"O}., \& {van Hoolst}, T. 2009,
  \nat, 459, 957

\bibitem[{{Lainey} {et~al.}(2007){Lainey}, {Dehant}, \&
  {P{\"a}tzold}}]{Lainey2007}
{Lainey}, V., {Dehant}, V., \& {P{\"a}tzold}, M. 2007, \aap, 465, 1075

\bibitem[{{Lainey} {et~al.}(2012){Lainey}, {Karatekin}, {Desmars}, {Charnoz},
  {Arlot}, {Emelyanov}, {Le Poncin-Lafitte}, {Mathis}, {Remus}, {Tobie}, \&
  {Zahn}}]{Lainey2012}
{Lainey}, V., {Karatekin}, {\"O}., {Desmars}, J., {et~al.} 2012, \apj, 752, 14

\bibitem[{{Laskar} {et~al.}(2012){Laskar}, {Bou{\'e}}, \&
  {Correia}}]{Laskar2012}
{Laskar}, J., {Bou{\'e}}, G., \& {Correia}, A.~C.~M. 2012, \aap, 538, A105

\bibitem[{{Leconte} \& {Chabrier}(2012)}]{Leconte2012}
{Leconte}, J. \& {Chabrier}, G. 2012, \aap, 540, A20

\bibitem[{{Leconte} {et~al.}(2010){Leconte}, {Chabrier}, {Baraffe}, \&
  {Levrard}}]{Leconte2010}
{Leconte}, J., {Chabrier}, G., {Baraffe}, I., \& {Levrard}, B. 2010, \aap, 516,
  A64

\bibitem[{{Love}(1911)}]{Love1911}
{Love}, A.~E.~H. 1911, {Some Problems of Geodynamics} (Cambridge University
  Press)

\bibitem[{{MacDonald}(1964)}]{MacDonald1964}
{MacDonald}, G.~J.~F. 1964, Reviews of Geophysics and Space Physics, 2, 467

\bibitem[{{Mathis} {et~al.}(2014{\natexlab{a}}){Mathis}, {Auclair-Desrotour},
  {Guenel}, \& {Le Poncin-Lafitte}}]{MADGLP2014}
{Mathis}, S., {Auclair-Desrotour}, P., {Guenel}, M., \& {Le Poncin-Lafitte}, C.
  2014{\natexlab{a}}, in SF2A-2014: Proceedings of the Annual meeting of the
  French Society of Astronomy and Astrophysics, ed. J.~{Ballet}, F.~{Martins},
  F.~{Bournaud}, R.~{Monier}, \& C.~{Reyl{\'e}}, 251--256

\bibitem[{{Mathis} \& {de Brye}(2011)}]{MdB2011}
{Mathis}, S. \& {de Brye}, N. 2011, \aap, 526, A65

\bibitem[{{Mathis} {et~al.}(2014{\natexlab{b}}){Mathis}, {Neiner}, \& {Tran
  Minh}}]{MNT2014}
{Mathis}, S., {Neiner}, C., \& {Tran Minh}, N. 2014{\natexlab{b}}, \aap, 565,
  A47

\bibitem[{{Mathis} \& {Remus}(2013)}]{MR2013}
{Mathis}, S. \& {Remus}, F. 2013, in Lecture Notes in Physics, Berlin Springer
  Verlag, Vol. 857, Lecture Notes in Physics, Berlin Springer Verlag, ed. J.-P.
  {Rozelot} \& C.~. {Neiner}, 111--147

\bibitem[{{Mayor} \& {Queloz}(1995)}]{MQ1995}
{Mayor}, M. \& {Queloz}, D. 1995, \nat, 378, 355

\bibitem[{{Ogilvie}(2005)}]{O2005}
{Ogilvie}, G.~I. 2005, Journal of Fluid Mechanics, 543, 19

\bibitem[{{Ogilvie}(2013)}]{Ogilvie2013}
{Ogilvie}, G.~I. 2013, \mnras, 429, 613

\bibitem[{{Ogilvie}(2014)}]{Ogilvie2014}
{Ogilvie}, G.~I. 2014, \araa, 52, 171

\bibitem[{{Ogilvie} \& {Lesur}(2012)}]{OL2012}
{Ogilvie}, G.~I. \& {Lesur}, G. 2012, \mnras, 422, 1975

\bibitem[{{Ogilvie} \& {Lin}(2004)}]{OL2004}
{Ogilvie}, G.~I. \& {Lin}, D.~N.~C. 2004, \apj, 610, 477

\bibitem[{{Ogilvie} \& {Lin}(2007)}]{OL2007}
{Ogilvie}, G.~I. \& {Lin}, D.~N.~C. 2007, \apj, 661, 1180

\bibitem[{{Pedlosky}(1982)}]{Pedlosky1982}
{Pedlosky}, J. 1982, {Geophysical fluid dynamics}

\bibitem[{{Perryman}(2011)}]{Perryman2011}
{Perryman}, M. 2011, {The Exoplanet Handbook}

\bibitem[{{Provost} {et~al.}(1981){Provost}, {Berthomieu}, \&
  {Rocca}}]{Provostetal1981}
{Provost}, J., {Berthomieu}, G., \& {Rocca}, A. 1981, \aap, 94, 126

\bibitem[{{Ray} {et~al.}(2001){Ray}, {Eanes}, \& {Lemoine}}]{Rayetal2001}
{Ray}, R.~D., {Eanes}, R.~J., \& {Lemoine}, F.~G. 2001, Geophysical Journal
  International, 144, 471

\bibitem[{{Remus} {et~al.}(2012{\natexlab{a}}){Remus}, {Mathis}, \&
  {Zahn}}]{RMZ2012}
{Remus}, F., {Mathis}, S., \& {Zahn}, J.-P. 2012{\natexlab{a}}, \aap, 544, A132

\bibitem[{{Remus} {et~al.}(2012{\natexlab{b}}){Remus}, {Mathis}, {Zahn}, \&
  {Lainey}}]{Remusanelastic2012}
{Remus}, F., {Mathis}, S., {Zahn}, J.-P., \& {Lainey}, V. 2012{\natexlab{b}},
  \aap, 541, A165

\bibitem[{{Rieutord} \& {Valdettaro}(2010)}]{RV2010}
{Rieutord}, M. \& {Valdettaro}, L. 2010, Journal of Fluid Mechanics, 643, 363

\bibitem[{{Sen}(2013)}]{Sen2013}
{Sen}, A. 2013, ArXiv e-prints

\bibitem[{{Sen} {et~al.}(2012){Sen}, {Mininni}, {Rosenberg}, \&
  {Pouquet}}]{Senetal2012}
{Sen}, A., {Mininni}, P.~D., {Rosenberg}, D., \& {Pouquet}, A. 2012, \pre, 86,
  036319

\bibitem[{Singer(1968)}]{Singer1968}
Singer, S. 1968, Geophysical Journal International, 15, 205

\bibitem[{{Tobie} {et~al.}(2005){Tobie}, {Mocquet}, \& {Sotin}}]{Tobie2005}
{Tobie}, G., {Mocquet}, A., \& {Sotin}, C. 2005, \icarus, 177, 534

\bibitem[{{Valsecchi} \& {Rasio}(2014)}]{VR2014}
{Valsecchi}, F. \& {Rasio}, F.~A. 2014, \apj, 786, 102

\bibitem[{{Webb}(1980)}]{Webb1980}
{Webb}, D.~J. 1980, Geophysical Journal, 61, 573

\bibitem[{{Webb}(1982)}]{Webb1982}
{Webb}, D.~J. 1982, Geophysical Journal, 70, 261

\bibitem[{{Williams} {et~al.}(2014){Williams}, {Konopliv}, {Boggs}, {Park},
  {Yuan}, {Lemoine}, {Goossens}, {Mazarico}, {Nimmo}, {Weber}, {Asmar},
  {Melosh}, {Neumann}, {Phillips}, {Smith}, {Solomon}, {Watkins}, {Wieczorek},
  {Andrews-Hanna}, {Head}, {Kiefer}, {Matsuyama}, {McGovern}, {Taylor}, \&
  {Zuber}}]{Williamsetal2014}
{Williams}, J.~G., {Konopliv}, A.~S., {Boggs}, D.~H., {et~al.} 2014, Journal of
  Geophysical Research (Planets), 119, 1546

\bibitem[{{Witte} \& {Savonije}(1999)}]{WS1999}
{Witte}, M.~G. \& {Savonije}, G.~J. 1999, \aap, 350, 129

\bibitem[{{Witte} \& {Savonije}(2001)}]{WS2001}
{Witte}, M.~G. \& {Savonije}, G.~J. 2001, \aap, 366, 840

\bibitem[{{Witte} \& {Savonije}(2002)}]{WS2002}
{Witte}, M.~G. \& {Savonije}, G.~J. 2002, \aap, 386, 222

\bibitem[{{Wu}(2005)}]{Wu2005}
{Wu}, Y. 2005, \apj, 635, 688

\bibitem[{{Zahn}(1966{\natexlab{a}})}]{Zahn1966a}
{Zahn}, J.~P. 1966{\natexlab{a}}, Annales d'Astrophysique, 29, 313

\bibitem[{{Zahn}(1966{\natexlab{b}})}]{Zahn1966b}
{Zahn}, J.~P. 1966{\natexlab{b}}, Annales d'Astrophysique, 29, 489

\bibitem[{{Zahn}(1966{\natexlab{c}})}]{Zahn1966c}
{Zahn}, J.~P. 1966{\natexlab{c}}, Annales d'Astrophysique, 29, 565

\bibitem[{{Zahn}(1975)}]{Zahn1975}
{Zahn}, J.-P. 1975, \aap, 41, 329

\bibitem[{{Zahn}(1977)}]{Zahn1977}
{Zahn}, J.-P. 1977, \aap, 57, 383

\bibitem[{{Zahn}(1989)}]{Zahn1989a}
{Zahn}, J.-P. 1989, \aap, 220, 112

\end{thebibliography}

\begin{appendix}

\section{Harmonic highest degree $ k_c $}

The expression of the highest harmonic degree, given in Eq.~(\ref{kc2}), is plotted as a function of the Ekman number and of the thermal diffusivity for various regimes from $ A \ll 1 $ to $ A \gg 1 $. It corresponds to the upper bound of harmonic degree that is sufficient to represent all the resonances of the spectrum.

\begin{figure*}[ht!]
 \centering
 \includegraphics[width=0.48\textwidth,clip]{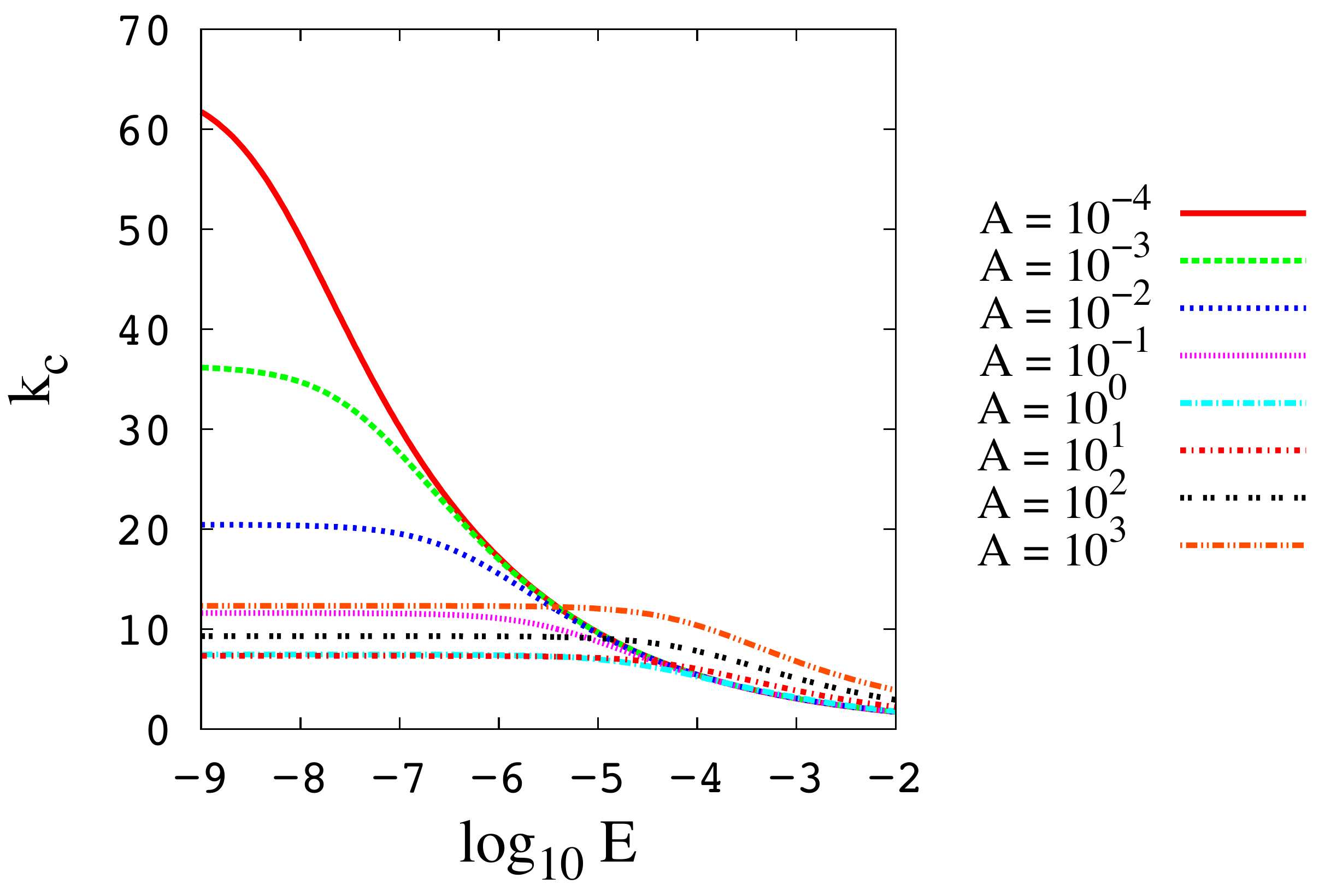}%
 \includegraphics[width=0.48\textwidth,clip]{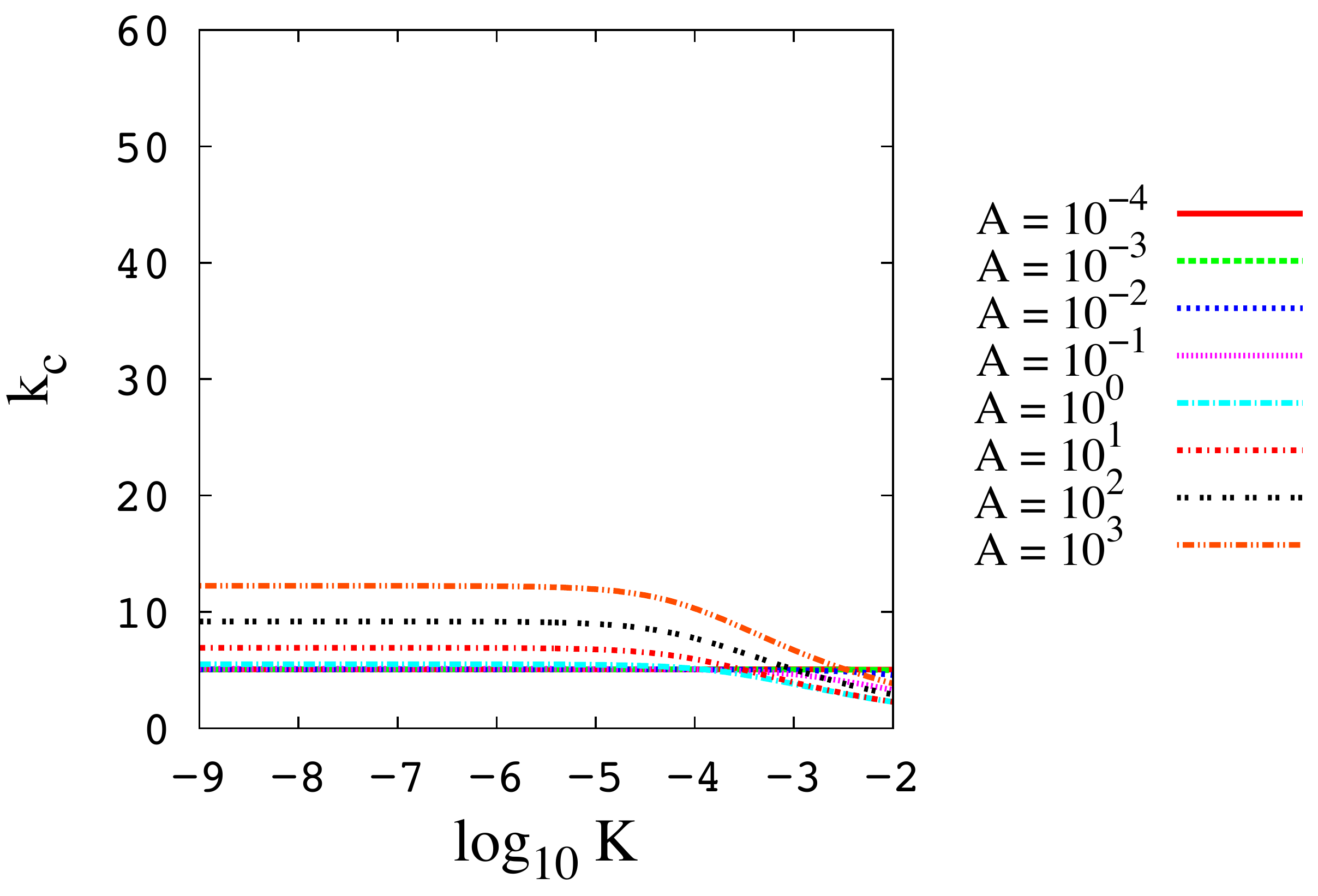}   
  \textsf{ \caption{\label{fig:kc_EA} Rank of the highest harmonic degree $ k_c $ as a function of $ E $ (with $ K = 10^{-4} $) and $ K $ (with $ E=10^{-4} $) for different values of $ A $ (in logarithmic scales). {\bf Left:} $ k_c - E $. {\bf Right:} $ k_c - K $.}}
\end{figure*}

\section{Relative differences $ \eta_{l} $ and $ \eta_{H}$}

\begin{figure*}[ht!]
 \centering
 \includegraphics[width=0.475\textwidth,clip]{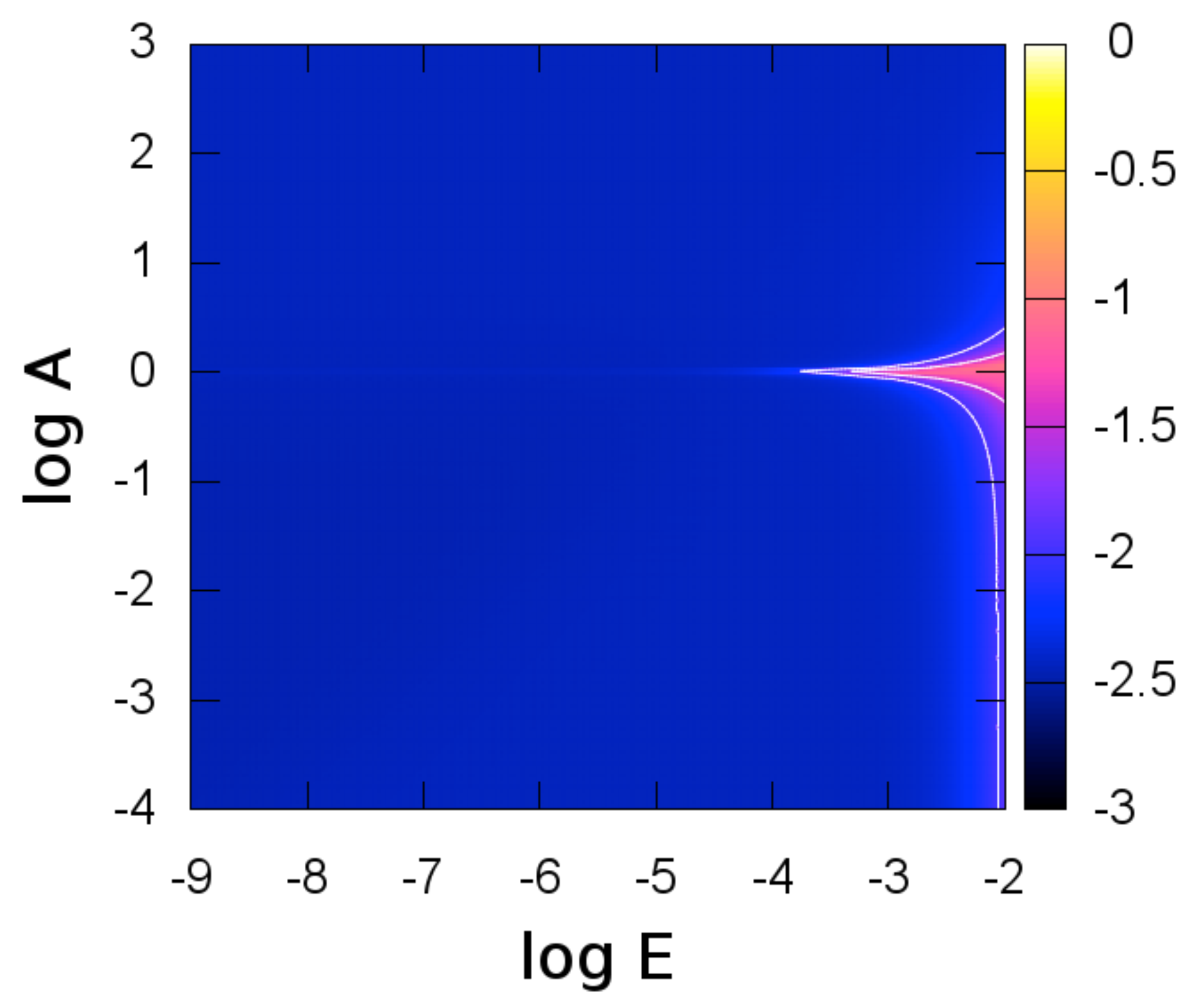}%
 \includegraphics[width=0.475\textwidth,clip]{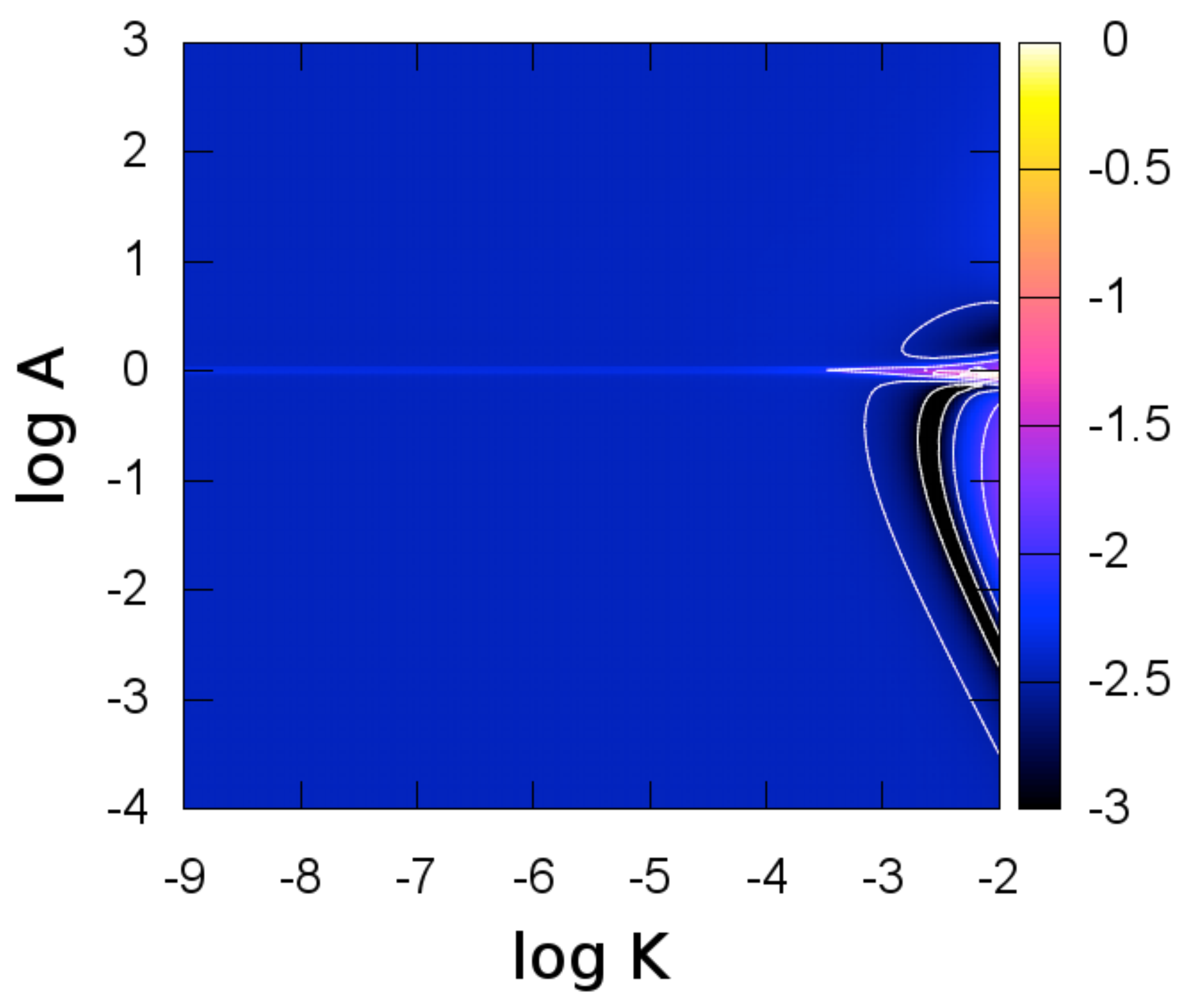} 
 \includegraphics[width=0.475\textwidth,clip]{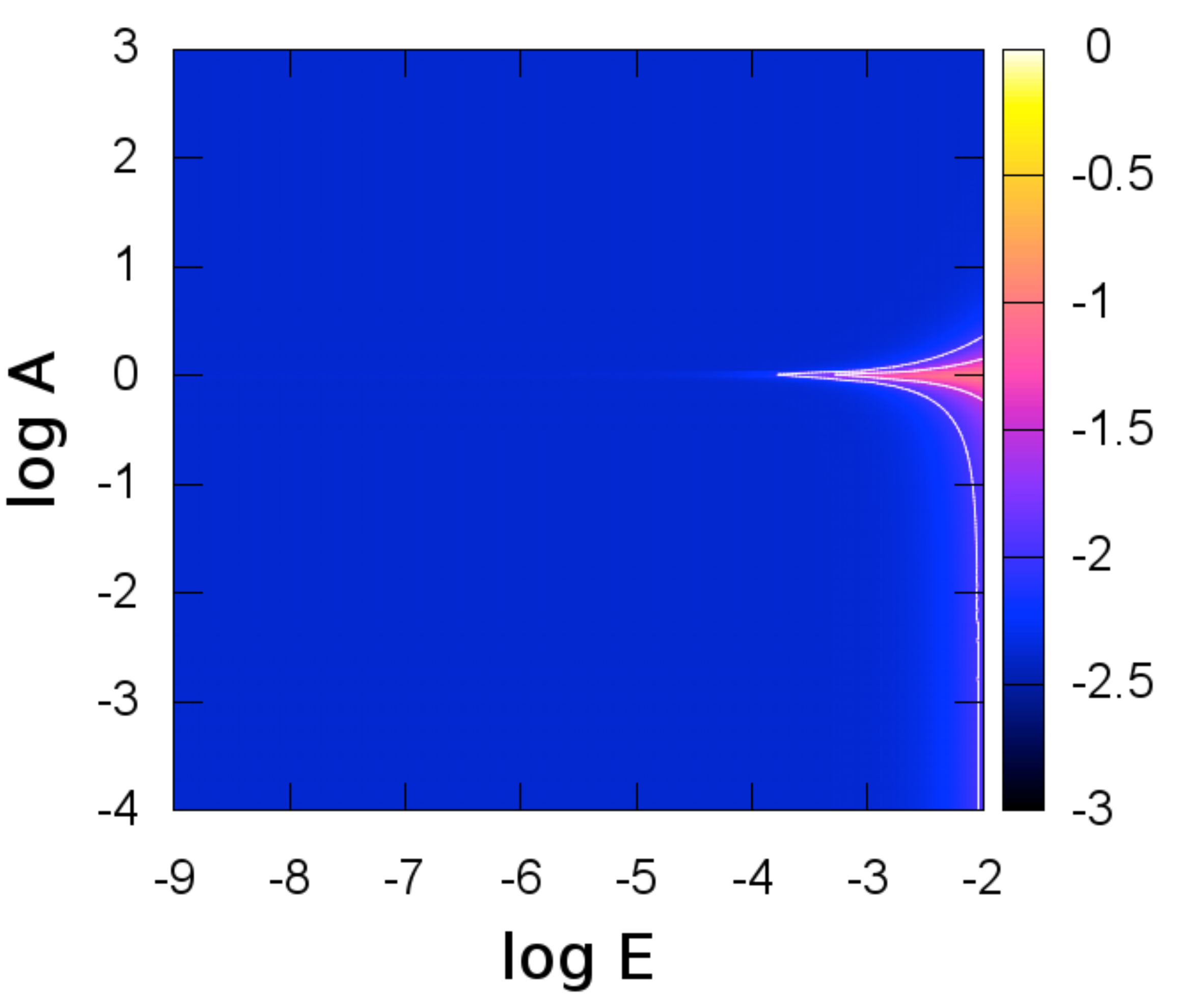}
 \includegraphics[width=0.475\textwidth,clip]{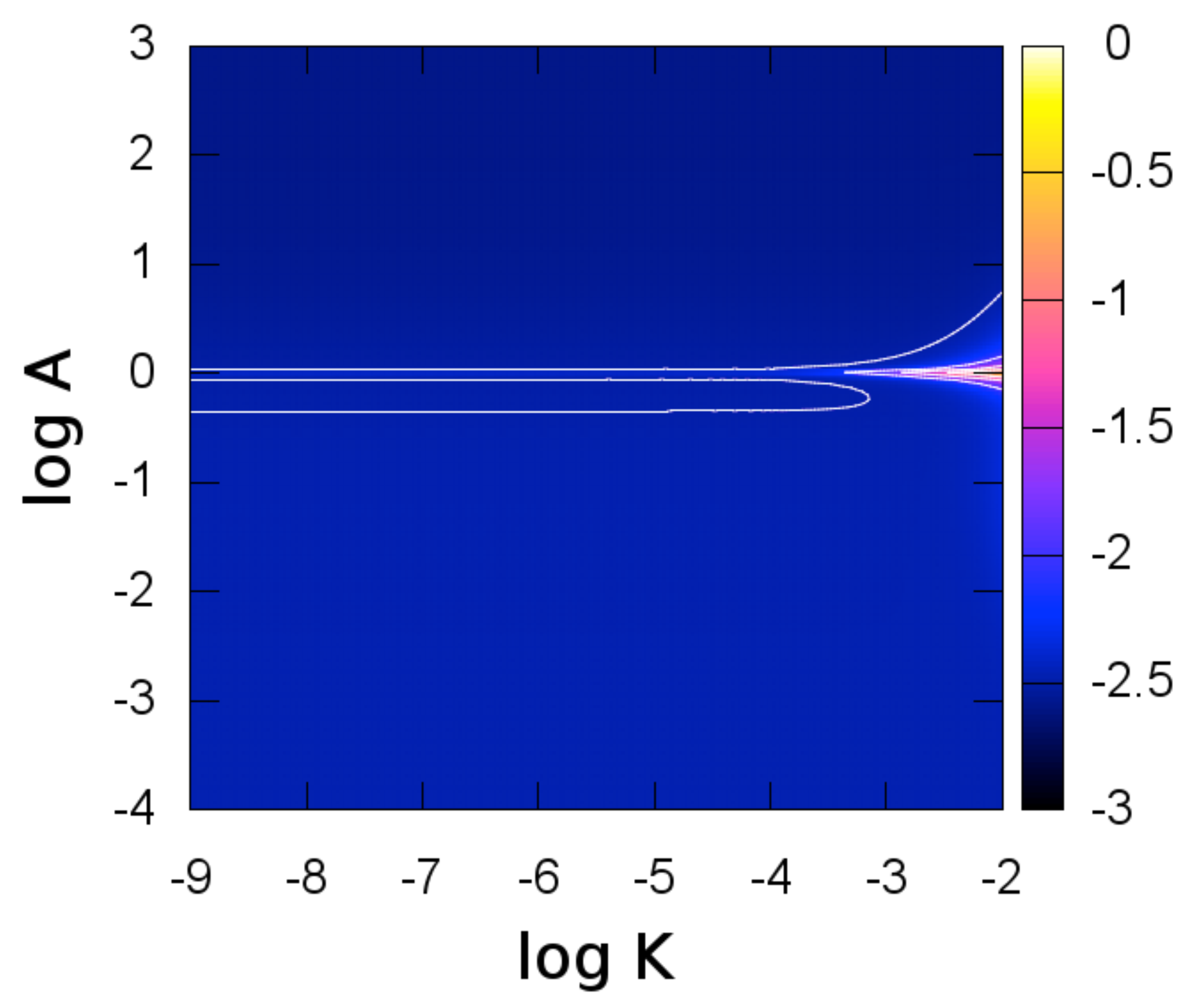}     
  \textsf{ \caption{\label{fig:ecarts} $ \eta_{l} $ and $ \eta_{H} $ as a function of $ A $, $ E $ and $ K $ in logarithm scales. White contours highlight critical zones where the quasi-adiabatic assumption ($ \left\{E,K\right\} \ll \left\{\sqrt{A},\cos \theta\right\} $) or the asymptotic condition ($ A \neq A_{11} $) are not satisfied. {\bf Top left:} $ \eta_{l} - \left( E,A \right) $. {\bf Top right:}  $ \eta_{l} - \left( K,A \right) $. {\bf Bottom left:} $ \eta_{H} - \left( E ,A  \right) $. {\bf Bottom right:} $ \eta_{H} - \left( K ,A  \right) $.}}
\end{figure*}

The results provided by analytical expressions of Eqs. (\ref{lmn}) and (\ref{Hmn_visc}) have been compared to the ones given by the complete formula of the energy dissipated by viscous friction (Eq. \ref{energy_visc}). The color maps of Fig. \ref{fig:ecarts} correspond to the computation of the width at mid-height $ l_{11} $ and height $ H_{11}^{\rm visc} $ of the main resonance, summarized in Fig. \ref{fig:lH_EA}. The relative differences $ \eta_{l} $ and $ \eta_H $ are calculated using the expressions:

\begin{equation}
\begin{array}{ccc}
  \eta_{l} = \displaystyle \frac{ \left| l_{\rm ana} - l_{\rm th}  \right| }{l_{\rm th}} & \mbox{and} & 
  \eta_{H} = \displaystyle \frac{\left| H_{\rm ana}^{\rm visc} - H_{\rm th}^{\rm visc} \right|}{H_{\rm th}^{\rm visc}},\\
\end{array}
\end{equation}

$ l_{\rm ana} $ and $ H_{\rm ana}^{\rm visc} $ coming from Eqs. (\ref{lmn}) and (\ref{Hmn_visc}), and $ l_{\rm th} $ and $ H_{\rm th}^{\rm visc} $ being computed with Eq. (\ref{energy_visc}). The plots highlight the asymptotic domains, in dark blue, where the analytical formulae constitute a relevant approximation for the width and height of a resonance, and the critical transition zones where they cannot be applied. The light blue horizontal line corresponds to $ A \sim A_{11} $, the hyper-resonant case. The color gradient in the regions near $ E \sim 10^{-2} $ and $ K \sim 10^{-2} $ indicates the values of $ A $, $ E $ and $ K $ for which the condition of the quasi-adiabatic approximation (Eq. \ref{condition_xi}) is not satisfied. The same can be done with the energy dissipated by thermal diffusion.

\end{appendix}

\end{document}